\documentclass{aa}  

\usepackage{graphicx}
\usepackage{txfonts}
\usepackage{longtable}
\usepackage[flushleft]{threeparttable}
\usepackage{threeparttablex}
\usepackage{afterpage}
\usepackage{enumitem}

\usepackage{breqn}
\usepackage{pdflscape}
\usepackage{natbib,twoopt}
\usepackage[breaklinks=true]{hyperref} 
\bibpunct{(}{)}{;}{a}{}{,} 
\makeatletter
\newcommandtwoopt{\citeads}[3][][]{\href{http://adsabs.harvard.edu/abs/#3}%
{\def\hyper@linkstart##1##2{}%
\let\hyper@linkend\@empty\citealp[#1][#2]{#3}}}
\newcommandtwoopt{\citepads}[3][][]{\href{http://adsabs.harvard.edu/abs/#3}%
{\def\hyper@linkstart##1##2{}%
\let\hyper@linkend\@empty\citep[#1][#2]{#3}}}
\newcommandtwoopt{\citetads}[3][][]{\href{http://adsabs.harvard.edu/abs/#3}%
{\def\hyper@linkstart##1##2{}%
\let\hyper@linkend\@empty\citet[#1][#2]{#3}}}
\newcommandtwoopt{\citeyearads}[3][][]%
{\href{http://adsabs.harvard.edu/abs/#3}
{\def\hyper@linkstart##1##2{}%
\let\hyper@linkend\@empty\citeyear[#1][#2]{#3}}}
\renewcommand*\aa@pageof{, page \thepage{} of \pageref*{LastPage}}
\makeatother

\newcommand{\msun}{M$_{\sun}$}

\newcommand\HI{\ion{H}{i}}
\newcommand\hi{\ion{H}{i}}
\newcommand\sofia{{\sc SoFiA}}
\newcommand\kms{km s$^{-1}$}

\newcommand{\mhi}{${\rm M_{\ion{H}{i}}}$}

\usepackage{xcolor}
\usepackage{ulem}

\begin{document}

   \title{HALOGAS: Strong Constraints on the Neutral Gas  Reservoir and Accretion Rate in Nearby Spiral Galaxies.  }
\titlerunning{HALOGAS: Strong Constraints on the Neutral Gas Reservoir and Accretion Rate.}

  \author{   P. Kamphuis \inst{1} \and 
  	      E. J\"utte \inst{1} \and 
  	      G. H. Heald \inst{2} \and 
          N. Herrera Ruiz \inst{1} \and 
          G. I. G. J{\'o}zsa \inst{3,4,5} \and 
          W. J. G. de Blok \inst{6,7,8} \and
          P. Serra \inst{9} \and 
          A. Marasco \inst{10} \and 
          R.-J. Dettmar \inst{1} \and
          N. M. Pingel \inst{11} \and
          T. Oosterloo \inst{6,7} \and
          R. J. Rand \inst{12} \and
          R. A. M. Walterbos \inst{13} \and
          J. M. van der Hulst \inst{7}
        }
\offprints{P. Kamphuis, \email{peter.kamphuis@astro.rub.de}} 
\institute{Ruhr University Bochum, Faculty of Physics and Astronomy, Astronomical Institute, 44780 Bochum, Germany \and
     CSIRO Astronomy and Space Science, PO Box 1130, Bentley WA 6102, Australia \and 
     South African Radio Astronomy Observatory, 2 Fir Street, Black River Park, Observatory, Cape Town, 7925, South Africa \and
     Department of Physics and Electronics, Rhodes University, PO Box 94, Makhanda, 6140, South Africa\and
     Max-Planck-Institut f\"ur Radioastronomie, Auf dem H\"ugel 69, D-53121 Bonn, Germany \and
     ASTRON, Netherlands Institute for Radio Astronomy, Oude Hoogeveensedijk 4, 7991 PD Dwingeloo, The Netherlands \and
     Kapteyn Astronomical Institute, University of Groningen, Landleven 12, 9747 AD Groningen, The Netherlands \and
     Department of Astronomy, University of Cape Town, Private Bag X3, Rondebosch 7701, South Africa \and
     INAF - Osservatorio Astronomico di Cagliari, Via della Scienza 5, I-09047 Selargius (CA), Italy \and
     INAF - Osservatorio Astrofisico di Arcetri, Largo E. Fermi 5, 50127 Firenze, Italy \and Research School of Astronomy and Astrophysics, The Australian National University, Canberra, ACT 2611, Australia \and
     Department of Physics and Astronomy, MSC07 4220, 1 University of New Mexico, Albuquerque NM 87131, U.S.A. \and 
     Department of Astronomy, New Mexico State University, Las Cruces, NM 88001, U.S.A. 
           }

   \date{}

  \abstract
   {Galaxies in the local Universe are thought to require ongoing replenishment of their gas reservoir in order to maintain the observed star formation rates. Cosmological simulations predict that such accretion can occur in both a dynamically hot and cold mode, depending on the redshift, halo mass and the details of the included feedback processes. However, until now observational evidence of the accretion required to match the observed star formation histories is lacking.}
   {Within the framework of the Hydrogen Accretion in LOcal GalaxieS (HALOGAS) survey, this paper attempts to determine whether galaxies in the local Universe possess a significant reservoir of cold neutral gas and what would be the accretion rates derived from such reservoirs. Additionally, with this moderately sized sample we can start to investigate whether the observed accretion is connected to intrinsic  properties of the galaxies like Hubble type, star formation rate or environment.}
   {We search the vicinity of 22 nearby galaxies for isolated \hi\ clouds or distinct streams, that are not yet connected to the galaxy's disk, in a systematic and automated manner. The HALOGAS observations were carried out with the Westerbork Synthesis Radio Telescope  and represent one of the most sensitive and detailed \hi\ surveys to date. These observations typically reach column density sensitivities of $\sim 10^{19}\,\rm{cm}^{-2}$ over a 20 \kms width. }
   {We find 14 secure \hi\ cloud candidates  without an observed optical counterpart in the entire HALOGAS sample. These cloud candidates appear to be analogues to the most massive clouds detected in the extensive cloud distributions around the Milky Way and M\,31. However, on average their numbers seem significantly reduced compared to these galaxies. Within the framework of cold accretion, we constrain upper limits for \hi\ accretion in the local Universe. The average \hi\ mass currently observed in a state suggestive of accretion onto the galaxies amounts to a rate of 0.05 \msun yr$^{-1}$ with a stringent upper limit of 0.22  \msun yr$^{-1}$, confirming previous estimates. This is much lower than the average star formation rate in this sample. Our best estimate, based on the Green Bank Telescope detection limits of several galaxies in the sample, suggests that at most another 0.04  \msun yr$^{-1}$ of neutral hydrogen could be accreted from clouds and streams that remain undetected.}
   { These results show that in nearby galaxies neutral hydrogen is not being accreted at the same rate as stars are currently being formed. Our study can not exclude that other forms of gas accretion, such as those provided by direct infall of ionized intergalactic gas or the condensation of coronal gas, triggered by galactic fountain activities, are at work. However,   these observations also do not reveal extended neutral gas reservoirs around most nearby spiral galaxies.}


   \maketitle
\section{Introduction}\label{intro}
It has been known for many years that spiral galaxies need an ongoing gas supply in order to explain several observational properties. Among others, the observed star formation rates (SFRs) and histories \citep{Daddi2007,Bothwell2011}, the chemical composition of the Galactic disk \citep{Chiosi1980} and large-scale disturbances of galactic disks, e.g., lopsidedness or warps \citep{vEymeren2011} are readily explained by a continuous accretion of gas. Further observational evidence for the continuous accretion of cold material, beyond the direct surroundings of the observed disks, comes from the fact that \hi-excess galaxies reside in \hi-overdense regions \citep{Wang2015} and that the \hi\ content in galaxies appears similar to their neighbors, the so-called \hi-conformity \citep{EWang2015}. Furthermore, on even larger scales, the decline of the Universal \hi\ mass density as a function of time is almost negligible compared to the build up of stars over the same period \citep{Putman2017}. This can only be explained with material from the intergalactic medium (IGM) and galaxy halos condensing onto the \hi\ disks \citep{Putman2012}. \\   
\indent The neutral gas accretion rates observed to date have been unable to match the levels required to explain these aforementioned observational properties. In the local Universe, a global \HI\ accretion rate of  0.2 \msun yr$^{-1}$ was estimated by \cite{Sancisi2008} based on observational findings. This accretion rate includes both the accretion of pure gas clouds as well as gas-rich dwarf galaxies and is based on the literature of single target studies.  \cite{DiTeodoro2014} refined this estimate by systematically searching the Westerbork \hi\ Survey of Irregular and Spiral sample \citep[WHISP,][]{vdHulst2002} for companion galaxies that could be accreted. The upper limit for gas accretion derived in this study (0.28 \msun yr$^{-1}$) was several times lower than the average SFR in the sample (1.29 \msun yr$^{-1}$), showing that minor mergers can not supply the required amounts of neutral hydrogen in this simple comparison.\\
\indent Unfortunately, in only a handful of objects signs of accretion have been studied in detail. It has been known since the early 1960's that the Milky Way (MW) halo contains significant amounts of neutral hydrogen in the form of high-velocity clouds (HVCs) \citep{Muller1963,Wakker1997} and the potentially more distant ultra-compact HVCs \citep{Adams2013}. Only in the last decade or so, analogous features have been observed around other nearby spiral galaxies as well, e.g., in M\,31 \citep{Westmeier2008}, NGC\,2403 \citep{Fraternali2002}, NGC\,891 \citep{Oosterloo2007} and NGC\,6946 \citep{Boomsma2008}, which indicates that such large and distinct \hi\ features are not uncommon.   However, typically only a few large-scale clouds or filaments are seen in the surroundings of a single galaxy and the bulk of these reside close to the disk. Thus, from these detailed studies and others, a picture has emerged in which only a fraction of this gas originates from outside the galaxy disks \citep{Fraternali2008, Fraternali2015}. \\
\indent From the theoretical perspective continuous accretion of gas onto the galaxy disks is also predicted, but it is unclear whether the majority of this accretion happens in a cold phase or a hot phase. Especially for lower redshifts (z $<$ 1) the debate is ongoing whether in lower mass halos (M$_{\rm vir} < 10^{11}$ \msun) the dominant mode is the accretion of gas that never passed through a hot phase \citep[i.e, cold mode accretion,][]{Nelson2013, Huang2019} or gas that did \citep[i.e., hot mode accretion,][]{Nelson2015,Schaller2015}. The answer depends crucially on the type of feedback included in the simulations and hence additional observational evidence can help constrain the simulations. In this it is important to note that some authors define the cold phase as gas that is never heated above 2.5$\times10^5$\,K \citep{Huang2019} whereas others define it as gas that has never exceeded the virial temperature of the halo it is being accreted on to \citep{Nelson2013}. Both definitions allow the cold phase to include significant fractions of ionized hydrogen, i.e. gas with temperatures >$10^4$\,K, which complicates a comparison of \hi\ observations to these predictions even further.   Additionally, simulations have not yet reached convergence, i.e various properties of the cold gas depend significantly on the resolution adopted, especially when considering the  circumgalactic medium \citep[CGM,][]{vdVoort2019}.\\
\indent In order to quantify the typical neutral gas accretion rate on a range of  spiral galaxies, we have embarked on the Westerbork Hydrogen Accretion in LOcal GAlaxieS (HALOGAS) survey to investigate general characteristics of gaseous halos and accretion in a significant sample \citep{Heald2011}.  For this purpose, we obtained deep (10 $\times$ 12 hours) \hi\ observations using the Westerbork Synthesis Radio Telescope (WSRT), reaching typical column density sensitivity levels of $\sim10^{18}$ cm$^{-2}$ in $\sim$ 4.12 \kms channels and a spatial resolution of {20"-30''} \citep[see][for further details]{Heald2011,Heald2020}. Here, we present the results of a systematic search for distinct intergalactic clouds and streams in the vicinity of the sample galaxies using the SOurce Finder  Application \citep[\sofia, ][]{Serra2015}. Using \sofia\ has distinct advantages over a visual inspection since it is unbiased, quantifiable and reproducible. As the HALOGAS sample includes a broad range of galaxy properties, our findings allow us to estimate global neutral gas accretion parameters.\\
\indent This paper is structured as follows. In $\S$ \ref{section:method} we describe the method that we use to analyze HALOGAS cubes in order to find cloud candidates in a robust manner. In $\S$ \ref{section:sourceresults} we present the tabulated result of our search and present ancillary data for the targets and how these relate to our search result. In $\S$ \ref{section:sampleresults} we present the results derived from the full sample and in $\S$ \ref{section:discussion} we discuss our findings. Finally in $\S$ \ref{section:conclusions} we summarize and conclude our analysis.\\  
\section{Methodology}\label{section:method}
In order to find  gas clouds around the galaxies in the HALOGAS sample we make use of the \hi\ source finder application \sofia\  \citep[v1.3.1, ][]{Serra2015,Wang2015,Westmeier2021,Jozsa2022}. \sofia\ is an extended package that has many different settings that can be fine tuned to optimize the detection of different types of sources. 
\subsection{Artificial Source Finding Test}
\begin{figure}[tbp]
\centering
\includegraphics[scale=0.5]{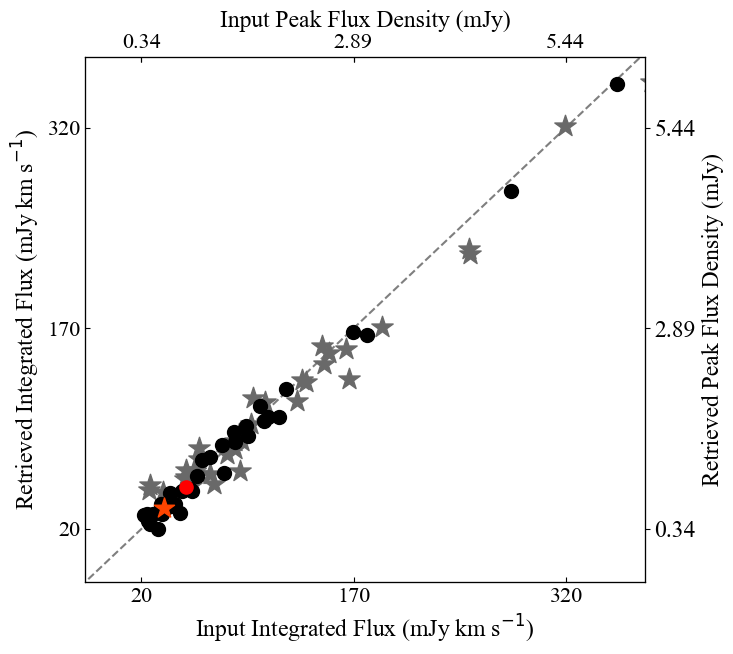}
\caption{Integrated flux (circles, mJy \kms, right and bottom axes) and spectral peak flux  densities (stars, mJy, left and top axes) from the input catalog versus \sofia\ retrieved values for the detected sources. Black/gray are the sources detected by \sofia, red is the false detection.}
\label{fig:flux}
\end{figure}

First we investigated the optimal \sofia\ parameter settings for finding small unresolved clouds by searching a data cube that contains  250 artificial point sources. This also provided us with a limiting flux threshold of our final source finding method. \\
\indent To create a data cube that resembles the HALOGAS observations as closely as possible, we first created an empty model cube with the same size, coordinates and spectral resolution as the actual full bandwidth data cube of the HALOGAS observations of NGC\,3198 \citep{Heald2011,Gentile2013}. NGC\,3198 was chosen as its data cube is representative for most of the data cubes in the survey.   It has the HALOGAS standard pixel size of 4"$\times$4" and channel width. This cube is populated with randomly distributed model \hi\ sources where each source is considered to be unresolved, i.e., a point source centered on a single pixel convolved with the point spread function (PSF,i.e. the synthesized beam). Although the point sources were injected randomly, care was taken that they were separated by  at least four PSF Full Widths at Half Maximum (FWHM, see Table \ref{table:cuts}) and five channels. This was done to prevent overlapping sources in the final artificial cube which potentially could confuse our detection statistics.\\
\indent In this test we focus on point sources as the main aim of this paper is to search the HALOGAS survey for clouds without optical counterparts. These clouds are expected to be of a physical size that will not be resolved by the WSRT synthesized beam in most cases. Additionally, experience taught us that in \sofia, even when settings are not optimized towards extended sources, it is unlikely to miss an extended source completely when settings are optimized for point sources. \\ 
\indent The \hi\ mass of each source was chosen randomly in the range 4 < log(\mhi) < 5.7 assuming a uniform distribution in log space. The systemic velocity was randomly selected in the range 100-1472 \kms, with the upper limit corresponding to the maximum velocity in the cube.  This velocity was converted to a distance by assuming a pure Hubble-Lema\^itre flow with H$_0=70$ \kms and subsequently the flux was calculated. Along the spectral axis the artificial sources had a Gaussian distribution with a FWHM in the range 15 - 30 \kms\ to represent the internal dispersion and some slight internal velocity structure in the clouds. If the initial mass of the artificial source was $>$  $5\times10^{4}$ \msun\ this FWHM was increased with a factor $\sqrt{\frac{{\rm Mass}}{5\times10^{4} }}$ to resemble the larger velocity spread of more massive sources. These settings lead to the peak flux densities being in the range of $\sim 0.003-30\sigma$ of the template cube, thus ensuring a range of sources from ones that could not be missed to others that are impossible to detect. The  sample of 250 artificial sources has the following mean values; Mass = $1.2\times10^5$ \msun, FWHM = 33.5 \kms and an average peak flux density of 0.4 mJy beam$^{-1}$, where the latter corresponds to 2.4$\sigma_{\rm template}$.\\
\indent The model cube, which is in the image domain, was then converted to the $uv$-plane with the task {\sc tclean} in the Common Astronomy Software Application \citep[{\sc CASA},][]{McMullin2007} for each observing run completed for NGC\,3198.
The model visibilities were doubled and then added to the XX correlation of the original calibrated and continuum-subtracted data but not to the YY correlation.
 Assuming that the original data does not contain any polarized emission or polarization errors, a Stokes Q data cube created from this data retains all the noise properties and instrumental characteristics of the real observations, as well as our artificial sources, without including any real emission. 
This artificial Stokes Q cube is produced in exactly the same manner as the original Stokes I data cube of NGC\,3198, i.e the inversion to the image domain and deconvolution was done with the same software and settings as in the real data. In both cubes any corrections required due to the primary beam response are ignored. When converting back to the image domain our point sources are effectively convolved with the dirty beam and thus no longer contained in a single pixel.\\
\indent We then ran \sofia\  on this mock data cube with different settings.   Among the tested parameters the noise determination, merging maximum separation, merging minimum size and reliability had a minimal impact on the final results. The following parameters are critical and hence we also list the investigated ranges:
\begin{itemize}
\renewcommand\labelitemi{--}
\item Detection threshold: The level below which the mask at each scale is clipped. We explored values between 3$\sigma$ and 6$\sigma$.
\item Kernels: The smoothing kernels used to search for emission in the data. We explored values in the spatial direction between 0 and 8 pixels;  and along the velocity axis from 0 to 16 channels.
\item Kernel scale: The scale used to smooth the  indicators that are used to determine the reliability of a detection. The full range from 0 to 1 was explored.
\item Integrated Signal-to-Noise cut: In the reliability evaluation it can transpire that positive sources are not real but lie in an area of the noise distribution that is sparsely sampled. This could lead to false positives. Hence a more absolute cutoff on the integrated signal to noise\footnote{  Defined  in \sofia\ as $\sum {\rm S}_{i}/(\sigma\times \sqrt{{\rm {N_{pix}}}})$ with S$_{i}$ the flux density in each pixel, N$_{\rm pix}$ the number of pixels in the source and $\sigma$ the noise in the cube. } is also applied. For our test set we sampled a range of 15 - 35.
\end{itemize}
\indent For more information on these parameters we refer to the \sofia\ website\footnote{\url{https://github.com/SoFiA-Admin/SoFiA/wiki}} and \cite{Serra2015}. We searched this grid of parameters by systematically varying one parameter in the search grid while keeping the others at the central value for the range. The best value was identified as the value which gave the highest ratio of real to false detections. Once all optimal values were determined in this way we ensured that the parameters could be optimized in this independent manner by once more running through the grid with all the optimal values set, albeit with smaller ranges. This did not lead to any change in the optimal parameters.\\ 
\indent Table \ref{tab:settings} gives the final parameters that resulted in the highest ratio of real to false detections. After source detection the sources' masks were dilated until the integrated flux in the source increased by a factor less than 0.02 compared to the previous mask with a maximum of 3 pixels increase around the original detection.\\
\begin{table} 
\caption{Parameters used for source finding in \sofia.} \label{tab:settings} 
\centering  
\begin{tabular}{lr} 
\hline 
Parameter & Value  \\ 
\hline 
Noise determination &  global\\
Detection threshold & 4$\sigma$ \\
Kernels (pixels) $^{\rm a}$ & [[0,0,0,'g'],[0,0,2,'g'],\\
 &[0,0,4,'g'],[2,2,0,'g'], \\
 &[2,2,2,'g'],[2,2,4,'g'],\\
 &[4,4,0,'g'],[4,4,2,'g'],\\
 &[4,4,4,'g']] \\
Merging radius (pixels) &  2,2,2 \\
Merging minimum size (pixels) & 2,2,2 \\
Reliability & 0.95 \\
Kernel scale &  0.30 \\
Integrated Signal-to-Noise cut & 27\\

\hline 
\end{tabular} 
	\tablefoot{ 
	\tablefoottext{a}{The kernels are build up as [x,y,z, 'shape'], where x, y and z are the kernel sizes in the corresponding dimensions and 'shape' stands for the type of smoothing. The g in the smoothing kernels indicates Gaussian smoothing.} 
} 
\end{table} 
\indent With these settings we retrieved  37  distinct model sources with \sofia.
Of these one was a false detection. A careful inspection of this false detection showed that it corresponds to residual emission from NGC\,3198, either due to small calibration errors in the data or imperfect continuum subtraction (possibly related to polarized non-thermal emission). This leads to an important point, our source finding is only as good as our data products, i.e. bright artifacts will be detected by \sofia. The 36 retrieved artificial sources all correspond to sources from the input catalog with an integrated flux larger than 21 mJy \kms, with a lowest spectral peak flux density of 0.43 mJy beam$^{-1}$,   which corresponds to $\sim2.7\sigma$ in a single channel in our mock cube. \\
\indent One issue here is the small merging radius adopted in our settings. Experiments have shown that this is necessary to not miss out on small sources as the noise statistics for the reliability calculations change when sources are merged, making \sofia\ less sensitive to the smaller sources. However, this means that for real extended sources there is a chance that they break up into several sources. Visual inspection is required to identify and merge extended sources that are broken into segments by \sofia. \\ 
\indent Another important point to investigate in our source finding is how well the flux of a source can be retrieved after it is identified. Figure \ref{fig:flux} shows the input integrated flux versus the retrieved integrated flux (circles) as well as the peak flux densities (stars). All sources lie closely to the line indicating identity, which means that once a source is found, \sofia\ has no issue retrieving most of the flux associated with that source. At the lower end, the scatter indicates that noise does affect our detections. There are four model sources that have an input integrated flux slightly above our nominal detection limit of  21 mJy \kms that are not detected by \sofia. A visual inspection shows that in the final cube these sources are indeed much less significant than our faintest detections, most likely due to the distribution of noise peaks on top of these sources, i.e. some lie on a positive noise peak while others lie on a negative one. \sofia\ was able to identify this reduced significance and exclude the sources as detections. \\
\subsection{Searching the Survey Data Cubes}
\indent The fact that \sofia\ will also detect artifacts means that a final visual inspection of all \sofia\ detections is required to determine whether they are artifacts. For this reason we decided to assign to each source identified by \sofia\ a numerical flag in the range 1-4 relating to our assessment of the reliability of the source. We used the following classification scheme:
\begin{itemize}
\renewcommand\labelitemi{ }
\item 1 - Real 
\item 2 - Likely real
\item 3 - Likely not real
\item 4 - Not real, Artifact
\end{itemize}
The visual inspection was performed by one of us (PK) and subsequently checked by several other authors. Once a source was classified as real or likely real, the NASA Extragalactic Database (NED)\footnote{\url{https://ned.ipac.caltech.edu/}} and Simbad \footnote{\url{http://simbad.u-strasbg.fr/simbad/}} were queried to search for optical counterparts classified as galaxies. Around each source we searched a cylinder with 500 \kms depth and a radius defined by the major axis of the 3$\sigma$ ellipse as fitted by \sofia. The maximum allowed radius for the cylinder was 2\arcmin, to avoid searching large catalogs of individual objects in well resolved sources.  When we found a source in this cylinder we deemed it the optical counterpart to our detection. These optical counterparts are listed in the counterpart column of Table~\ref{table:detec}. If no source was found we inspected available  optical images and queried the databases manually. Based on these we estimated whether optical emission  is present and, if so, whether it is a counterpart or a background source. These detections without spectroscopic confirmation are indicated with a * in Table~\ref{table:detec}.\\   
\indent To avoid complications by varying noise levels within a single cube we ran our source finding on data cubes that had not been corrected for the primary beam response. Instead we performed this correction on the integrated flux of the detected sources based on the distance of their central coordinates to the phase center. For this correction we applied the standard correction for the WSRT: cos$^6(c\times \nu \times r)$ with $\nu$ the frequency in Ghz, $r$ the distance to the phase center in radians, and $c$ a constant. For $c$ we did not use the standard $c$=68 but instead used $c$ = 63 as it has been shown to better match the non-parameterized WSRT beam \citep{EWang2015}.  We list both the corrected and uncorrected flux in Table \ref{table:detec}.\\
\indent Finally, to be able to run \sofia\ on the real sample we needed to apply some specific constraints to the source finding that differ from data cube to data cube.  As the integrated SNR is based on a flux level in Jy beam$^{-1}$, the integrated flux SNR threshold needs to be corrected for the beam size. This is done by scaling this threshold with the ratio of the square root of the beam in the cube under investigation and the mock data cube. The thresholds used for each cube are shown in Table \ref{table:cuts}. Before running \sofia\ on the real data we first identified bad channels. Such channels can occur at the edges of the cube or due to MW emission. We identified these by calculating a mean rms level for the cube and then checking each channel against this mean. Channels where the noise increases by more than 1\% in three of the four corners were flagged. This typically resulted in a good identification of channels containing MW emission and increased noise. However, in some cases even this stringent criterion still misses channels with MW emission. In these cases the channel range being flagged was manually adjusted. \\
\indent With these setting we now run \sofia\ on the 22 datacubes of the HALOGAS survey that cover their full 10 MHz bandwidth ($\sim 2000$ \kms), i.e. not only the channels close to the target galaxy. These extended cubes are not publicly available due to their large size but can be shared on reasonable request to the corresponding author. The normal HALOGAS cubes are available at \url{https://www.astron.nl/halogas/data.php}. The use of the extended cubes means that NGC\,891 and NGC\,2403 are not included in this analysis as only data cubes with a narrow range of velocities around the target are available.\\
\begin{table} 
\caption{Noise cut limits and beam sizes used in \sofia for each data cube.} \label{table:cuts} 
\centering  
\begin{tabular}{lcl} 
\hline 
Data cube & Int. SNR Threshold & FWHM$_{maj}\times$FWHM$_{min}$ \\ 
\hline 
NGC\,0672 & 32.5 & 31.6"$\times$14.3" \\ 
NGC\,0925 & 30.5 & 26.6"$\times$14.9" \\ 
NGC\,0949 & 28.9 & 24.2"$\times$14.7" \\ 
NGC\,1003 & 27.9 & 22.6"$\times$14.7" \\ 
NGC\,2541 & 26.3 & 19.1"$\times$15.4" \\ 
NGC\,3198 & 26.9 & 21.0"$\times$14.8" \\ 
NGC\,4062 & 30.9 & 27.6"$\times$14.8" \\ 
NGC\,4244 & 29.2 & 23.9"$\times$15.2" \\ 
NGC\,4258 & 26.9 & 20.9"$\times$14.7" \\ 
NGC\,4274 & 31.7 & 29.5"$\times$14.5" \\ 
NGC\,4414 & 31.7 & 28.6"$\times$15.0" \\ 
NGC\,4448 & 32.1 & 30.5"$\times$14.4" \\ 
NGC\,4559 & 32.4 & 31.2"$\times$14.3" \\ 
NGC\,4565 & 34.1 & 33.5"$\times$14.8" \\ 
NGC\,4631 & 31.6 & 27.8"$\times$15.3" \\ 
NGC\,5023 & 27.1 & 21.3"$\times$14.7" \\ 
NGC\,5055 & 27.4 & 22.1"$\times$14.5" \\ 
NGC\,5229 & 26.8 & 20.1"$\times$15.2" \\ 
NGC\,5585 & 24.9 & 17.5"$\times$15.0" \\ 
UGC\,2082 & 32.0 & 31.8"$\times$13.8" \\ 
UGC\,4278 & 26.9 & 20.7"$\times$14.9" \\ 
UGC\,7774 & 28.4 & 23.3"$\times$14.8" \\ 
\hline 
\end{tabular} 
\end{table} 
\section{Source Finding Results} \label{section:sourceresults}
From the 22 Stokes I data cubes of the HALOGAS sample we initially obtained 102 individual sources detected with \sofia. After combining artificially separated segments of single galaxies, a total of 80 sources were left. From these, we have 54 real sources, 7 likely real detections, 1 likely not real and 18 artifacts.\\
\indent In order to not only judge the reliability of our detections but also the type of detection we also assigned a class identifier to each source during the visual inspection. These are defined as follows: 
\begin{itemize}
    \item Targets:
    \indent The main body of the galaxies that were selected for the survey and are located in the center of the observations.
    \item Companions: 
    \indent These are detections with an optical counterpart separated from the target by more than one beam in all channels.
    \item  Cloud Candidates:
    \indent The same as companions but without an optical counterpart.
    \item Artifacts:
   \indent Sources that can be identified as artifacts of the data reduction.
\end{itemize}
\indent Of the sources that are real or likely real,  22 are the target galaxies, 25 are companions and 14 are cloud candidates. Of the latter seven are classified real and seven as likely real. The constructed catalog of the \sofia\ detected sources is presented in Table~\ref{table:clouds} for the cloud candidates and Table~\ref{table:detec} for all other sources. The column entries are the following: 
\begin{itemize}[leftmargin=2.5cm]
\item[{\it {Column (1) }} --] Target galaxy.
\item[{\it {Column (2) }} --] \sofia\ assigned source number within the target field.
\item[{\it {Column (3) }} --] Right ascension of flux-weighted centroid in J2000.
\item[{\it {Column (4) }} --] Declination of flux-weighted centroid in J2000.
\item[{\it {Column (5) }} --] Class: T - Target galaxy, C - Companion, CC - Cloud Candidate, A - Artifact 
\item[{\it {Column (6) }} --] Peak flux density, in mJy beam$^{-2}$.
\item[{\it {Column (7) }} --] Integrated flux, in Jy \kms.
\item[{\it {Column (8) }} --] Integrated flux, in Jy \kms, primary beam corrected.
\item[{\it {Column (9)}} --] Line width at 20\% of the peak flux density in \kms.
\item[{\it {Column (10)}} --] Systemic velocity of flux-weighted centroid, in \kms.
\item[{\it {Column (11)}} --] Projected distance to main target in kpc.
\item[{\it {Column (12)}} --] Flag: 1 -  Real;  2  - Likely real; 3 - Likely not real; 4 - Not real, artifact.
\item[{\it {Column (13)}} --] Identified optical counterpart (only applies to Table~\ref{table:detec}).  
\end{itemize}
\afterpage{
\onecolumn
\begin{landscape} 
\begin{table}\label{table:clouds} 
\caption{\sofia\ detected Cloud Candidates.} \label{table:cloud} 
\centering 
\begin{tabular}{lllllllllllll}
\hline 
Galaxy & ID & RA & Dec & Class  & S$_{p}$ & $F_{i}$  & $F_{i}$,pbcorr & W20  & Vel. & Dist. to target & Flag \\ 
  &   &  [hms] & [dms] & &   [mJy  bm$^{-1}$] & [Jy \kms] & [Jy \kms] & [\kms] & [\kms] & [kpc] &  \\ 
   (1) & (2) & (3) & (4) & (5) & (6) & (7) & (8) & (9) & (10) & (11) & (12)   \\  
\hline 
NGC\,0925  & 2          & 2h27m5.62s   & 33d43m13.73s & CC     & 0.871      & 0.048     & 0.056      & 45.1       & 579.1      & 22.6       & 2      \\ 
NGC\,1003  & 2          & 2h39m12.94s  & 40d55m51.93s & CC     & 0.893      & 0.086     & 0.088      & 36.3       & 593.9      & 12.3       & 1      \\ 
NGC\,4258  & 3          & 12h19m2.16s  & 47d6m52.02s  & CC     & 0.830      & 0.067     & 0.088      & 41.5       & 252.9      & 24.5       & 1      \\ 
NGC\,4258  & 8          & 12h18m34.75s & 47d32m46.60s & CC     & 1.174      & 0.088     & 0.141      & 42.0       & 570.7      & 33.8       & 2      \\ 
NGC\,4274  & 9          & 12h20m18.55s & 29d18m48.96s & CC     & 1.082      & 0.032     & 0.069      & 30.3       & 791.7      & 108.4      & 1      \\ 
NGC\,4274  & 10         & 12h20m18.47s & 29d22m2.91s  & CC     & 0.962      & 0.045     & 0.077      & 32.9       & 979.4      & 91.3       & 1      \\ 
NGC\,4274  & 11         & 12h20m0.86s  & 29d34m26.72s & CC     & 1.224      & 0.110     & 0.112      & 42.2       & 998.7      & 19.8       & 1      \\ 
NGC\,4414  & 2          & 12h26m26.93s & 31d21m20.80s & CC     & 0.888      & 0.045     & 0.051      & 36.0       & 622.0      & 42.8       & 1      \\ 
NGC\,4565  & 3          & 12h36m49.28s & 25d55m49.89s & CC     & 0.799      & 0.058     & 0.065      & 50.4       & 1017.6     & 24.6       & 2      \\ 
NGC\,4565  & 6          & 12h35m39.69s & 25d55m58.35s & CC     & 1.092      & 0.025     & 0.031      & 25.9       & 1327.0     & 30.2       & 1      \\ 
NGC\,4565  & 7          & 12h35m59.96s & 26d8m16.66s  & CC     & 0.785      & 0.057     & 0.070      & 37.3       & 1381.8     & 30.0       & 2      \\ 
NGC\,4565  & 8          & 12h38m11.61s & 25d36m44.94s & CC     & 0.991      & 0.030     & 0.411      & 98.2       & 1576.6     & 107.5      & 2      \\ 
NGC\,5055  & 2          & 13h16m59.56s & 42d6m39.26s  & CC     & 0.981      & 0.309     & 0.462      & 48.6       & 534.6      & 36.3       & 2      \\ 
NGC\,5585  & 1          & 14h21m24.29s & 57d4m47.70s  & CC     & 0.916      & 0.020     & 0.076      & 63.1       & 111.0      & 62.0       & 2      \\ 
\hline 
\end{tabular} 
\end{table} 
\end{landscape}
\twocolumn 
}
\indent In the online Appendix of this paper we present various visualizations of the detections. In Appendix \ref{Overview_Source_Images} we present an overview image for every field with a detection besides the target galaxy and discuss some of the specifics of individual detections and targets. In these the column density contours of all detections are overlaid on the $R$-band images of HALOSTARS \citep[See,][]{Heald2011} or $R$-band images taken at Kitt Peak National Observatory (KPNO, described in the atlas paper). If no image was available the background is a DSS 2 $R$-band image. Additionally, in Figures \ref{fig:n0925_image_src_2.pdf} - \ref{fig:n5585_image_src_1.pdf} we show, from left to right, the \HI\ column density map and intensity weighted velocity field (also known as moment 0 and moment 1 maps), a Position-Velocity (PV) Diagram, along the morphological major axis as determined by \sofia, and a line profile for all cloud candidates\footnote{integrated within the \sofia\ mask}. Finally,  figures \ref{fig:n0925_image_src_4.pdf} - \ref{fig:u7774_image_src_2.pdf}, show moment maps of all other reliable sources except the target galaxies.  All data products relating to the targets will be presented in \citep{Heald2020} and not reproduced here. These images also indicate the contour levels used in the overview images as they vary from source to source.\\
\indent The overview images give an impression of the distribution of the  retrieved sources around the targets. An excellent example of this is Figure \ref{fig:Overview_n4565.pdf}. In this figure all the different classes are represented. This image also shows that some detections are so close to the target galaxies that they can be considered as part of the galaxy's halo (Sources 3, 5 and 7) whereas others are further removed and might not be associated with the target (Sources 1, 4, 6, 8). For the purpose of this paper though we only differentiate between detections based on the existence of a known optical counter part.\\
\indent As we have optimized the source finding for point sources at this point we also check that no extended sources have been missed in the source finding by comparing the sources found in this analysis to those found in the HALOGAS galaxies that have undergone detailed individual studies \citep[See Appendix \ref{Overview_Source_Images}][]{Heald2011b,Zschaechner2011,Zschaechner2012,Gentile2013,Kamphuis2013,deBlok2014,Vargas2017} and a visual inspection of the cubes by several of the authors.\\
\indent In order to facilitate the discussion later in this paper, Table \ref{table:hosts} shows an overview of properties for the target galaxies in the HALOGAS survey. The table does not list the quantities for NGC\,672 and NGC\,4631 as these systems are interacting and hence they will not be included in the final analysis. The distances come from the original HALOGAS paper \citep{Heald2011}.
Systemic velocity, W$_{20}$, R$_{\rm max}$ and R$_{\rm min}$ are all as retrieved by our \sofia\ run. Where R$_{\rm max}$ and R$_{\rm min}$ are half the major and minor axis of the 3$\sigma$ ellipse fitted by \sofia. For consistency we use the values from our \sofia\ run. These values are slightly different from the ones published in the data release and the Atlas paper \citep{Heald2020} due to our cubes being higher resolution and the \sofia\ parameters being optimized for point sources.  These differences are typically very small, e.g., the mean difference in W$_{20}$ is 1.6 \kms. The stellar mass is calculated as the average from the methods and catalogues listed in the notes. The \hi\ mass in solar masses is calculated from the corrected integrated flux as listed in Table \ref{table:detec} and the distance (D [Mpc]) with \mhi =  $F_{i}$,pbcorr $\times2.36\times10^{5}\times{\rm D}^2$ (\msun).  The SFRs are taken from the pilot paper \citep{Heald2011} and the depletion time ($\tau$) is given by \mhi/SFR. The calculation of the dynamical mass and virial radius is described in Section \ref{section:accretion_rate}.\\ 
\indent The HALOGAS sample currently does not appear in a homogeneous sample of SFR determinations. The values listed here are pre-dominantly compiled from a set of  H$\alpha$ and infrared surveys allowing the determination of dust attenuation corrected SFRs. The pilot paper does not list errors on the SFRs and such errors can be significant due to the different methods, errors in the distances or even the timescales probed in different bands. Here we follow \cite{Kennicutt2009} who determined that dust attenuation corrected SFRs should at maximum differ by $\pm$ 0.3 dex from their calculated value. Assuming that to correspond to a 3$\sigma$ limit, the error on the values listed in \ref{table:hosts} would be at most 30\% . \\
\begin{table*}[h!] 
\caption{Properties of the galaxies in the HALOGAS sample.} \label{table:hosts} 
\centering  
\begin{tabular}{l l l l l l l l l l l l l l} 
\hline 
Galaxy & Dist. & $\rm{ v_{sys}}$ & W$_{20}$ & R$_{\rm max}$ & R$_{\rm min}$& M$_*$ &  M$_{HI}$ & M$_{\rm dyn}$  &r$_{\rm vir}$\tablefootmark{d} & SFR\tablefootmark{e} & $\tau$ & D$_{25}$ & $i$  \\ 
        &  Mpc &  \kms  &  \kms  & kpc & kpc & $10^{9}$\,\msun & $10^{8}$\,\msun & $10^{10}$\,\msun &kpc  & \msun yr$^{-1}$ & $10^9\,yr$ & kpc & $^{\circ}$  \\ 
\hline 
NGC\,0925  &        9.1 &      550.8&      222.0 &       28.1 &       19.8 &       7.48 \tablefootmark{a),b),c)  } &      54.6 &       12.3 &      103.0 &       0.77 &        7.1 &       29.9 & 54\\
NGC\,0949  &       11.3 &      607.7&      207.2 &       13.3 &        6.0 &       3.43 \tablefootmark{c)        } &       5.6 &        5.3 &       78.0 &       0.31 &        1.8 &       11.5 & 52\\
NGC\,1003  &       11.6 &      621.2&      229.6 &       35.8 &        7.4 &       3.04 \tablefootmark{c)        } &      59.3 &       12.9 &      104.7 &       0.40 &       14.8 &       21.3 & 67\\
NGC\,2541  &       12.0 &      555.1&      209.4 &       28.3 &       12.5 &       1.65 \tablefootmark{c)        } &      47.5 &        8.5 &       91.0 &       0.35 &       13.6 &       25.1 & 67\\
NGC\,3198  &       14.5 &      659.6&      317.4 &       48.1 &       12.9 &      22.13 \tablefootmark{a),b),c)  } &     112.4 &       31.5 &      140.9 &       1.10 &       10.2 &       37.1 & 71\\
NGC\,4062  &       16.9 &      767.7&      310.8 &       18.0 &        7.2 &       0.11 \tablefootmark{c)        } &      17.5 &       11.7 &      101.4 &       0.67 &        2.6 &       22.1 & 68\\
NGC\,4244  &        4.4 &      245.8&      216.3 &       17.7 &        3.7 &       2.33 \tablefootmark{a),b),c)  } &      18.0 &        4.8 &       75.3 &       0.12 &       15.0 &       20.2 & 90\\
NGC\,4258  &        7.6 &      444.6&      435.8 &       28.9 &       12.4 &      65.88 \tablefootmark{b),c)     } &      57.1 &       35.6 &      146.8 &       1.70 &        3.4 &       37.8 & 71\\
NGC\,4274  &       19.4 &      933.0&      463.0 &       19.6 &        7.2 &      71.94 \tablefootmark{c)        } &       7.9 &       27.0 &      133.8 &       1.20 &        0.7 &       36.7 & 72\\
NGC\,4414  &       17.8 &      718.6&      395.6 &       37.5 &       26.0 &      65.55 \tablefootmark{c)        } &      50.1 &       58.2 &      172.8 &       4.20 &        1.2 &       23.3 & 50\\
NGC\,4448  &        9.7\tablefootmark{f} &      647.5&      382.8 &        4.4 &        2.4 &      8.760 \tablefootmark{c)        } &       0.4 &        4.2 &       71.8 &       0.06 &        0.8 &       10.7 & 71\\
NGC\,4559  &        7.9 &      809.6&      253.1 &       21.8 &        9.1 &       6.32 \tablefootmark{a),b),c)  } &      43.6 &        9.3 &       93.9 &       0.69 &        6.3 &       26.0 & 69\\
NGC\,4565  &       10.8 &     1237.8&      517.4 &       34.7 &        6.0 &      71.18 \tablefootmark{a),b),c)  } &      69.8 &       54.0 &      168.6 &       0.67 &       10.4 &       50.9 & 90\\
NGC\,5023  &        6.6 &      405.9&      192.2 &        9.8 &        2.5 &       1.51 \tablefootmark{a),b),c)  } &       5.8 &        2.1 &       57.2 &       0.04 &       14.8 &       13.1 & 90\\
NGC\,5055  &        8.5 &      498.1&      393.5 &       45.4 &       27.8 &      69.03 \tablefootmark{a),b),c)  } &      73.5 &       60.9 &      175.5 &       2.10 &        3.5 &       32.1 & 55\\
NGC\,5229  &        5.1 &      356.9&      138.9 &        5.1 &        1.6 &      0.004 \tablefootmark{c)        } &       1.4 &        0.6 &       37.1 &       0.01 &       16.8 &        5.2 & 90\\
NGC\,5585  &        8.7 &      306.9&      161.7 &       19.6 &       15.2 &       1.49 \tablefootmark{c)        } &      26.1 &        4.9 &       76.0 &       0.41 &        6.4 &       13.9 & 51\\
UGC\,2082  &       14.1 &      706.4&      208.0 &       23.6 &        5.8 &       2.40 \tablefootmark{c)        } &      23.4 &        5.9 &       80.8 &       0.04 &       57.0 &       23.8 & 89\\
UGC\,4278  &       13.6 &      559.6&      192.7 &       13.5 &        5.3 &       1.50 \tablefootmark{a),b),c)  } &      20.0 &        2.9 &       63.8 &       0.18 &       11.1 &       17.0 & 90\\
UGC\,7774  &       24.4 &      521.7&      205.5 &       33.6 &        9.0 &        nan \tablefootmark{          } &      35.4 &        8.3 &       90.2 &       0.10 &       37.3 &       24.8 & 90\\
\hline 
\end{tabular} 
\tablefoot{  
\tablefoottext{a}{Galaxy integrated flux  taken from \cite{Clark2018}, mass calculated using 3.6 and 4.5 $\mu$m following \cite{Eskew2012}.}  
\tablefoottext{b}{Galaxy integrated flux taken from \cite{Clark2018},  calculated using 3.4 and 4.6 $\mu$m following \cite{Cluver2014}.}  
\tablefoottext{c}{Galaxy integrated flux taken from \cite{Skrutskie2006},  mass calculated using 3.6 $\mu$m following \cite{McGaugh2014}.}  
\tablefoottext{d}{Derived from dynamical mass.}  
\tablefoottext{e}{SFR from the pilot paper.}  
\tablefoottext{f}{This distance is likely an underestimate by a factor of two.}  
} 
 \end{table*}  
\indent All-sky surveys such as WISE or DSS are not sufficiently sensitive to provide counterpart detections or stellar mass estimates for our \sofia\ detections. This means that we can not estimate the ratio of \mhi to M$_{*}$ for the cloud candidates. Thus we are relying on our NED search and our deep $R$-band images to verify these cloud candidates are without an optical counterpart.\\
\indent Finally, in order to ensure that our cloud candidates are not related to the Milky Way system of high velocity clouds (HVCs) we compare the  location and systemic velocity of the targets to the Leiden/Argentine/Bonn Survey \citep[LAB, ][]{Kalberla2005}. We do not find any emission from the Galactic system which is near to any of our detections, neither in projected location on the sky nor velocity. Therefore, we conclude that our HALOGAS findings are not caused by Galactic halo clouds which lie along the same line-of-sight. \\
\section{Accretion Rates}\label{section:sampleresults}
\subsection{Accretion Rate of Cloud Candidates and Companions.}\label{section:accretion_rate}
\begin{figure}[tbp]
\centering
\includegraphics[width=9cm]{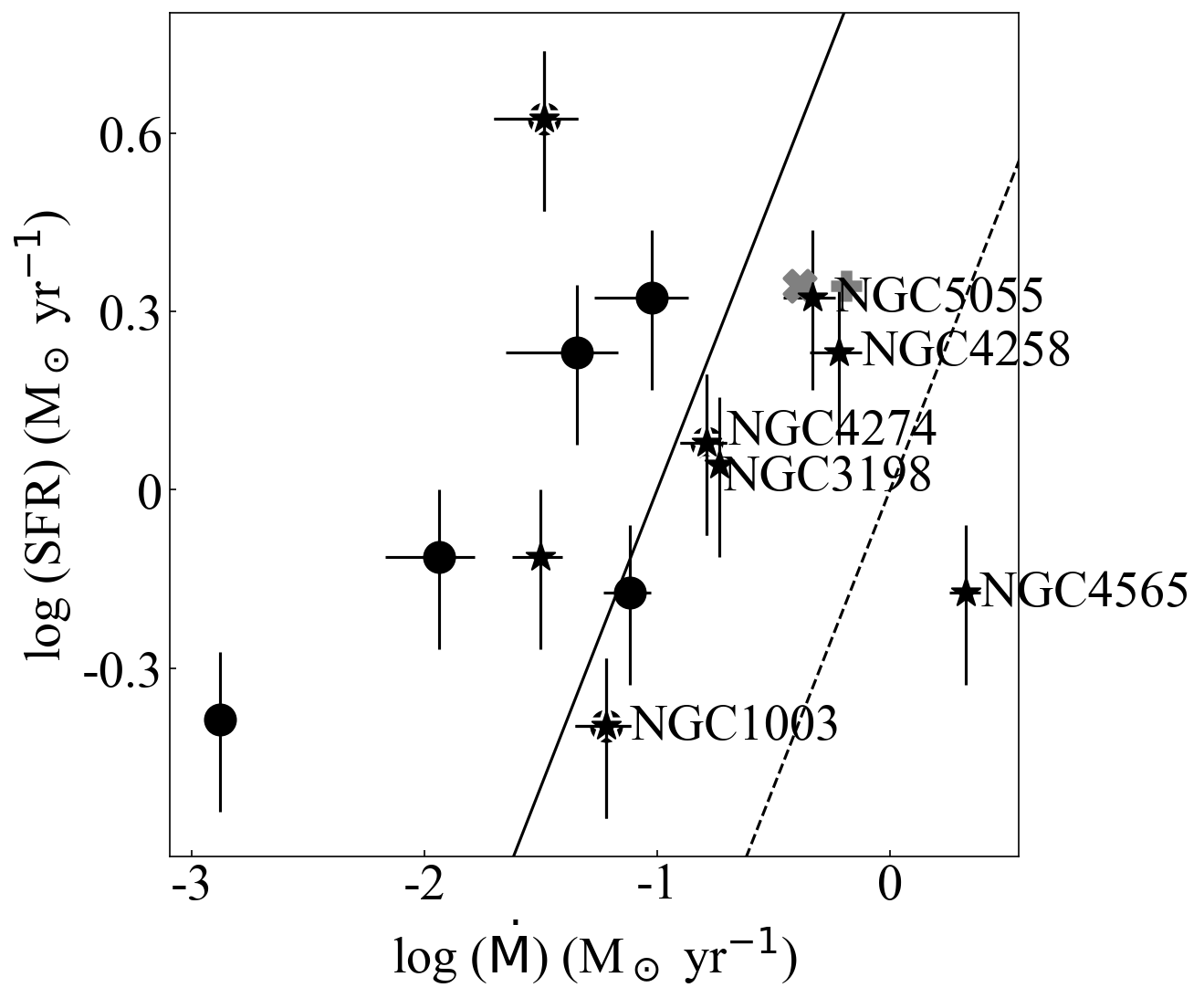}
\caption{ ${\rm \dot{M}}$ vs SFR of host galaxies. Solid circles represent the average of accretion rates with a timescale of $t_{\rm ff1}$ (infall on center) and a timescale of $t_{\rm ff2}$ (infall on edge of disk) for galaxies with cloud candidates, with the horizontal error indicating the distance to these upper limits. The errors on the SFR indicate the assumed 30\% error.} Stars show the same but now for all sources around the target within the virial radius for the whole sample barring the merging galaxies. The solid line represents ${\rm \dot{M} = 0.1 \times SFR}$ (galaxies to the right of this line are labeled with their name) and the  dashed line ${\rm \dot{M} = SFR}$. The gray cross and plus symbol indicate where NGC\,891 would fall according to our method and when individual \sofia\ detections are not merged with the disk, respectively.(See $\S$\ref{implicatcold}) 
\label{fig:mdotvsSFR}
\end{figure}

One of the goals of the HALOGAS Survey is to determine the \hi\ accretion rate in a sample of star forming galaxies that spans a broad range of galaxy characteristics such as mass and environment. As we have  identified all \hi\ sources around the sample galaxies in a systematic manner we can now approximate the accretion rate for the full sample. In this section we calculate the actual observed accretion rate for both the cloud candidates as well as all observed \hi\ within the virial radius of the HALOGAS targets. For this we assume that the cloud candidates fall freely onto the disk and as such these rates are upper limits. \\  
\indent To estimate the accretion rates due to  \hi\ clouds, we use the projected distance ($d$)  the cloud candidate needs to travel to be accreted onto the disk of the host galaxy and their  free fall time (t$_{\rm ff}$). The latter is calculated as:
\begin{equation}\label{eq:fftime}
t_{ff}=\frac{\pi}{2}\times \sqrt{\frac{d^3}{2{\rm{G(M_{dyn}+m_{cloud})}}}}
\end{equation}
with m$_{\rm cloud}$ the \hi\ mass of the cloud candidate and M$_{\rm dyn}$ the dynamical mass which is calculated using the line-width W$_{20}$ and the spherical approximation as:\\
\begin{equation}\label{eq:mdyn}
{\rm M_{dyn}}=\frac{(0.5\times{\rm W_{20}})^2\times {\rm R}}{{\rm sin}(i)^2\times{\rm G}}
\end{equation}
with W$_{20}$ from our \sofia\ runs and R as half the major axis of the  3$\sigma$ ellipse fitted to the target by \sofia. Even though these are not the most accurate estimates of the rotational velocity and size of the disk that are available, their accuracy is not expected to be a significant factor in the error on the estimated accretion rates. Using these values allows for a consistent approach for all galaxies in the sample. This is also the reason we do not use the rotation curves from \cite{Marasco2019}, as these are only derived for the intermediately inclined galaxies. The individual dynamical masses and inclinations for the targets are  listed in Table \ref{table:hosts}. \\
\indent The time scale for accretion, and thus the estimated accretion rate, depends on the trajectory of the assumed infall. We establish a range by considering infall to the center ($t_{ff1}$) and to the edge of the galactic disk ($t_{ff2}$) as defined  by the 3$\sigma$ ellipse fitted by \sofia.
The accretion rates are then ${\rm \dot{M}} = \frac{\rm m_{cloud}}{t_{ff}}$ (see Table\,\ref{table:accretion_clouds}). The difference between the two calculated rates give an indication of the uncertainty of these accretion rates.\\
\indent The average accretion rates of cloud candidates on each target galaxy are shown as the circles in Figure \ref{fig:mdotvsSFR}. Besides calculating these accretion rates for our cloud candidates (Table\,\ref{table:accretion_clouds}), we also calculate the total accretion rates for all the targets, i.e. including companions, in the sample. This is done in the same manner as for the cloud candidates however now we only consider objects within the projected virial radius, as defined by the dynamical mass, of the target and for which the velocity difference between the source and host is less than the host's escape velocity at the projected distance of the companion. The latter is derived from the host's dynamical mass combined with the companions projected distance to the galaxy in the usual manner. These rates are shown in Figure  \ref{fig:mdotvsSFR} as star symbols. The dashed line in this figure shows ${\rm \dot{M}}$ = SFR whereas the solid line shows  ${\rm \dot{M}}$ = 0.1 $\times$ SFR. Note that these values simply set the range of times-scales and are stringent upper limits on the accretion rates as we are using projected distances and free fall times. \\ 
\begin{table*}[ht!] 
\caption{Accretion rate estimates of the cloud candidates.} \label{table:accretion_clouds} 
\centering  
\begin{tabular}{c c c c c c c } 
\hline 
Host Galaxy & Source & v$_{esc}$&  $t_{\rm ff1}$ & $t_{\rm ff2}$ & \.M$_1$ & \.M$_2$  \\ 
        &  & \kms& $10^{7}$ yr &  $10^{7}$ yr & \msun yr$^{-1}$& \msun yr$^{-1}$ \\ 
\hline 
NGC\,0925  & 2 &      216.4 &         16 &          6 &      0.007 &      0.016  \\ 
NGC\,1003  & 2 &      300.7 &          6 &          3 &       0.04 &       0.08  \\ 
NGC\,4258  & 3 &      353.7 &         10 &          2 &       0.01 &       0.04  \\ 
NGC\,4258  & 8 &      301.1 &         17 &          7 &       0.01 &       0.03  \\ 
NGC\,4274  & 9 &      146.3 &        113 &        107 &      0.005 &      0.006  \\ 
NGC\,4274  & 10 &      159.5 &         87 &         81 &      0.008 &      0.008  \\ 
NGC\,4274  & 11 &      342.4 &          8 &          5 &       0.11 &       0.18  \\ 
NGC\,4414  & 2 &      341.9 &         19 &          8 &       0.02 &       0.05  \\ 
NGC\,4565  & 3 &      434.6 &          8 &          4 &       0.02 &       0.04  \\ 
NGC\,4565  & 6 &      392.3 &         11 &         10 &      0.007 &      0.009  \\ 
NGC\,4565  & 7 &      393.6 &         11 &          6 &       0.02 &       0.03  \\ 
NGC\,4565  & 8 &      207.9 &         79 &         64 &       0.01 &       0.02  \\ 
NGC\,5055  & 2 &      379.7 &         14 &          5 &       0.05 &       0.14  \\ 
NGC\,5585  & 1 &       82.8 &        115 &         91 &      0.001 &      0.001  \\ 
\hline 
\end{tabular} 
	\tablefoot{ The difference between \.M$_1$ and \.M$_2$ can be considered as an indication of the uncertainty on these upper limits.
} 
\end{table*} 
\begin{table*}[ht!] 
\caption{Accretion rate estimates of the companion candidates.} \label{table:accretion_companions} 
\centering  
\begin{tabular}{c c c c c c c } 
\hline 
Host Galaxy & Source & v$_{esc}$&  $t_{\rm ff1}$ & $ t_{\rm ff2}$ & \.M$_1$ & \.M$_2$  \\ 
        &  & \kms& $10^{7}$ yr &  $10^{7}$ yr & \msun yr$^{-1}$& \msun yr$^{-1}$ \\ 
\hline 
NGC\,0925  & 2 &      216.4 &         16 &          6 &      0.007 &      0.016  \\ 
NGC\,0925  & 4 &      132.9 &         69 &         51 &       0.02 &       0.02  \\ 
NGC\,1003  & 2 &      300.7 &          6 &          3 &       0.04 &       0.08  \\ 
NGC\,2541  & 1 &       79.1 &        227 &        198 &       0.02 &       0.02  \\ 
NGC\,3198  & 2 &      140.6 &        149 &        137 &       0.18 &       0.19  \\ 
NGC\,4258  & 3 &      353.7 &         10 &          2 &       0.01 &       0.04  \\ 
NGC\,4258  & 4 &      240.3 &         33 &         24 &       0.13 &       0.18  \\ 
NGC\,4258  & 6 &      329.7 &         13 &          8 &       0.28 &       0.45  \\ 
NGC\,4258  & 8 &      301.1 &         17 &          7 &       0.01 &       0.03  \\ 
NGC\,4258  & 10 &      501.2 &          3 &          1 &       0.02 &       0.06  \\ 
NGC\,4274  & 9 &      146.3 &        113 &        107 &      0.005 &      0.006  \\ 
NGC\,4274  & 10 &      159.5 &         87 &         81 &      0.008 &      0.008  \\ 
NGC\,4274  & 11 &      342.4 &          8 &          5 &       0.11 &       0.18  \\ 
NGC\,4414  & 2 &      341.9 &         19 &          8 &       0.02 &       0.05  \\ 
NGC\,4565  & 1 &      389.1 &         12 &         10 &       0.02 &       0.03  \\ 
NGC\,4565  & 3 &      434.6 &          8 &          4 &       0.02 &       0.04  \\ 
NGC\,4565  & 4 &      335.0 &         18 &         16 &       0.92 &       1.03  \\ 
NGC\,4565  & 5 &      529.1 &          4 &          3 &       0.82 &       1.30  \\ 
NGC\,4565  & 6 &      392.3 &         11 &         10 &      0.007 &      0.009  \\ 
NGC\,4565  & 7 &      393.6 &         11 &          6 &       0.02 &       0.03  \\ 
NGC\,5055  & 2 &      379.7 &         14 &          5 &       0.05 &       0.14  \\ 
NGC\,5055  & 4 &      300.1 &         29 &         19 &       0.29 &       0.44  \\ 
\hline 
\end{tabular} 
	\tablefoot{The difference between \.M$_1$ and \.M$_2$ can be considered as an indication of the uncertainty on these upper limits.
} 
\end{table*} 
\indent It is quite obvious from Figure \ref{fig:mdotvsSFR} that none of the galaxies here have an observed accretion rate anywhere near their SFRs when only the cloud candidates are considered, in fact it appears that the current observed neutral gas accretion rate is uncorrelated with the SFR. Even when we consider all observed gas within the virial radius there is only one galaxy (NGC\,4565) in the sample that has sufficient \hi\ in its surroundings to exceed its current SFR. Even considering the significant errors on the SFRs no other galaxy would break this barrier. This implies that we are not fully tracing the gas reservoir that is fueling the SF in these galaxies. \cite{Sancisi2008} found that the average total accretion rate amounted to $\sim$10\% of the SFR, which is, broadly speaking, consistent with the results presented in this paper.  However, out of 20 galaxies in our sample there are only six galaxies with observed accretion rate upper limits in excess of 10\% of their SFR, these are labeled in Figure \ref{fig:mdotvsSFR}. Simply assuming that the galaxies in the sample are all forming stars at some rate already poses a problem as in more than half our sample we do not detect any \hi\ with in the virial radius and in six observations no \hi\ is detected at all in addition to the target. \\
\indent Calculating the accretion rate as described above we find mean neutral gas accretion rates onto the disk of the host galaxies (${\rm \dot{M}_2}$) of 0.03 \msun\ yr$^{-1}$ for all cloud candidates, 0.18  \msun\ yr$^{-1}$  for all companions within the virial radius, and   0.22  \msun\ yr$^{-1}$ for all gas detected within the virial radius of our galaxies.  Although the virial radius is likely to be an underestimate as we derive it from the dynamical mass in the galaxy, we find a negligible effect on the accretion rate estimates by increasing the virial radius by 50-100\%, probably due to the fact that in most cases our sensitivity is severely reduced outside these radii.\\
\indent These averages are consistent with previous results in the literature \citep[e.g.,][]{Sancisi2008, DiTeodoro2014}. On the other hand, the average SFR in the HALOGAS sample (0.8 \msun\ yr$^{-1}$) is slightly lower than that typically assumed or measured in spiral galaxies \citep{Sancisi2008, Bothwell2011} but still well above the observed accretion rates.\\
\indent  As previously said, these accretion rates are stringent upper limits because of the use of projected distances. For the total survey we can obtain more realistic values for the   accretion rates within the virial radius through a statistical correction. If we assume that on average the distance of the detection along the line of sight is half the distance to the boundary of the virial sphere, i.e.\footnote{For $\dot{\rm M}_2$ the depth along the line of sight is calculated using the projected distance to the center and not $d$ hence this formula is only exact for $\dot{\rm M}_1$. Additionally, for $\dot{\rm M}_2$ {we do not correct the projected size of the disk}.} $d_{\rm corr} = \sqrt{\frac{3}{4}d^2 + \frac{1}{4}{\rm r_{vir}}^2 }$, we find an accretion rate of 0.05 (0.04) \msun\ yr$^{-1}$ for $\dot{\rm M}_2$ ($\dot{\rm M}_1$) thus increasing the discrepancy with the average SFR.  \\  
\indent Most of the sample-averaged accretion rate comes from a single galaxy, NGC\, 4565 (see Table \ref{table:accretion_companions}) giving the impression that a significant part  of accretion often happens through minor mergers while most of the time little to no neutral gas is being accreted.
\subsection{The Upper Limit on the Accretion Rate.}\label{section:upper_limits}
\begin{figure}[tbp]
\centering
\includegraphics[width=9cm]{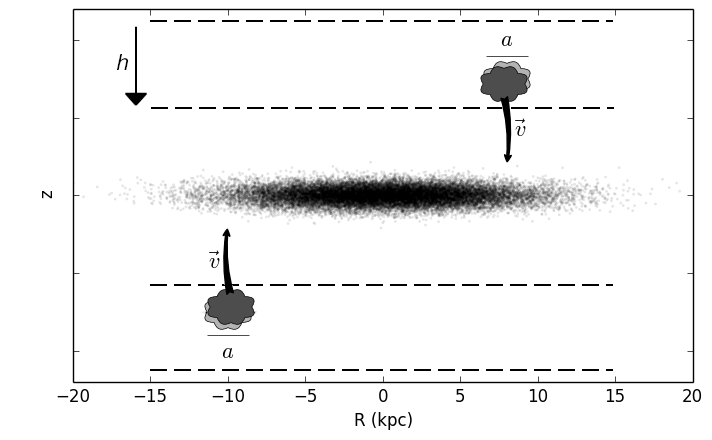}
\caption{Cartoon indicating two accreting clouds of radius $a$ and infall velocity $v_{acc}$, in a volume of height $h$ at an arbitrary distance
from the midplane of the galaxy (here, D$_{25}$ = 30 kpc). }
\label{fig:Upper_Limit_Input}
\end{figure}

 Although HALOGAS provides the deepest resolved survey of nearby galaxies to date, there is still potentially a large amount of undetected gas at low column density. Such a reservoir was recently claimed to have been detected in sensitive Green Bank Telescope (GBT) measurements for NGC\,891 and NGC\,4565 \citep{Das2020}. Here we attempt to estimate the neutral hydrogen gas reservoir that could still be hiding below our sensitivity limits. For this calculation we deviate from the concept presented in the previous section where we considered a proxy for the virial volume around our targets. Even though such volumes present a maximal amount of accretion, the time scales involved are long. As we want to know whether the observed SFR can be fueled by \hi\ accretion, in this section we consider an area closer to the disk to estimate the current accretion rate that can be hidden below our sensitivity limits.\\
\indent In order to get a realistic estimate of the amount of gas that can fuel current SF we need to estimate the time it takes a certain  \hi\ mass to cross the boundary of the SF disk, i.e. a crossing time (${\rm t_{cross}}$) and a total mass in the boundary area (${\rm M_{boundary}}$) such that:
\begin{equation}
   {\rm \dot{M}_{upper}= \frac{M_{boundary}}{t_{cross}}}
\end{equation}
with ${\rm \dot{M}_{upper}}$ the upper limit accretion rate.    \\
\indent First we consider M$_{\rm boundary}$. If we want  to estimate it from our detection limits we will require the number of clouds in the boundary region and multiply it with their mass (M$_{\rm cl}$). The latter is simply set by by our detection limit obtained in $\S$2. To estimate the number of clouds we consider the volume of the boundary region to be two cylinders sitting right above and below the disk of the galaxies (Figure \ref{fig:Upper_Limit_Input}) as defined by D$_{25}$ (see Table \ref{table:hosts} and \cite{Heald2011}) and an arbitrary height $h$. Even though gas can be accreted  outside the optical disk onto the \hi\ disk this gas would need to lose significant amounts of angular momentum before it would form stars. As such it would only become available on timescales much longer than typically considered in SF.\\
The mass in this boundary volume (V$_{\rm boundary}$) is then:
\begin{equation}\label{eq:boundary}
{\rm M_{boundary}}= \frac{{\rm V_{boundary}}}{{\rm V_{cl}}}{\rm \times M_{cl} \times} f
\end{equation}
 where $f$ is a filling factor describing how much of the volume is filled with our clouds and V$_{\rm cl}$ the volume of our clouds. For V$_{\rm cl}$ we consider spherical clouds with a diameter ${\rm FWHM_{maj}}$. The filling factor for \hi\ clouds is not well established. If we calculate this from the clouds we do detect within the total virial volume of the survey we find $f$= 3$\times10^{-5}$. This is likely to be a gross underestimate as smaller clouds will be more numerous and thus take up a larger fraction of the volume considered. From the literature it seems a secure upper limit would be $f=0.1-0.2$ \citep[][and references therein]{Dutta2019}. Therefore we use the range f=3$\times10^{-5}-0.1$ as bracketing values for our upper limit on the accretion rates. For these calculations we assume that the product of mass of the cloud and the filling factor at a given cloud size is constant.\\
 \indent The time required to accrete this mass would be the time it takes the entire volume to sink into the disk, i.e:
 \begin{equation}
{\rm t_{cross}} = \frac{h}{v_{cross}}  
 \end{equation}
 with $v_{cross}$ the speed of the accretion flow. Particles that are ejected from the disk in ballistic fountain models \citep{Collins2002, Fraternali2008} have typical vertical return speeds $\sim50-100$ \kms. In this calculation we assume that the accretion of external gas happens at similar speeds as the flows would mix and the ballistic model provides an upper limit for clouds ejected from the disc. As we are looking for an upper limit we take the maximum speed, $v_{cross}$ = 100  \kms, and we get an accretion rate of: 
\begin{equation} 
{\rm \dot{M}_{upper}} \leq 3\frac{{\rm D_{25}}^2}{a^3} \times {\rm M_{cl}}\times f \times  v_{cross}.
\end{equation}
in which the height of the cylinders cancels out against the same height in the accretion time and $a$ is the diameter of the clouds under consideration, i.e FWHM$_{\rm maj}$. As the accretion rate is a mass flux rate the height of the cylinder should cancel out and hence this height does not affect the final calculations. Using the aforementioned range of $f$, this leads to an upper limit range on the accretion rate of 0.003-7.5 \msun yr$^{-1}$ from the HALOGAS Survey alone.\\
\indent This range is not very informative due to the uncertainty in the filling factor. Therefore we turn to the aforementioned GBT observations. These observations hardly detect any additional \hi\  compared to HALOGAS \citep{Pingel2018,Pingel2018PhD} and gas that is possibly detected is at very low column densities \citep[$\sim4\times10^{17}$ cm$^{-2}$,][]{Das2020}. Taking $5\times10^{17}$ cm$^{-2}$ with a line width of 20 \kms as an indicative upper limit on the detection from the GBT we can obtain a limiting maximum mass, by multiplying this limit with the area of the GBT beam and the mass of the hydrogen atom, that can still be present in the area of the GBT beam.  We obtain the mass in our boundary area by replacing the cloud mass in Eq. \ref{eq:boundary} with the aforementioned limiting GBT mass and V$_{\rm cl}$ with the volume of the cylinder described by the GBT beam area with a depth of $2\times$R$_{\rm max}$ (See Table \ref{table:hosts}). As the GBT volume is of the order of V$_{\rm boundary}$, we use $f$=1 and obtain:\\
\begin{equation}
{\rm \dot{M}_{upper}} \leq \frac{\pi}{4}\times \sigma_{\rm N_{HI}}\times m_{\rm HI} \times \sqrt{\frac{\rm W_{20}}{20.}} \times \frac{\rm D_{25}^2}{{\rm R_{max}}}\times v_{cross}     
\end{equation}
as the upper limit on the accretion rate. In this $\sigma_{\rm N_{HI}}$ is the limiting column density, $m_{\rm HI}$ the mass of a neutral hydrogen atom and ${\rm W_{20}}$ the line width of the galaxy. The latter is to scale $\sigma_{\rm N_{HI}}$ to the linewidth expected in the GBT cylinder under consideration \citep{Wolfe2015}. From this we find a upper limit of 0.04  \msun yr$^{-1}$ showing that the galaxies of the sample would not accrete enough neutral gas to maintain their SFR and are not embedded in extended neutral hydrogen gas reservoirs.\\
\begin{table*}[ht!] 
\caption{Accretion rate upper limits for the sample.} \label{table:accretion_upper} 
\centering  
\begin{tabular}{c c c c c c } 
\hline 
Host Galaxy & Detect. Limit & Detect. Limit& ${\rm \dot{M}}_{f=0.1}$ & ${\rm \dot{M}}_{f=3\times10^{-5}}$ &  ${\rm \dot{M}}_{\rm GBT}$ \\ 
        &  mJy \kms & 10$^5$\msun & \msun yr$^{-1}$ & \msun yr$^{-1}$ & \msun yr$^{-1}$\\ 
\hline 
NGC0925    &       24.2 &        4.7 &       8.07 &      0.002 &      0.034 \\ 
NGC0949    &       23.5 &        7.1 &       1.23 &      0.000 &      0.010 \\ 
NGC1003    &       24.4 &        7.8 &       5.24 &      0.002 &      0.014 \\ 
NGC2541    &       26.9 &        9.1 &      12.89 &      0.004 &      0.023 \\ 
NGC3198    &       22.1 &       11.0 &      14.53 &      0.004 &      0.037 \\ 
NGC4062    &       26.8 &       18.1 &       2.36 &      0.001 &      0.034 \\ 
NGC4244    &       21.4 &        1.0 &       9.20 &      0.003 &      0.024 \\ 
NGC4258    &       29.6 &        4.0 &      38.66 &      0.012 &      0.074 \\ 
NGC4274    &       22.1 &       19.6 &       3.81 &      0.001 &      0.106 \\ 
NGC4414    &       24.8 &       18.5 &       2.06 &      0.001 &      0.021 \\ 
NGC4448    &       23.4 &        5.2 &       0.62 &      0.000 &      0.037 \\ 
NGC4559    &       23.2 &        3.4 &       4.13 &      0.001 &      0.035 \\ 
NGC4565    &       22.5 &        6.2 &       9.14 &      0.003 &      0.122 \\ 
NGC5023    &       27.5 &        2.8 &       4.64 &      0.001 &      0.017 \\ 
NGC5055    &       25.0 &        4.3 &      17.98 &      0.005 &      0.032 \\ 
NGC5229    &       23.8 &        1.5 &       0.98 &      0.000 &      0.004 \\ 
NGC5585    &       23.5 &        4.2 &       6.15 &      0.002 &      0.009 \\ 
UGC2082    &       27.0 &       12.6 &       2.14 &      0.001 &      0.025 \\ 
UGC4278    &       23.4 &       10.2 &       3.54 &      0.001 &      0.021 \\ 
UGC7774    &       27.5 &       38.6 &       3.49 &      0.001 &      0.019 \\ 
\hline 
\end{tabular} 
\end{table*} 
\begin{figure}[tbp]
\centering
\includegraphics[width=8cm]{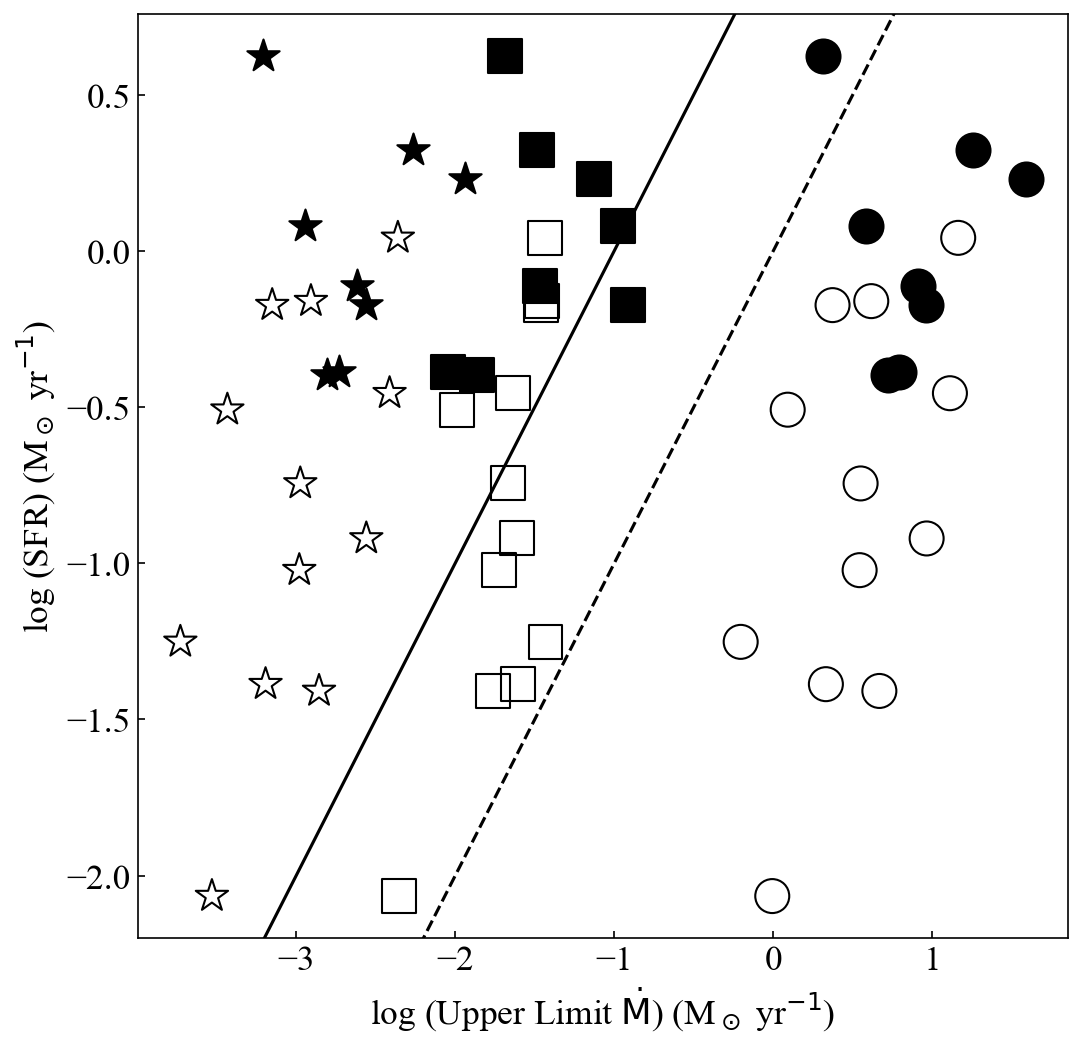}
\caption{  Upper limits on ${\rm \dot{M}}$ vs SFR of host galaxies. Circles are as calculated with $f$ =0.1, stars with $f=3\times10^{-5}$ and squares from scaling the GBT limits. Solid symbols represent hosts with cloud candidate detections and open symbols without. The solid line indicates  ${\rm \dot{M} = 0.1 \times SFR}$ and the dashed line marks ${\rm  \dot{M} = SFR }$. }
\label{fig:upperacc_sphere}
\end{figure}
\indent The values for the individual galaxies are shown in Figure \ref{fig:upperacc_sphere} and Table \ref{table:accretion_upper}. From this figure it is obvious that all accretion rate upper limits from the GBT limit are below the host's SFR, and only for galaxies with a low SFR the limits surpass 0.1 $\times$ SFR.\\
\subsection{Cloud Kinematics}\label{sec:cloud_kinematics}
\begin{figure}[tbp]
\centering
\includegraphics[width=8cm]{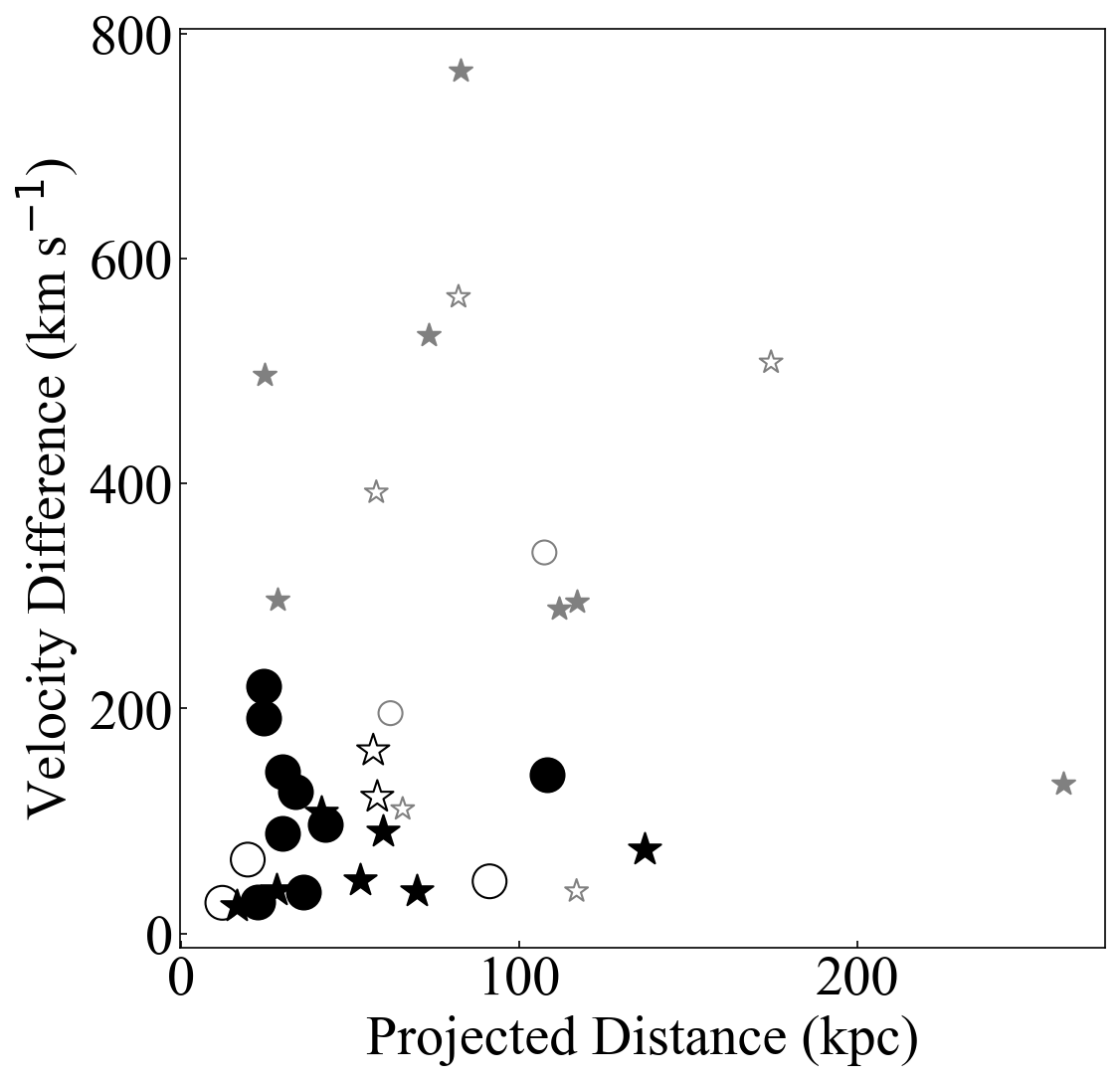}
\caption{Absolute velocity difference between the cloud or companion and the target plotted against projected distance of the cloud or companion from the target center. Open symbols counter-rotating, Filled symbols co-rotating. Large black symbols inside virial radius, small gray symbols outside  virial radius. Stars are companions, circles are the cloud candidates.}
\label{fig:kinematics}
\end{figure}

When investigating the individual fields we have marked every detection as co- or counter-rotating compared to the rotation in the disk of the target galaxy. A detection is considered co-rotating when its location and systemic velocity puts it on the same side of the target's central coordinates as the disk rotation, and counter-rotating when not.\\
\indent Figure \ref{fig:kinematics} shows these kinematics for all detections  in the sample. Here the cloud candidates (circles) and companions (stars) are separated and we differentiate between sources within the virial radius (As defined in $\S$ \ref{section:accretion_rate}) of the target and outside this radius (small gray symbols). Inside the virial radius both companions and cloud candidates appear to preferentially co-rotate. Out of 21 sources, 16 co-rotate. These 16 sources are almost equally split between cloud candidates and companions (9 and 7 respectively), making it likely that the cloud candidates and companions have a similar origin.\\
\section{Discussion}\label{section:discussion}
We have investigated  many different parameters (number of group members and group luminosity (as listed in \cite{Heald2011}), SFR, depletion time,  distance, mass, \HI\ richness) of our sample to see if we could find any difference between the targets with cloud candidates and those without. However, none of these parameters showed significant evidence of the two groups coming from different parent samples.\\
\subsection{Comparison with High Velocity Clouds.}
\begin{figure}[tbp]
\centering
\includegraphics[width=8cm]{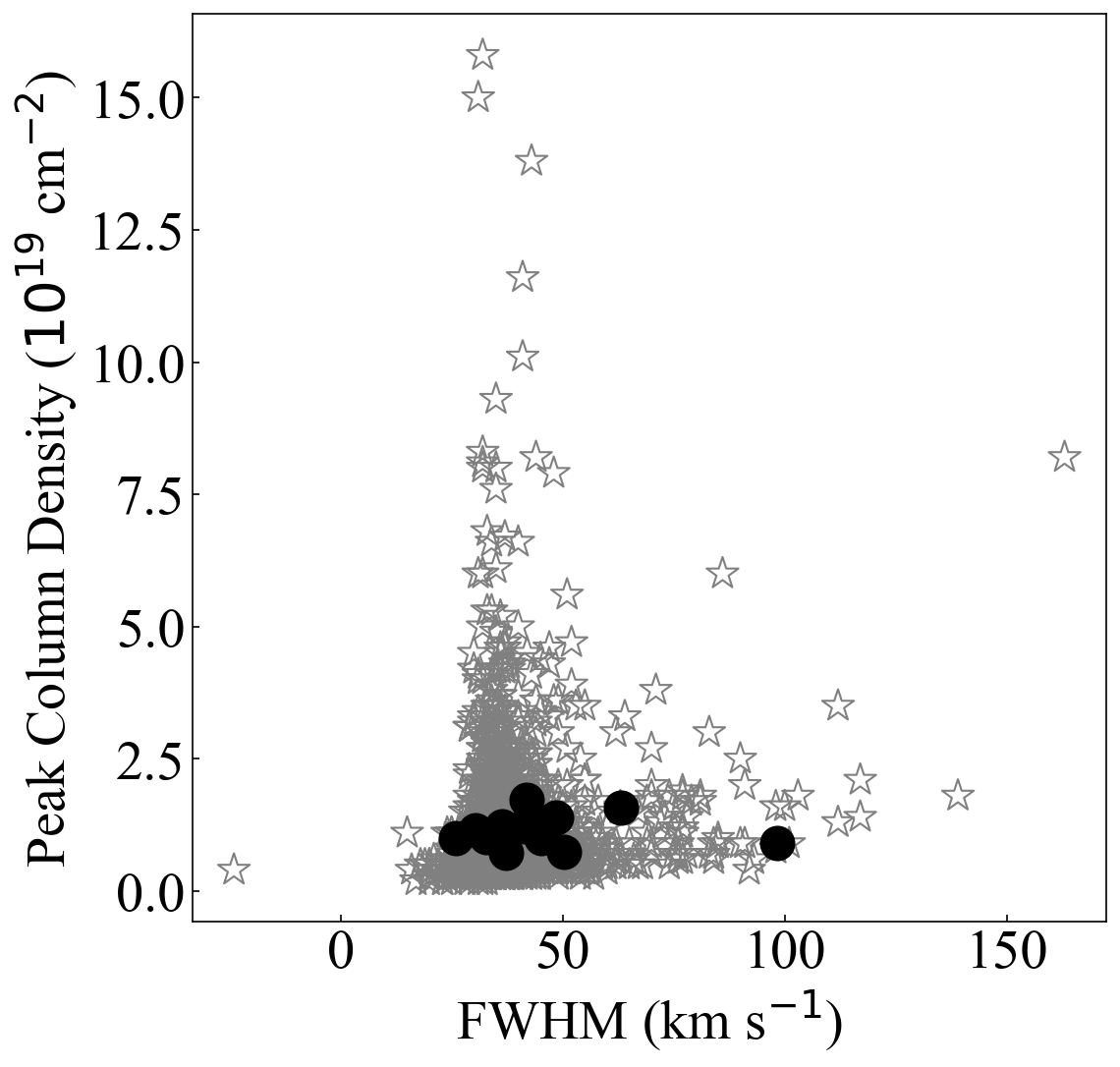}
\caption{FWHM vs. Peak Column for the Milky Way HVCs observed in HIPASS \citep[stars, ][]{Putman2002} and our detections (circles).} 
\label{fig:hvcvsNHI}
\end{figure}
\begin{figure}[tbp]
\centering
\includegraphics[width=8cm]{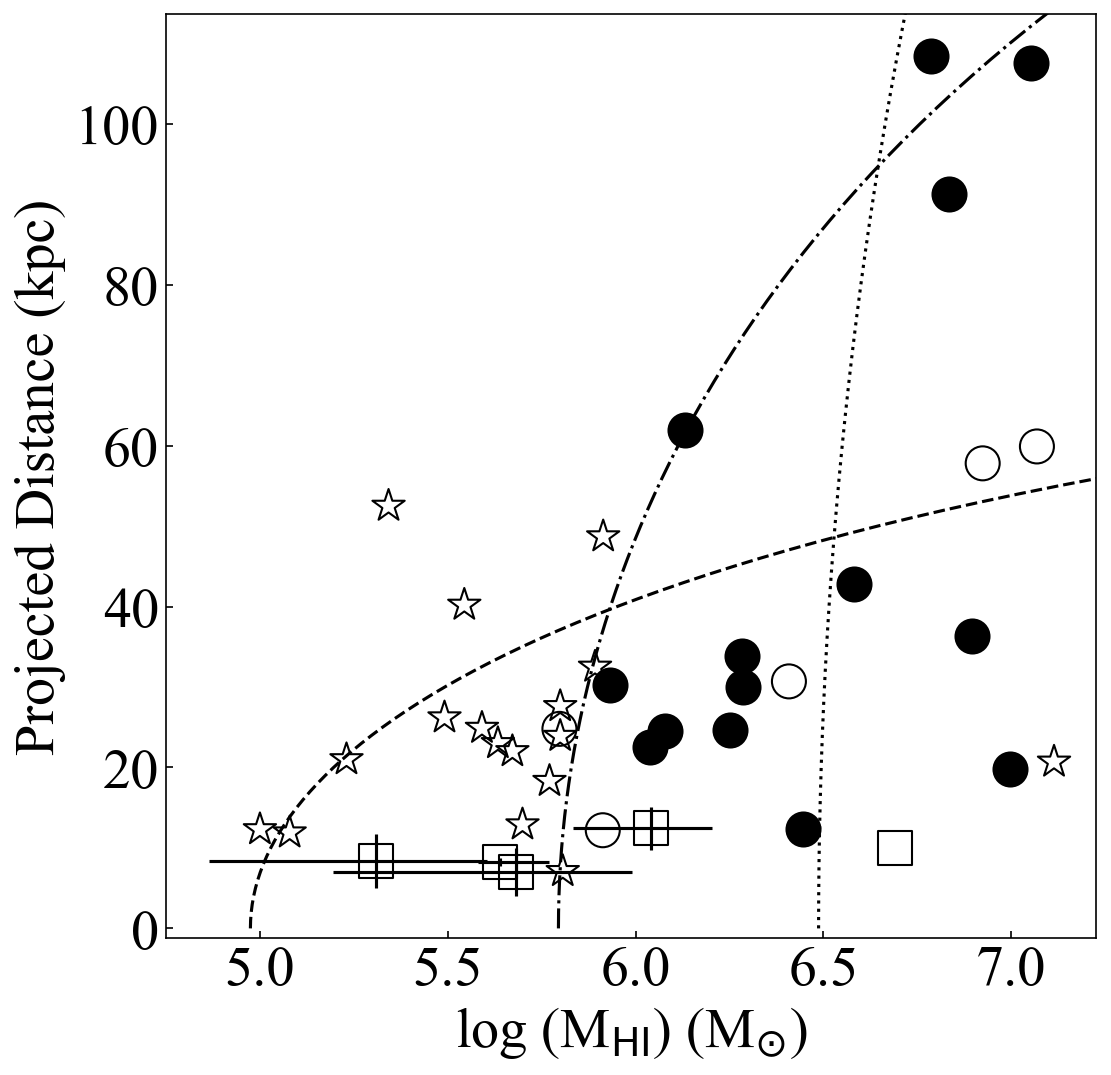}
\caption{Mass vs Projected Distance to host for the clouds surrounding M\,31 \citep[stars, ][]{Westmeier2008} and our detections. Filled circles are the cloud candidates, open circles are our detections with a stellar counterpart. Open squares indicate the Milky Way HVC complexes with their errors the actual detected ranges. The dashed, dot-dashed and dotted lines indicate the minimum, median and maximum detection limits in the survey galaxies. The limits are primary beam corrected and hence increase with increased distance to the host. The axes are not adjusted to show the companions hence only a few are visible.}
\label{fig:hvcvsMass}
\end{figure}
\indent We now investigate whether our cloud candidates can be analogues to the HVCs found in the MW and M\,31. We compare their velocity width and peak column densities to observations of HVCs as observed by the \hi\ Parkes All Sky Survey \citep[HIPASS, ][]{Putman2002} in Figure \ref{fig:hvcvsNHI}. Our detections occupy the same region in this diagram as the HVC population of the Milky Way. However, there are many more detections of clouds in the Milky Way than in our full sample. This is to be expected due to sensitivity effects and makes it questionable whether peak column density is a useful comparison as it is heavily dependent on how well the cloud in question is resolved. A better comparison is one with HVC mass. \\
\indent We compare our detections to the masses of the Milky Way HVC complexes, Complex C, Complex GCP, Complex WB, Chain A and the Cohen Stream \citep{vanWoerden1999,Thom2006,Thom2008,Wakker2008} where we take the mass at the mean of the lower and upper distance limits. As it is easier to obtain \hi\ masses for clouds in external galaxies, we add to the mass comparison the cloud complex/HVC analogues that were found in the M\,31 system \citep{Thilker2004, Westmeier2008}. This comparison is shown in Figure \ref{fig:hvcvsMass}. This figure shows that our detections are more massive than most of the detections in M\,31  and the MW. Additionally, far more clouds are detected around M\,31 than in the galaxies in our sample. This is not simply a sensitivity issue as we will explain below.\\
\indent The dashed, dot-dashed and dotted lines in Figure \ref{fig:hvcvsMass} indicate  the minimum, median and maximum detection limit in our sample of galaxies. The minimum and maximum correspond to the galaxies NGC\,4244 and UGC\,7774 respectively and the median limit is calculated using the median distance of 11.0 Mpc. The differences in sensitivity are pre-dominantly due to the projected distance to the host and the corresponding primary beam correction when projected distances and fluxes are converted to physical units. \\
\indent These sensitivity limits show that in our most sensitive observation we expect to detect the majority of a population of clouds as seen in M\,31. However, we find no cloud candidates in the cube for NGC\,4244. In this specific case this might still be due to imaging artifacts (see section \ref{Overview_Source_Images}) or low number statistics. However, also in the two closest, and hence  with the most sensitive detection limits in the center, galaxies after NGC\,4244 (NGC\,5229 and NGC\,5023) no cloud candidates are detected. In fact, we find few detections below the median detection limit. \\
\indent As we have done extensive testing on our source finding sensitivity limit in $\S$ \ref{section:method} it is unlikely that our sensitivity estimates should be increased by a factor of two. Figure \ref{fig:hvcvsMass} also shows that resolution is not an issue here as the projected distances of the M\,31 sample and our detections are in a similar range. Furthermore, the 4-5 most massive clouds of the M\,31 population fall securely in our detection range for several galaxies as detections for the sample overlap with clouds in M\,31. If we consider the mass range 7$\times10^5$ \msun < \mhi < 1.2$\times10^6$ \msun we find two M\,31 clouds  and at least one MW cloud. In the same mass bin in our sample there are three cloud candidates associated with three different galaxies. This means that we should have detected many more cloud candidates in several targets. However, the most cloud candidates we detect in a single galaxy is four.  \\
\indent In all targets of the survey we are sensitive to clouds like the most massive \hi\ complexes found in the MW and M\,31, i.e. Complex C and Davies Cloud (M$_{\rm HI}$ > 5$\times$10$^6$ \msun). However,  we only detect five cloud candidates with a mass $> 5\times 10^6$ \msun\ in a sample of 20 galaxies, two of which are associated with interacting galaxies (See $\S$ \ref{ind:4274}). This lack of massive cloud complexes such as present in the MW and M\,31 has already been previously seen towards the isolated galaxy NGC\,2903 \citep{Irwin2009}, but is now for the first time confirmed in a sample of galaxies and shows that the distribution of HVCs around M\,31 and the MW is not typical for other spiral galaxies. This poses the question what makes our sample so different from the MW and M\,31 or NGC\,891 for that matter? In our sample there is a hint of galaxies with more close-by companions having more cloud candidates but the current sample is too small to draw any strong conclusions.
\subsection{Implications on Cold Accretion Mode}\label{implicatcold}
Observations have shown that the required ongoing gas supply, of disk galaxies in the local Universe, cannot be satisfied by minor mergers with gas-rich companions \citep[this work, ][]{Sancisi2008, DiTeodoro2014}. Thus, accretion from intergalactic (primordial) gas potentially plays an important role for sustaining the star formation rates of large disk galaxies. From a theoretical perspective it is unclear whether this accretion should happen in a cold or hot mode in local galaxies \citep{Nelson2013, Nelson2015,Schaller2015,Huang2019}. Therefore collecting observational evidence on this accretion is crucial in understanding galaxy evolution.\\
\indent \cite{Fernandez2012} investigated  the \hi\ distribution around a simulated MW sized galaxy and found that the neutral gas being accreted consists of both compact clouds and filaments with peak column densities of $\sim 10^{19}\,{\rm cm}^{-2}$ out to at least 90\,kpc from the accreting galaxy. Our results do not show any significant sign of ongoing accretion of cold neutral gas at these levels despite being sensitive to them. Recent GBT observations suggest that there is neutral gas at even lower column densities \citep{Das2020}. However, as shown in $\S$ \ref{section:upper_limits}, this is not enough material to fuel current SFRs. Hence, if the cold-mode accretion seen in the simulations exists in reality, it has to be in the ionized state.\\  
\indent The accretion rates, both observed and upper limits, presented in this paper rule out the possibility that galaxies are accreting neutral hydrogen gas at the same rate as they are forming stars. However, it is still necessary to confirm this in even more sensitive observations since our main conclusion comes with several caveats.\\
\indent First, we do not identify accreting gas that is already connected to the main disk of its host. We have tested this by applying our source-finding method to NGC\,891 \citep{Oosterloo2007}. Even though most of the gas previously identified is detected, the bulk of it is connected to the disk. This means that in our analysis it would not be counted as accreting gas but as part of the main disk. In the case of NGC\,891, \sofia\ initially detects the filament and some clouds as separate sources that are then grown to merge with the disk. If we do not combine these with the host, this would result in an additional 0.25 \msun\ yr$^{-1}$ (See Figure \ref{fig:mdotvsSFR}), a significant amount. However, a visual inspection of the data and masks in our sample shows only few such structures in the HALOGAS sample (see section \ref{Overview_Source_Images} and \cite{Heald2020}).\\
\indent  A second caveat is that our calculation assumes that the infall timescale is well characterized by the free fall time which is likely to result in an underestimate of the timescale. Furthermore, the free fall time itself is likely to be over estimated due to us calculating it from the dynamical mass as measured from the \hi\  observation of the host galaxy (see $\S$ \ref{eq:mdyn}). Hence, our observed accretion rate limits are uncertain in both directions.  \\
\indent The upper limits calculated in this paper are calculated by assuming that accreted gas only becomes available for SF if it is accreted within the optical disk. This assumption is made because there are only few observations where large scale radial inflows are observed. Even in galaxies where such flows are detected \citep[e.g., NGC\,2403][]{Fraternali2002} the radial velocities are such that the migration of this gas to the inner disk would happen on timescales an order of magnitude larger than those sampled by the current SF.\\
\indent In our calculations we assume that the gas flows predominantly parallel to the angular momentum vector of the disks. This assumption assures a high accretion rate, as any other path would be longer. The thick disks around intermediately inclined HALOGAS galaxies show radial inflows of the order 20-30 \kms \citep{Marasco2019} which would result in longer infall times. On the other hand, the gas can be expected to follow the orientation of the magnetic fields in the halo \citep{Kwak2009} which are often observed to have a significant component perpendicular to the disk \citep[e.g.,][]{Stein2019}.  \\
\indent For clarity, we have not incorporated a helium component to the considered masses. However, in the literature, gas masses typically are assumed to be 1.3-1.4 times the neutral hydrogen mass, to account for the presence of helium. This correction would raise our combined average accretion limit to 0.13 \msun yr$^{-1}$), still significantly below the average SFR in the sample (0.8 \msun yr$^{-1}$). On average the galaxies in this sample are observed to accrete only 9\% of their current SFR in neutral gas  when the accretion rates are corrected for the helium contribution. Another 7\% of neutral hydrogen gas accretion could be hiding below current GBT sensitivity limits.  \\
\indent Finally, even though the HALOGAS collaboration has attempted to collect a set of SFRs as homogeneous and accurately as possible, it should be noted that SFRs for an individual galaxy are rather uncertain as they were derived from many different sources of information and include several assumptions and corrections. \\
\indent The above discussion shows that the limits derived from the HALOGAS sample are, and can only be, rough estimates.  Even though the survey shows that extended neutral gas reservoirs, such as found in NGC\,891, M\,31 or the MW, are exceptional rather than common place, single dish observations are still required to establish a  meaningful upper limit on the possible remaining gas reservoir that can hide below the sensitivity limits. Future observations such as the MeerKAT HI Observations of Nearby Galactic Objects - Observing Southern Emitters (MHONGOOSE) Survey \citep{deBlok2016} on MeerKAT will reach column density sensitivity limits  equal to the current GBT observations with a spatial resolution $\sim$ 8 times higher. This will significantly improve the statistics and self-consistency of the observations and reduce uncertainties in the analysis presented here.  
\section{Conclusions}\label{section:conclusions}
We have performed an automated search with \sofia\ \citep{Serra2015} of 22 galaxies in the HALOGAS survey \citep{Heald2011}. A mock data cube  has shown that with this method we can detect sources down to a limit of 21 mJy \kms\ integrated flux. Depending on the distance of the cloud, this translates to a detection threshold  in the range $\sim$ 0.5-3$\times10^6$ \msun.\\
\indent In our analysis we find  14 detections for which we are unable to find an optical counterpart. These detections account for an average accretion rate of 0.03 \msun\ yr$^{-1}$ over the whole sample. If we simply compare this to the average SFR in the sample, 0.8  \msun\ yr$^{-1}$, this would not be sufficient to replenish the gas used in star formation. When we consider all detections within the virial radius of our host galaxies, we confirm previous estimates from the literature and find a stringent mean upper limit of the neutral gas (\hi) accretion rate onto the disks of the galaxies of 0.22 \msun\ yr$^{-1}$. This upper limit is based on projected distances and can statistically be corrected to a limit for actual distances for the average. This correction lowers the observed accretion rate to 0.05  \msun\ yr$^{-1}$. \\
\indent Our cloud candidates have similar peak column densities and FWHM in their velocity distributions as the HVCs in our Galaxy. However, they are far less numerous. For the HVCs this could be merely a sensitivity issue as it is difficult to estimate the masses of this population. However, also when we compare our results to the HVC analogues found around M\,31 we find fewer cloud candidates than expected, indicating that this type of population is  missing or much less abundant in the HALOGAS galaxies. From our limited sample size we can not identify a definite cause for this difference but there are hints that galaxies with closer companions have more clouds. If this trend could be confirmed in a larger sample it is likely that our cloud candidates have a tidal origin instead of being primordial accretion. For now the question of why some galaxies display a HVC population and others do not remains open.\\
\indent The HALOGAS observations are some of the most sensitive observations at these resolutions, and the upper limits set by these observations on neutral gas accretion exclude that all accretion occurs in the form of neutral hydrogen down to column densities of a few times $10^{19}$ cm$^{-2}$. Due to uncertainties in the filling factor of the gas these limits still allow for a significant amount of neutral hydrogen to go undetected. However, current GBT limits are more stringent and based on these, we find an upper limit on the still-undetected gas accretion rate of 0.04 \msun yr$^{-1}$ for the HALOGAS sample. Combined with the observed accretion rate of other gas detections this leads to a total possible accretion rate of neutral hydrogen of ${\rm \dot{M}}$ = 0.09 \msun yr$^{-1}$ or ${\rm \dot{M}}$ = 0.13 \msun yr$^{-1}$ when accounting for the presence of helium.\\
\indent Thus the neutral hydrogen gas accretion onto $z$ = 0 galaxies is lower than their current SFR, and this probably means that this type of accretion no longer occurs on large scales or that it happens in another phase than the neutral gas phase. More importantly though, the HALOGAS survey shows that most nearby galaxies are not embedded in an extended neutral hydrogen reservoir as detected around some galaxies, e.g. NGC\,891. The more sensitive observations of the MHONGOOSE project \citep{deBlok2016} will be able to provide better insights what role neutral hydrogen plays in the accretion of gas onto galaxies  and why some galaxies are embedded in extended gas reservoirs while most are not.\\

\begin{acknowledgements}
The work at RUB is partially supported by the BMBF project 05A17PC2 for D-MeerKAT. N.H.R. further acknowledges support from the BMBF through the project D-LOFAR IV (FKZ: 05A17PC1). This project has received funding from the European Research Council (ERC) under the European Union’s Horizon 2020 research and innovation program grant agreement no. 882793, project name MeerGas. This research has made use of the VizieR catalog access tool, CDS, Strasbourg, France (DOI : 10.26093/cds/vizier). The original description 
 of the VizieR service was published in A\&AS 143, 23. This publication makes use of data products from the Wide-field Infrared Survey Explorer, which is a joint project of the University of California, Los Angeles, and the Jet Propulsion Laboratory/California Institute of Technology, funded by the National Aeronautics and Space Administration. This research has made use of the NASA/IPAC Extragalactic Database (NED) which is operated by the Jet Propulsion Laboratory, California Institute of Technology, under contract with the National Aeronautics and Space Administration
\end{acknowledgements}

\bibliographystyle{aa}
\bibliography{references}

\newcommand{\noop}[1]{}
\begin{thebibliography}{65}
\expandafter\ifx\csname natexlab\endcsname\relax\def\natexlab#1{#1}\fi

\bibitem[{{Adams} {et~al.}(2013){Adams}, {Giovanelli}, \& {Haynes}}]{Adams2013}
{Adams}, E. A.~K., {Giovanelli}, R., \& {Haynes}, M.~P. 2013, \apj, 768, 77

\bibitem[{{Boomsma} {et~al.}(2008){Boomsma}, {Oosterloo}, {Fraternali}, {van
  der Hulst}, \& {Sancisi}}]{Boomsma2008}
{Boomsma}, R., {Oosterloo}, T.~A., {Fraternali}, F., {van der Hulst}, J.~M., \&
  {Sancisi}, R. 2008, \aap, 490, 555

\bibitem[{{Bothwell} {et~al.}(2011){Bothwell}, {Kenicutt}, {Johnson}, {Wu},
  {Lee}, {Dale}, {Engelbracht}, {Calzetti}, \& {Skillman}}]{Bothwell2011}
{Bothwell}, M.~S., {Kenicutt}, R.~C., {Johnson}, B.~D., {et~al.} 2011, \mnras,
  415, 1815

\bibitem[{{Chiosi}(1980)}]{Chiosi1980}
{Chiosi}, C. 1980, \aap, 83, 206

\bibitem[{{Clark} {et~al.}(2018){Clark}, {Verstocken}, {Bianchi}, {Fritz},
  {Viaene}, {Smith}, {Baes}, {Casasola}, {Cassara}, {Davies}, {De Looze}, {De
  Vis}, {Evans}, {Galametz}, {Jones}, {Lianou}, {Madden}, {Mosenkov}, \&
  {Xilouris}}]{Clark2018}
{Clark}, C.~J.~R., {Verstocken}, S., {Bianchi}, S., {et~al.} 2018, \aap, 609,
  A37

\bibitem[{{Cluver} {et~al.}(2014){Cluver}, {Jarrett}, {Hopkins}, {Driver},
  {Liske}, {Gunawardhana}, {Taylor}, {Robotham}, {Alpaslan}, {Baldry}, {Brown},
  {Peacock}, {Popescu}, \& {Tuffs}}]{Cluver2014}
{Cluver}, M.~E., {Jarrett}, T.~H., {Hopkins}, A.~M., {et~al.} 2014, \apj, 782,
  90

\bibitem[{{Collins} {et~al.}(2002){Collins}, {Benjamin}, \&
  {Rand}}]{Collins2002}
{Collins}, J.~A., {Benjamin}, R.~A., \& {Rand}, R.~J. 2002, \apj, 578, 98

\bibitem[{{Daddi} {et~al.}(2007){Daddi}, {Dickinson}, {Morrison}, {Chary},
  {Cimatti}, {Elbaz}, {Frayer}, {Renzini}, {Pope}, {Alexander}, {Bauer},
  {Giavalisco}, {Huynh}, {Kurk}, \& {Mignoli}}]{Daddi2007}
{Daddi}, E., {Dickinson}, M., {Morrison}, G., {et~al.} 2007, \apj, 670, 156

\bibitem[{{Das} {et~al.}(2020){Das}, {Sardone}, {Leroy}, {Mathur}, {Gallagher},
  {Pingel}, {Pisano}, \& {Heald}}]{Das2020}
{Das}, S., {Sardone}, A., {Leroy}, A.~K., {et~al.} 2020, \apj, 898, 15

\bibitem[{{de Blok} {et~al.}(2016){de Blok}, {Adams}, {Amram}, {Athanassoula},
  {Bagetakos}, {Balkowski}, {Bershady}, {Beswick}, {Bigiel}, {Blyth}, {Bosma},
  {Booth}, {Bouchard}, {Brinks}, {Carignan}, {Chemin}, {Combes}, {Conway},
  {Elson}, {English}, {Epinat}, {Frank}, {Fiege}, {Fraternali}, {Gallagher},
  {Gibson}, {Heald}, {Henning}, {Holwerda}, {Jarrett}, {Jerjen}, {J{\'o}zsa},
  {Kapala}, {Kl{\"o}ckner}, {Koribalski}, {Kraan-Korteweg}, {Leon}, {Leroy},
  {Loubser}, {Lucero}, {McGaugh}, {Meurer}, {Meyer}, {Mogotsi}, {Namumba},
  {Oh}, {Oosterloo}, {Pisano}, {Popping}, {Ratcliffe}, {Sellwood},
  {Schinnerer}, {Schr{\"o}der}, {Sheth}, {Smith}, {Sorgho}, {Spekkens},
  {Stanimirovic}, {van der Heyden}, {Verdes-Montenegro}, {van Driel}, {Walter},
  {Westmeier}, {Wilcots}, {Williams}, {Wong}, {Woudt}, \&
  {Zijlstra}}]{deBlok2016}
{de Blok}, W.~J.~G., {Adams}, E.~A.~K., {Amram}, P., {et~al.} 2016, in
  Proceedings of MeerKAT Science: On the Pathway to the SKA. 25-27 May, 7

\bibitem[{{de Blok} {et~al.}(2014){de Blok}, {J{\'o}zsa}, {Patterson},
  {Gentile}, {Heald}, {J{\"u}tte}, {Kamphuis}, {Rand}, {Serra}, \&
  {Walterbos}}]{deBlok2014}
{de Blok}, W.~J.~G., {J{\'o}zsa}, G.~I.~G., {Patterson}, M., {et~al.} 2014,
  \aap, 566, A80

\bibitem[{{Di Teodoro} \& {Fraternali}(2014)}]{DiTeodoro2014}
{Di Teodoro}, E.~M. \& {Fraternali}, F. 2014, \aap, 567, A68

\bibitem[{{Dutta}(2019)}]{Dutta2019}
{Dutta}, R. 2019, Journal of Astrophysics and Astronomy, 40, 41

\bibitem[{{Eskew} {et~al.}(2012){Eskew}, {Zaritsky}, \& {Meidt}}]{Eskew2012}
{Eskew}, M., {Zaritsky}, D., \& {Meidt}, S. 2012, \aj, 143, 139

\bibitem[{{Fern{\'a}ndez} {et~al.}(2012){Fern{\'a}ndez}, {Joung}, \&
  {Putman}}]{Fernandez2012}
{Fern{\'a}ndez}, X., {Joung}, M.~R., \& {Putman}, M.~E. 2012, \apj, 749, 181

\bibitem[{{Fraternali} \& {Binney}(2008)}]{Fraternali2008}
{Fraternali}, F. \& {Binney}, J.~J. 2008, \mnras, 386, 935

\bibitem[{{Fraternali} {et~al.}(2015){Fraternali}, {Marasco}, {Armillotta}, \&
  {Marinacci}}]{Fraternali2015}
{Fraternali}, F., {Marasco}, A., {Armillotta}, L., \& {Marinacci}, F. 2015,
  \mnras, 447, L70

\bibitem[{{Fraternali} {et~al.}(2002){Fraternali}, {van Moorsel}, {Sancisi}, \&
  {Oosterloo}}]{Fraternali2002}
{Fraternali}, F., {van Moorsel}, G., {Sancisi}, R., \& {Oosterloo}, T. 2002,
  \aj, 123, 3124

\bibitem[{{Gentile} {et~al.}(2013){Gentile}, {J{\'o}zsa}, {Serra}, {Heald}, {de
  Blok}, {Fraternali}, {Patterson}, {Walterbos}, \& {Oosterloo}}]{Gentile2013}
{Gentile}, G., {J{\'o}zsa}, G.~I.~G., {Serra}, P., {et~al.} 2013, \aap, 554,
  A125

\bibitem[{{Heald} {et~al.}(2011{\natexlab{a}}){Heald}, {Allan}, {Zschaechner},
  {Kamphuis}, {Rand}, {J{\'o}zsa}, \& {Gentile}}]{Heald2011b}
{Heald}, G., {Allan}, J., {Zschaechner}, L., {et~al.} 2011{\natexlab{a}}, in
  IAU Symposium, Vol. 277, Tracing the Ancestry of Galaxies, ed. C.~{Carignan},
  F.~{Combes}, \& K.~C. {Freeman}, 59--62

\bibitem[{{Heald} \& {HALOGAS Team}(\noop{3001}in prep.)}]{Heald2020}
{Heald}, G. \& {HALOGAS Team}. \noop{3001}in prep., xxx, 000

\bibitem[{{Heald} {et~al.}(2011{\natexlab{b}}){Heald}, {J{\'o}zsa}, {Serra},
  {Zschaechner}, {Rand}, {Fraternali}, {Oosterloo}, {Walterbos}, {J{\"u}tte},
  \& {Gentile}}]{Heald2011}
{Heald}, G., {J{\'o}zsa}, G., {Serra}, P., {et~al.} 2011{\natexlab{b}}, \aap,
  526, A118+

\bibitem[{{Huang} {et~al.}(2019){Huang}, {Katz}, {Dav{\'e}}, {Fardal},
  {Kollmeier}, {Oppenheimer}, {Peeples}, {Roberts}, {Weinberg}, {Hopkins}, \&
  {Thompson}}]{Huang2019}
{Huang}, S., {Katz}, N., {Dav{\'e}}, R., {et~al.} 2019, \mnras, 484, 2021

\bibitem[{{Irwin} {et~al.}(2009){Irwin}, {Hoffman}, {Spekkens}, {Haynes},
  {Giovanelli}, {Linder}, {Catinella}, {Momjian}, {Koribalski}, {Davies},
  {Brinks}, {de Blok}, {Putman}, \& {van Driel}}]{Irwin2009}
{Irwin}, J.~A., {Hoffman}, G.~L., {Spekkens}, K., {et~al.} 2009, \apj, 692,
  1447

\bibitem[{{J{\'o}zsa} {et~al.}(2022){J{\'o}zsa}, {Jarrett}, {Cluver}, {Wong},
  {Havenga}, {Yao}, {Marchetti}, {Taylor}, {Kamphuis}, {Maccagni}, {Ramaila},
  {Serra}, {Smirnov}, {White}, {Kilborn}, {Holwerda}, {Hopkins}, {Brough},
  {Pimbblet}, {Driver}, \& {Kuijken}}]{Jozsa2022}
{J{\'o}zsa}, G.~I.~G., {Jarrett}, T.~H., {Cluver}, M.~E., {et~al.} 2022, \apj,
  926, 167

\bibitem[{{Kalberla} {et~al.}(2005){Kalberla}, {Burton}, {Hartmann}, {Arnal},
  {Bajaja}, {Morras}, \& {P{\"o}ppel}}]{Kalberla2005}
{Kalberla}, P.~M.~W., {Burton}, W.~B., {Hartmann}, D., {et~al.} 2005, \aap,
  440, 775

\bibitem[{{Kamphuis} {et~al.}(2013){Kamphuis}, {Rand}, {J{\'o}zsa},
  {Zschaechner}, {Heald}, {Patterson}, {Gentile}, {Walterbos}, {Serra}, \& {de
  Blok}}]{Kamphuis2013}
{Kamphuis}, P., {Rand}, R.~J., {J{\'o}zsa}, G.~I.~G., {et~al.} 2013, \mnras,
  434, 2069

\bibitem[{{Karachentsev} {et~al.}(2013){Karachentsev}, {Makarov}, \&
  {Kaisina}}]{KMK2013}
{Karachentsev}, I.~D., {Makarov}, D.~I., \& {Kaisina}, E.~I. 2013, \aj, 145,
  101

\bibitem[{{Kennicutt} {et~al.}(2009){Kennicutt}, {Hao}, {Calzetti},
  {Moustakas}, {Dale}, {Bendo}, {Engelbracht}, {Johnson}, \&
  {Lee}}]{Kennicutt2009}
{Kennicutt}, Robert~C., J., {Hao}, C.-N., {Calzetti}, D., {et~al.} 2009, \apj,
  703, 1672

\bibitem[{{Kwak} {et~al.}(2009){Kwak}, {Shelton}, \& {Raley}}]{Kwak2009}
{Kwak}, K., {Shelton}, R.~L., \& {Raley}, E.~A. 2009, \apj, 699, 1775

\bibitem[{{Marasco} {et~al.}(2019){Marasco}, {Fraternali}, {Heald}, {de Blok},
  {Oosterloo}, {Kamphuis}, {Jozsa}, {Vargas}, {Winkel}, {Walterbos}, {Dettmar},
  \& {Jutte}}]{Marasco2019}
{Marasco}, A., {Fraternali}, F., {Heald}, G., {et~al.} 2019, arXiv e-prints,
  arXiv:1909.04048

\bibitem[{{McGaugh} \& {Schombert}(2014)}]{McGaugh2014}
{McGaugh}, S.~S. \& {Schombert}, J.~M. 2014, \aj, 148, 77

\bibitem[{{McMullin} {et~al.}(2007){McMullin}, {Waters}, {Schiebel}, {Young},
  \& {Golap}}]{McMullin2007}
{McMullin}, J.~P., {Waters}, B., {Schiebel}, D., {Young}, W., \& {Golap}, K.
  2007, in Astronomical Society of the Pacific Conference Series, Vol. 376,
  Astronomical Data Analysis Software and Systems XVI, ed. R.~A. {Shaw},
  F.~{Hill}, \& D.~J. {Bell}, 127

\bibitem[{{Muller} {et~al.}(1963){Muller}, {Oort}, \& {Raimond}}]{Muller1963}
{Muller}, C.~A., {Oort}, J.~H., \& {Raimond}, E. 1963, Academie des Sciences
  Paris Comptes Rendus, 257, 1661

\bibitem[{{Nelson} {et~al.}(2015){Nelson}, {Genel}, {Vogelsberger}, {Springel},
  {Sijacki}, {Torrey}, \& {Hernquist}}]{Nelson2015}
{Nelson}, D., {Genel}, S., {Vogelsberger}, M., {et~al.} 2015, \mnras, 448, 59

\bibitem[{{Nelson} {et~al.}(2013){Nelson}, {Vogelsberger}, {Genel}, {Sijacki},
  {Kere{\v s}}, {Springel}, \& {Hernquist}}]{Nelson2013}
{Nelson}, D., {Vogelsberger}, M., {Genel}, S., {et~al.} 2013, \mnras, 429, 3353

\bibitem[{{Oosterloo} {et~al.}(2007){Oosterloo}, {Fraternali}, \&
  {Sancisi}}]{Oosterloo2007}
{Oosterloo}, T., {Fraternali}, F., \& {Sancisi}, R. 2007, \aj, 134, 1019

\bibitem[{{Pingel}(2018)}]{Pingel2018PhD}
{Pingel}, N.~M. 2018, PhD thesis, West Virginia University

\bibitem[{{Pingel} {et~al.}(2018){Pingel}, {Pisano}, {Heald}, {Jarrett}, {de
  Blok}, {J{\'o}zsa}, {J{\"u}tte}, {Rand}, {Oosterloo}, \&
  {Winkel}}]{Pingel2018}
{Pingel}, N.~M., {Pisano}, D.~J., {Heald}, G., {et~al.} 2018, \apj, 865, 36

\bibitem[{{Putman}(2017)}]{Putman2017}
{Putman}, M.~E. 2017, in Astrophysics and Space Science Library, Vol. 430, Gas
  Accretion onto Galaxies, ed. A.~{Fox} \& R.~{Dav{\'e}}, 1

\bibitem[{{Putman} {et~al.}(2002){Putman}, {de Heij}, {Staveley-Smith},
  {Braun}, {Freeman}, {Gibson}, {Burton}, {Barnes}, {Banks}, {Bhathal}, {de
  Blok}, {Boyce}, {Disney}, {Drinkwater}, {Ekers}, {Henning}, {Jerjen},
  {Kilborn}, {Knezek}, {Koribalski}, {Malin}, {Marquarding}, {Minchin},
  {Mould}, {Oosterloo}, {Price}, {Ryder}, {Sadler}, {Stewart}, {Stootman},
  {Webster}, \& {Wright}}]{Putman2002}
{Putman}, M.~E., {de Heij}, V., {Staveley-Smith}, L., {et~al.} 2002, \aj, 123,
  873

\bibitem[{{Putman} {et~al.}(2012){Putman}, {Peek}, \& {Joung}}]{Putman2012}
{Putman}, M.~E., {Peek}, J.~E.~G., \& {Joung}, M.~R. 2012, \araa, 50, 491

\bibitem[{{Rand}(1994)}]{Rand1994}
{Rand}, R.~J. 1994, \aap, 285, 833

\bibitem[{{Sancisi} {et~al.}(2008){Sancisi}, {Fraternali}, {Oosterloo}, \& {van
  der Hulst}}]{Sancisi2008}
{Sancisi}, R., {Fraternali}, F., {Oosterloo}, T., \& {van der Hulst}, T. 2008,
  \aapr, 15, 189

\bibitem[{{Schaller} {et~al.}(2015){Schaller}, {Dalla Vecchia}, {Schaye},
  {Bower}, {Theuns}, {Crain}, {Furlong}, \& {McCarthy}}]{Schaller2015}
{Schaller}, M., {Dalla Vecchia}, C., {Schaye}, J., {et~al.} 2015, \mnras, 454,
  2277

\bibitem[{{Serra} {et~al.}(2015){Serra}, {Westmeier}, {Giese}, {Jurek},
  {Fl{\"o}er}, {Popping}, {Winkel}, {van der Hulst}, {Meyer}, {Koribalski},
  {Staveley-Smith}, \& {Courtois}}]{Serra2015}
{Serra}, P., {Westmeier}, T., {Giese}, N., {et~al.} 2015, \mnras, 448, 1922

\bibitem[{{Skrutskie} {et~al.}(2006){Skrutskie}, {Cutri}, {Stiening},
  {Weinberg}, {Schneider}, {Carpenter}, {Beichman}, {Capps}, {Chester},
  {Elias}, {Huchra}, {Liebert}, {Lonsdale}, {Monet}, {Price}, {Seitzer},
  {Jarrett}, \& et~al.}]{Skrutskie2006}
{Skrutskie}, M.~F., {Cutri}, R.~M., {Stiening}, R., {et~al.} 2006, \aj, 131,
  1163

\bibitem[{{Stein} {et~al.}(2019){Stein}, {Dettmar}, {We{\.z}gowiec}, {Irwin},
  {Beck}, {Wiegert}, {Krause}, {Li}, {Heesen}, {Miskolczi}, {MacDonald}, \&
  {English}}]{Stein2019}
{Stein}, Y., {Dettmar}, R.~J., {We{\.z}gowiec}, M., {et~al.} 2019, \aap, 632,
  A13

\bibitem[{{Thilker} {et~al.}(2004){Thilker}, {Braun}, {Walterbos}, {Corbelli},
  {Lockman}, {Murphy}, \& {Maddalena}}]{Thilker2004}
{Thilker}, D.~A., {Braun}, R., {Walterbos}, R. A.~M., {et~al.} 2004, \apjl,
  601, L39

\bibitem[{{Thom} {et~al.}(2008){Thom}, {Peek}, {Putman}, {Heiles}, {Peek}, \&
  {Wilhelm}}]{Thom2008}
{Thom}, C., {Peek}, J.~E.~G., {Putman}, M.~E., {et~al.} 2008, \apj, 684, 364

\bibitem[{{Thom} {et~al.}(2006){Thom}, {Putman}, {Gibson}, {Christlieb},
  {Flynn}, {Beers}, {Wilhelm}, \& {Lee}}]{Thom2006}
{Thom}, C., {Putman}, M.~E., {Gibson}, B.~K., {et~al.} 2006, \apjl, 638, L97

\bibitem[{{van de Voort} {et~al.}(2019){van de Voort}, {Springel}, {Mandelker},
  {van den Bosch}, \& {Pakmor}}]{vdVoort2019}
{van de Voort}, F., {Springel}, V., {Mandelker}, N., {van den Bosch}, F.~C., \&
  {Pakmor}, R. 2019, \mnras, 482, L85

\bibitem[{{van der Hulst}(2002)}]{vdHulst2002}
{van der Hulst}, J.~M. 2002, in Astronomical Society of the Pacific Conference
  Series, Vol. 276, Seeing Through the Dust: The Detection of HI and the
  Exploration of the ISM in Galaxies, ed. A.~R. {Taylor}, T.~L. {Landecker}, \&
  A.~G. {Willis}, 84

\bibitem[{{van Eymeren} {et~al.}(2011){van Eymeren}, {J{\"u}tte}, {Jog},
  {Stein}, \& {Dettmar}}]{vEymeren2011}
{van Eymeren}, J., {J{\"u}tte}, E., {Jog}, C.~J., {Stein}, Y., \& {Dettmar},
  R.-J. 2011, \aap, 530, A29

\bibitem[{{van Woerden} {et~al.}(1999){van Woerden}, {Schwarz}, {Peletier},
  {Wakker}, \& {Kalberla}}]{vanWoerden1999}
{van Woerden}, H., {Schwarz}, U.~J., {Peletier}, R.~F., {Wakker}, B.~P., \&
  {Kalberla}, P. M.~W. 1999, \nat, 400, 138

\bibitem[{{Vargas} {et~al.}(2017){Vargas}, {Heald}, {Walterbos}, {Fraternali},
  {Patterson}, {Rand}, {J{\'o}zsa}, {Gentile}, \& {Serra}}]{Vargas2017}
{Vargas}, C.~J., {Heald}, G., {Walterbos}, R. A.~M., {et~al.} 2017, \apj, 839,
  118

\bibitem[{{Wakker} \& {van Woerden}(1997)}]{Wakker1997}
{Wakker}, B.~P. \& {van Woerden}, H. 1997, Annual Review of Astronomy and
  Astrophysics, 35, 217

\bibitem[{{Wakker} {et~al.}(2008){Wakker}, {York}, {Wilhelm}, {Barentine},
  {Richter}, {Beers}, {Ivezi{\'c}}, \& {Howk}}]{Wakker2008}
{Wakker}, B.~P., {York}, D.~G., {Wilhelm}, R., {et~al.} 2008, \apj, 672, 298

\bibitem[{{Wang} {et~al.}(2015{\natexlab{a}}){Wang}, {Wang}, {Kauffmann},
  {J{\'o}zsa}, \& {Li}}]{EWang2015}
{Wang}, E., {Wang}, J., {Kauffmann}, G., {J{\'o}zsa}, G. I.~G., \& {Li}, C.
  2015{\natexlab{a}}, \mnras, 449, 2010

\bibitem[{{Wang} {et~al.}(2015{\natexlab{b}}){Wang}, {Serra}, {J{\'o}zsa},
  {Koribalski}, {van der Hulst}, {Kamphuis}, {Li}, {Fu}, {Xiao}, {Overzier},
  {Wieringa}, \& {Wang}}]{Wang2015}
{Wang}, J., {Serra}, P., {J{\'o}zsa}, G. I.~G., {et~al.} 2015{\natexlab{b}},
  \mnras, 453, 2399

\bibitem[{{Westmeier} {et~al.}(2008){Westmeier}, {Br{\"u}ns}, \&
  {Kerp}}]{Westmeier2008}
{Westmeier}, T., {Br{\"u}ns}, C., \& {Kerp}, J. 2008, \mnras, 390, 1691

\bibitem[{{Westmeier} {et~al.}(2021){Westmeier}, {Kitaeff}, {Pallot}, {Serra},
  {van der Hulst}, {Jurek}, {Elagali}, {For}, {Kleiner}, {Koribalski},
  {Lee-Waddell}, {Mould}, {Reynolds}, {Rhee}, \&
  {Staveley-Smith}}]{Westmeier2021}
{Westmeier}, T., {Kitaeff}, S., {Pallot}, D., {et~al.} 2021, \mnras, 506, 3962

\bibitem[{{Wolfe} {et~al.}(2015){Wolfe}, {Pisano}, \& {Lockman}}]{Wolfe2015}
{Wolfe}, S., {Pisano}, D.~J., \& {Lockman}, F.~J. 2015, {Mapping diffuse
  neutral hydrogen with the GBT}, Tech. rep., Green Bank Telescope, NRAO, gBT
  Memo 289

\bibitem[{{Zschaechner} {et~al.}(2012){Zschaechner}, {Rand}, {Heald},
  {Gentile}, \& {J{\'o}zsa}}]{Zschaechner2012}
{Zschaechner}, L.~K., {Rand}, R.~J., {Heald}, G.~H., {Gentile}, G., \&
  {J{\'o}zsa}, G. 2012, \apj, 760, 37

\bibitem[{{Zschaechner} {et~al.}(2011){Zschaechner}, {Rand}, {Heald},
  {Gentile}, \& {Kamphuis}}]{Zschaechner2011}
{Zschaechner}, L.~K., {Rand}, R.~J., {Heald}, G.~H., {Gentile}, G., \&
  {Kamphuis}, P. 2011, \apj, 740, 35

\end{thebibliography}


\begin{appendix}\label{App}
\section{Source Catalog}\label{App_Source}
\onecolumn
\begin{landscape}\begin{longtable}{lllllllllllll}
\caption{\sofia\ detected sources.} \label{table:detec}\\ 
\hline 
Galaxy & ID & RA & Dec & Class & S$_{p}$ & $F_{i}$ & $F_{i}$,pbcorr& W20  & Vel. & Dist. to target & Flag & Counterpart \\ 
  &   &  [hms] & [dms] & & [mJy  bm$^{-1}$] & [Jy \kms]  & [Jy \kms]& [\kms] & [\kms] & [kpc] &  & \\ 
   (1) & (2) & (3) & (4) & (5) & (6) & (7) & (8) & (9) & (10) & (11) & (12) & (13)  \\  
\hline 
\endfirsthead 
\caption{continued.}\\ 
\hline 
Galaxy & ID & RA & Dec & Class  & S$_{p}$ & $F_{i}$ & $F_{i}$-pbcorr& W20  & Vel. & Dist. to target & Flag & Counterpart  \\ 
  &   &  [hms] & [dms] & &  [mJy bm$^{-1}$] & [Jy \kms] & [Jy \kms] & [\kms] & [\kms] & [kpc] &  &  \\ 
   (1) & (2) & (3) & (4) & (5) & (6) & (7) & (8) & (9) & (10) & (11) & (12) & (13)  \\  
\hline 
\endhead 
\hline 
\endfoot 
 
NGC\,0672 \tablefootmark{a} & 1          & 1h47m46.75s  & 27d24m9.62s  & T      & 53.474     & 249.158    & 252.323   & 276.9      & 397.1      & 0.0        & 1     & NGC\,672+IC\,1727              \\ 
NGC\,0672  & 7          & 1h46m22.55s  & 27d44m7.89s  & A      & 0.998      & 0.057      & 0.298     & 43.7       & 357.9      & 60.4       & 4     &                                \\ 
NGC\,0925 \tablefootmark{a} & 1          & 2h27m14.86s  & 33d34m53.99s & T      & 36.194     & 279.347    & 279.462   & 222.0      & 550.8      & 0.0        & 1     & NGC\,925                       \\ 
NGC\,0925  & 4          & 2h27m19.97s  & 33d57m30.11s & C      & 2.963      & 0.196      & 0.599     & 35.5       & 641.8      & 59.9       & 1     & Halogas                        \\ 
NGC\,0949 \tablefootmark{a} & 1          & 2h30m49.08s  & 37d8m18.34s  & T      & 13.686     & 18.522     & 18.522    & 207.2      & 607.7      & 0.0        & 1     & NGC\,949                       \\ 
NGC\,0949  & 2          & 2h28m1.99s   & 37d3m38.76s  & A      & 1.165      & 0.089      & 1.252     & 534.4      & 848.4      & 110.6      & 4     & Bad\,continuum\,subtraction.   \\ 
NGC\,1003 \tablefootmark{a} & 1          & 2h39m17.82s  & 40d52m20.74s & T      & 36.422     & 186.582    & 186.594   & 229.6      & 621.2      & 0.0        & 1     & NGC\,1003                      \\ 
NGC\,2541  & 1          & 8h12m39.68s  & 48d36m41.05s & C      & 1.369      & 0.096      & 1.347     & 47.0       & 517.4      & 117.0      & 1     & WISEA\,J081239.49+483645.3     \\ 
NGC\,2541 \tablefootmark{a} & 2          & 8h14m39.75s  & 49d3m46.30s  & T      & 29.115     & 139.790    & 139.793   & 209.4      & 555.1      & 0.0        & 1     & NGC\,2541                      \\ 
NGC\,3198 \tablefootmark{a} & 1          & 10h19m55.03s & 45d32m55.04s & T      & 24.100     & 226.582    & 226.586   & 317.4      & 659.6      & 0.0        & 1     & NGC\,3198                      \\ 
NGC\,3198  & 2          & 10h18m43.05s & 46d2m55.56s  & C      & 2.486      & 0.463      & 5.322     & 76.8       & 584.9      & 137.2      & 1     & VV\,834\,NED02                 \\ 
NGC\,4062 \tablefootmark{a} & 1          & 12h4m4.04s   & 31d53m43.28s & T      & 17.084     & 25.936     & 25.936    & 310.8      & 767.7      & 0.0        & 1     & NGC\,4062                      \\ 
NGC\,4244 \tablefootmark{a} & 1          & 12h17m28.97s & 37d48m20.46s & T      & 83.074     & 393.036    & 393.079   & 216.3      & 245.8      & 0.0        & 1     & NGC\,4244                      \\ 
NGC\,4244  & 2          & 12h18m41.54s & 37d41m40.03s & A      & 0.724      & 0.299      & 0.501     & 64.1       & 176.2      & 20.3       & 4     &                                \\ 
NGC\,4244  & 3          & 12h17m8.30s  & 38d1m27.46s  & A      & 1.095      & 1.384      & 2.038     & 62.2       & 170.4      & 17.6       & 4     &                                \\ 
NGC\,4244  & 4          & 12h16m43.51s & 37d46m18.88s & A      & 1.023      & 0.069      & 0.083     & 41.6       & 224.6      & 11.8       & 4     &                                \\ 
NGC\,4244  & 5          & 12h16m31.94s & 37d53m54.14s & A      & 0.953      & 3.781      & 5.265     & 77.0       & 308.9      & 16.1       & 4     &                                \\ 
NGC\,4244  & 6          & 12h17m50.07s & 38d4m58.72s  & A      & 0.603      & 0.055      & 0.100     & 46.7       & 265.3      & 22.0       & 4     &                                \\ 
NGC\,4244  & 7          & 12h18m3.79s  & 37d35m3.38s  & A      & 0.937      & 0.489      & 0.781     & 69.5       & 324.4      & 19.2       & 4     &                                \\ 
NGC\,4244  & 8          & 12h18m9.03s  & 37d40m6.91s  & A      & 0.927      & 0.582      & 0.760     & 52.9       & 328.3      & 14.6       & 4     &                                \\ 
NGC\,4258 \tablefootmark{a} & 1          & 12h18m55.27s & 47d17m52.04s & T      & 36.034     & 418.909    & 419.152   & 435.8      & 444.6      & 0.0        & 1     & MESSIER\,106                   \\ 
NGC\,4258  & 4          & 12h18m11.10s & 46d55m3.52s  & C      & 5.577      & 0.851      & 3.148     & 79.0       & 397.4      & 53.1       & 1     & SDSS\,J121811.04+465501.2      \\ 
NGC\,4258  & 5          & 12h18m56.49s & 47d5m22.48s  & A      & 1.041      & 0.133      & 0.187     & 35.9       & 441.0      & 27.6       & 3     &                                \\ 
NGC\,4258  & 6          & 12h17m50.73s & 47d24m25.38s & C      & 8.142      & 1.922      & 2.709     & 84.9       & 483.5      & 28.2       & 1     & NGC\,4248                      \\ 
NGC\,4258  & 7          & 12h18m13.34s & 47d13m21.86s & A      & 0.873      & 0.090      & 0.106     & 37.7       & 514.0      & 18.6       & 4     &                                \\ 
NGC\,4258  & 9          & 12h15m53.26s & 47d5m51.19s  & C      & 2.020      & 1.668      & 23.763    & 388.4      & 976.9      & 73.3       & 1     & NGC\,4217                      \\ 
NGC\,4258  & 10         & 12h19m27.48s & 47d18m42.60s & C*     & 1.811      & 0.057      & 0.060     & 39.0       & 811.3      & 12.2       & 1     & WISEA\,J121927.11+471844.3     \\ 
NGC\,4258  & 11         & 12h19m33.39s & 47d27m1.63s  & C      & 1.329      & 0.037      & 0.046     & 29.5       & 941.1      & 24.8       & 1     & WISEA\,J121933.24+472705.1     \\ 
NGC\,4274  & 1          & 12h20m9.23s  & 29d16m39.95s & C      & 1.011      & 2.319      & 5.725     & 472.3      & 637.8      & 117.1      & 1     & NGC\,4278                      \\ 
NGC\,4274  & 6          & 12h20m42.02s & 29d20m47.03s & C      & 1.969      & 0.384      & 0.865     & 128.6      & 644.1      & 111.7      & 1     & NGC\,4286                      \\ 
NGC\,4274 \tablefootmark{a} & 8          & 12h19m49.73s & 29d36m58.74s & T      & 7.013      & 8.883      & 8.884     & 463.0      & 933.0      & 0.0        & 1     & NGC\,4274                      \\ 
NGC\,4414 \tablefootmark{a} & 1          & 12h26m26.47s & 31d13m4.40s  & T      & 19.786     & 66.938     & 66.956    & 395.6      & 718.6      & 0.0        & 1     & NGC\,4414                      \\ 
NGC\,4414  & 7          & 12h24m13.24s & 31d31m9.96s  & C      & 2.333      & 1.629      & 23.095    & 196.8      & 1226.5     & 174.5      & 1     & NGC\,4359                      \\ 
NGC\,4448 \tablefootmark{a} & 1          & 12h28m15.33s & 28d37m13.98s & T      & 4.307      & 1.917      & 1.917     & 382.8      & 647.5      & 0.0        & 1     & NGC\,4448                      \\ 
NGC\,4448  & 2          & 12h26m9.77s  & 28d27m57.87s & C      & 1.769      & 0.236      & 1.577     & 53.3       & 1213.6     & 82.1       & 1     & IC\,3334                       \\ 
NGC\,4559 \tablefootmark{a} & 1          & 12h35m58.54s & 27d57m31.09s & T      & 41.176     & 296.186    & 296.216   & 253.1      & 809.6      & 0.0        & 1     & NGC\,4559                      \\ 
NGC\,4559  & 2          & 12h35m21.34s & 27d33m44.64s & C      & 2.300      & 0.143      & 0.570     & 50.3       & 1201.9     & 57.8       & 1     & WISEA\,J123521.10+273342.9     \\ 
NGC\,4565  & 1          & 12h37m1.18s  & 26d2m8.30s   & C*     & 2.790      & 0.078      & 0.093     & 25.7       & 856.2      & 30.7       & 1     & SDSS\,J123701.22+260208.5      \\ 
NGC\,4565 \tablefootmark{a} & 2          & 12h36m19.01s & 25d59m43.72s & T      & 63.489     & 253.191    & 253.430   & 517.4      & 1237.8     & 0.0        & 1     & HIJASS\,J1236+25               \\ 
NGC\,4565  & 4          & 12h35m34.84s & 25d51m3.95s  & C      & 10.837     & 4.418      & 6.326     & 140.7      & 1345.2     & 41.4       & 1     & NGC\,4562                      \\ 
NGC\,4565  & 5          & 12h36m20.09s & 26d5m1.24s   & C      & 10.387     & 1.337      & 1.431     & 43.9       & 1262.2     & 16.6       & 1     & IC\,3571                       \\ 
NGC\,4631 \tablefootmark{a} & 1          & 12h42m7.55s  & 32d32m14.76s & T      & 74.905     & 614.285    & 614.367   & 319.0      & 602.6      & 0.1        & 1     & NGC\,4631                      \\ 
NGC\,4631  & 2          & 12h43m55.95s & 32d10m28.93s & C      & 4.939      & 41.146     & 415.921   & 200.3      & 640.6      & 69.9       & 1     & 2MASS\,J12435666+3210138       \\ 
NGC\,4631  & 3          & 12h43m39.49s & 32d33m2.54s  & A      & 1.762      & 4.448      & 9.770     & 44.0       & 644.9      & 43.0       & 4     &                                \\ 
NGC\,4631  & 6          & 12h41m47.46s & 32d51m11.44s & A+C    & 5.648      & 1.089      & 2.373     & 60.2       & 682.3      & 42.9       & 2     & SDSS\,J124146.99+325124.8      \\ 
NGC\,4631  & 7          & 12h43m59.45s & 32d31m49.21s & A      & 1.189      & 0.083      & 0.274     & 47.9       & 681.2      & 52.2       & 4     &                                \\ 
NGC\,4631  & 8          & 12h41m26.89s & 32d20m51.36s & A      & 1.357      & 0.286      & 0.442     & 37.3       & 705.9      & 31.5       & 4     &                                \\ 
NGC\,4631  & 9          & 12h40m9.81s  & 32d39m20.10s & C      & 5.240      & 1.114      & 4.811     & 85.9       & 765.2      & 56.9       & 1     & SDSS\,J124010.08+323930.4      \\ 
NGC\,4631  & 10         & 12h43m7.01s  & 32d29m27.92s & C      & 7.686      & 0.929      & 1.302     & 66.3       & 899.6      & 28.5       & 1     & MCG\,+06-28-022                \\ 
NGC\,4631  & 11         & 12h42m1.52s  & 32d24m4.83s  & C*     & 2.377      & 0.082      & 0.095     & 35.1       & 908.3      & 18.2       & 1     & SDSS\,J124201.68+322356.8      \\ 
NGC\,5023 \tablefootmark{a} & 1          & 13h12m11.73s & 44d2m18.91s  & T      & 40.568     & 56.017     & 56.022    & 192.2      & 405.9      & 0.0        & 1     & NGC\,5023                      \\ 
NGC\,5055 \tablefootmark{a} & 1          & 13h15m45.64s & 42d1m26.13s  & T      & 22.779     & 430.681    & 431.171   & 393.5      & 498.1      & 0.0        & 1     & MESSIER\,63                    \\ 
NGC\,5055  & 4          & 13h13m54.32s & 42d12m39.61s & C      & 5.069      & 1.474      & 5.117     & 123.4      & 619.5      & 58.1       & 1     & UGC\,8313                      \\ 
NGC\,5055  & 6          & 13h18m44.29s & 41d56m57.63s & C      & 1.479      & 0.551      & 6.761     & 137.6      & 1265.3     & 82.8       & 1     & UGC\,8365                      \\ 
NGC\,5229 \tablefootmark{a} & 1          & 13h34m3.22s  & 47d54m49.51s & T      & 31.071     & 23.533     & 23.534    & 138.9      & 356.9      & 0.0        & 1     & NGC\,5229                      \\ 
NGC\,5585 \tablefootmark{a} & 2          & 14h19m49.28s & 56d44m1.81s  & T      & 24.749     & 146.362    & 146.392   & 161.7      & 306.9      & 0.0        & 1     & NGC\,5585                      \\ 
UGC\,2082 \tablefootmark{a} & 1          & 2h36m15.97s  & 25d25m25.56s & T      & 43.440     & 49.840     & 49.841    & 208.0      & 706.4      & 0.0        & 1     & UGC\,2082                      \\ 
UGC\,4278  & 1          & 8h14m30.44s  & 45d56m42.79s & A      & 1.958      & 0.127      & 0.183     & 179.9      & -124.8     & 52.5       & 4     &                                \\ 
UGC\,4278  & 2          & 8h14m30.45s  & 45d56m43.22s & A      & 1.791      & 0.070      & 0.101     & 77.2       & 64.2       & 52.5       & 4     &                                \\ 
UGC\,4278  & 3          & 8h14m30.49s  & 45d56m43.28s & A      & 1.822      & 0.143      & 0.206     & 197.6      & 194.1      & 52.5       & 4     &                                \\ 
UGC\,4278  & 4          & 8h13m14.79s  & 45d59m22.38s & C      & 6.853      & 12.274     & 21.774    & 113.2      & 448.9      & 65.6       & 1     & NGC\,2537                      \\ 
UGC\,4278 \tablefootmark{a} & 6          & 8h13m58.62s  & 45d44m39.78s & T      & 57.354     & 45.789     & 45.789    & 192.7      & 559.6      & 0.0        & 1     & IC\,2233                       \\ 
UGC\,4278  & 7          & 8h12m42.27s  & 45d36m54.45s & A      & 1.218      & 0.092      & 0.151     & 311.8      & 914.8      & 61.0       & 4     &                                \\ 
UGC\,4278  & 8          & 8h14m30.47s  & 45d56m42.78s & A      & 2.645      & 0.159      & 0.228     & 130.7      & 1319.9     & 52.5       & 4     &                                \\ 
UGC\,7774 \tablefootmark{a} & 1          & 12h36m23.16s & 40d0m18.06s  & T      & 31.820     & 25.198     & 25.198    & 205.5      & 521.7      & 0.0        & 1     & UGC\,7774                      \\ 
UGC\,7774  & 2          & 12h33m53.14s & 39d37m28.21s & C      & 1.177      & 0.220      & 5.888     & 40.8       & 654.9      & 260.9      & 1     & MCG\,+07-26-024                \\ 
\end{longtable} 
 \tablefoot{ 
\tablefoottext{*}{These cloud candidates have a counterpart identification that can not be confirmed in velocity.} 
\tablefoottext{a}{These are the values for the cloud searching \sofia\ run performed in this study.} 
} 
\end{landscape}
\twocolumn 
\section{Source Visualisation}
\subsection{Overview of the Target Fields}\label{Overview_Source_Images}
\subsubsection{NGC 0925} 
\begin{figure*}[htbp] 
\centering 
\includegraphics[width = 18 cm]{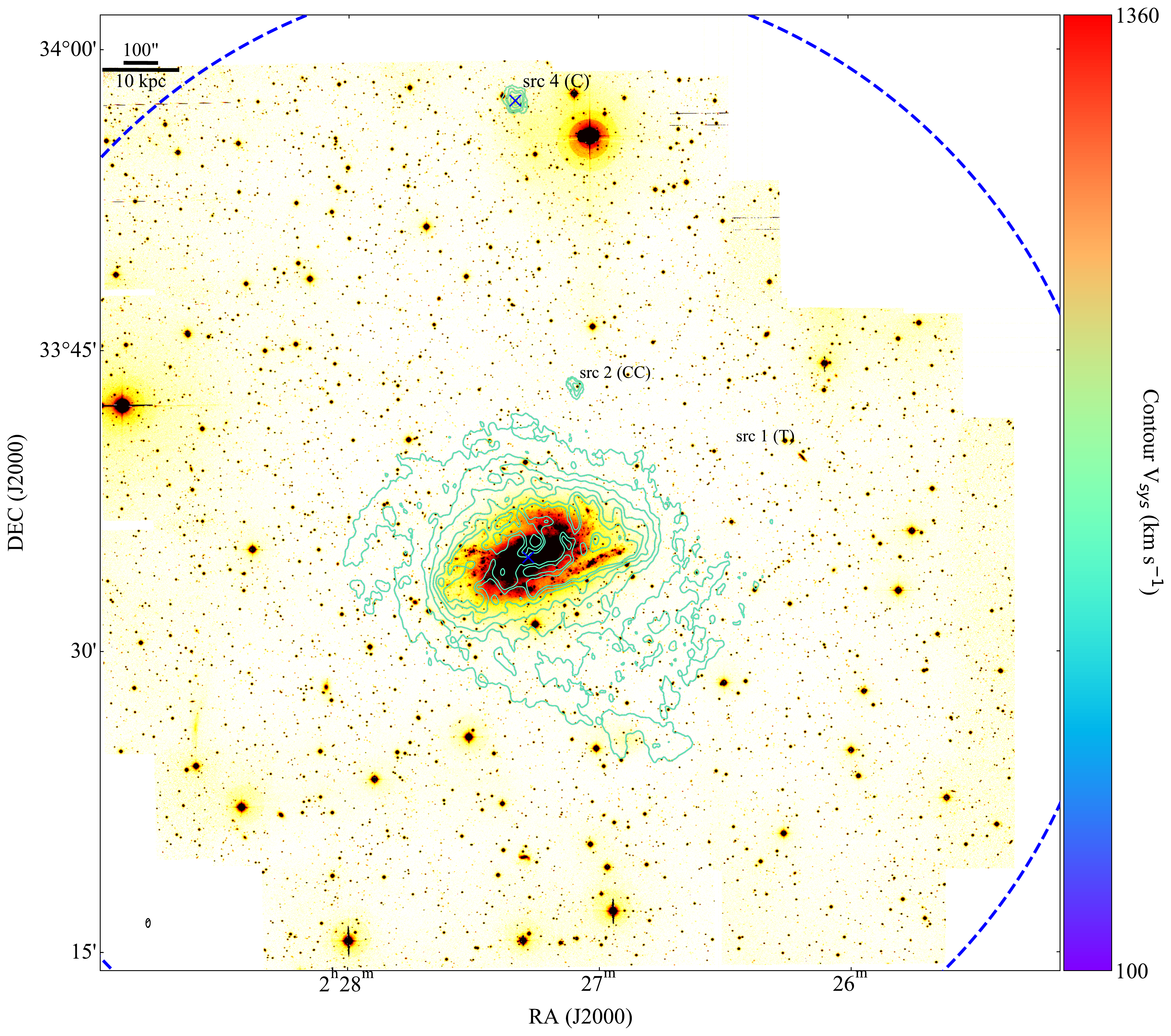} 
\caption{HALOSTARS image for the NGC\,0925 field overlaid with contours from the detected sources. Contour colours correspond to the systemic velocity of each source as determined by \sofia\ and indicated by the colorbar on the right. Contour levels vary per source and are listed in the captions of the individual images. The contours for the target correspond to 5.1, 20.5, 40.9, 81.8, 122.7, 162.3$\times 10^{19} {\rm cm}^{-2}$. The blue solid ellipse indicates the nominal virial radius (which can fall outside the field of view) and the dashed ellipse the FWHM of the WSRT primary beam. Labels indicate source number and class as listed in Tables \ref{table:accretion_clouds} and \ref{table:detec}. Blue crosses indicate galaxies spectroscopically confirmed to be in the velocity range of the cube.} 
\label{fig:Overview_n0925.pdf}
\end{figure*}
\begin{figure} 
\centering 
\includegraphics[width = 8 cm]{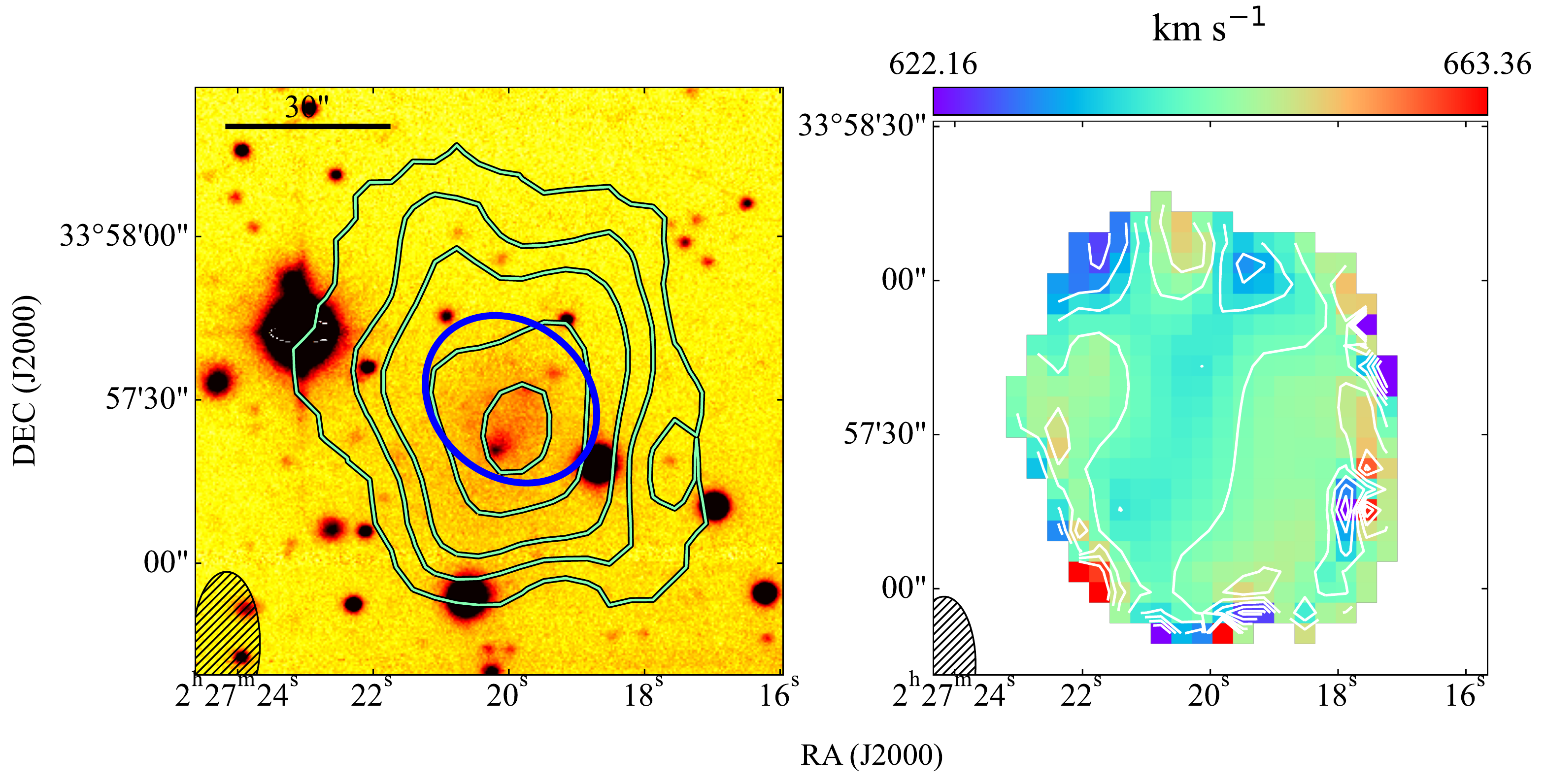} 
\caption{Moment maps for source no. 4 in the cube of NGC\,0925. Contours for moment 0 are: 1.9, 7.6, 15.2, 30.4, 45.6$\times 10^{19} {\rm cm}^{-2}$  overlaid on our HALOSTARS $R$-band image. These are corrected with the same factor as the integrated flux in the various tables to account for the primary beam response. The blue ellipse indicates the SoFiA fitted ellipse. Contours for the velocity field start at 622.16 km\,s$^{-1}$ and increase with 5.00 km s$^{-1}$.} 
\label{fig:n0925_image_src_4.pdf}
\end{figure}

NGC\,925 is a spiral galaxy with an extended gas distribution and a XUV disk. Besides the target we find two more \hi\ sources. One detection (source 2) is found relatively close to the target galaxy and could be part of the extended low surface brightness disk. The projected distance to the center of the target galaxy  is $\sim$ 23 kpc  and its systemic velocity only differs by 30 \kms\ from NGC\,925. We could find no optical counterpart and hence it is classified as a cloud candidate. Technically the cloud is counter rotating. However, it is very close to the kinematic minor axis of the galaxy making the kinematical assessment difficult.\\
\indent The other detection (source 4) is separated from the target galaxy. It has an UV counterpart, as stated in \cite{KMK2013}, and also in our deep $R$-band image an optical counterpart can be seen (See Figure \ref{fig:n0925_image_src_4.pdf}). Its systemic velocity deviates about 90 \kms  from  NGC\,925  and it is at a projected distance of $\sim$ 62 kpc from the center of NGC\,925 and is co-rotating with its disk. 
\subsubsection{NGC 1003} 
\begin{figure*}[htbp] 
\centering 
\includegraphics[width = 18 cm]{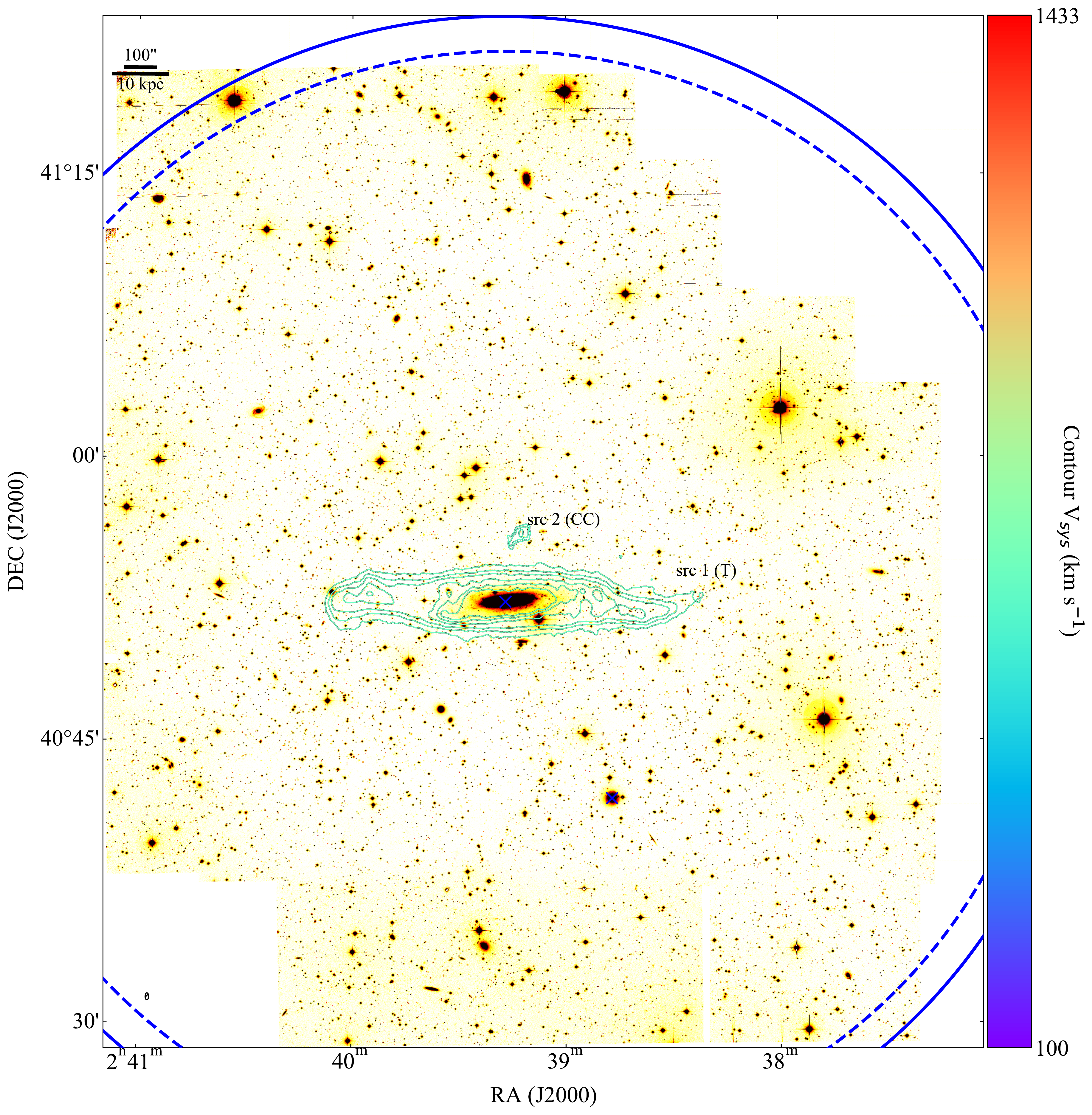} 
\caption{As Figure \ref{fig:Overview_n0925.pdf} but for the NGC\,1003 field.} 
\label{fig:Overview_n1003.pdf}
\end{figure*}
This is an edge-on galaxy of which the \hi\ disk extends far beyond the visible optical disk. In the outer parts of the disk there are hints of warping and several clouds, considered to be HVC analogues, can be seen around the main disk \citep{Heald2011b}. These clouds are all detected however only one of them is separated from the emission in the disk (see section \ref{implicatcold} on how this affects our results).\\
\indent Even though this counter rotating  source is separated by more than a FWHM from the mask encompassing the emission of NGC\,1003, it is still close to the disk. The HALOSTARS image (Figure \ref{fig:n1003_image_src_2.pdf}) does not show an optical counterpart at its location, but the field is crowded with foreground stars making identification of low level optical emission difficult. 
\subsubsection{NGC 2541} 
\begin{figure*}[htbp] 
\centering 
\includegraphics[width = 18 cm]{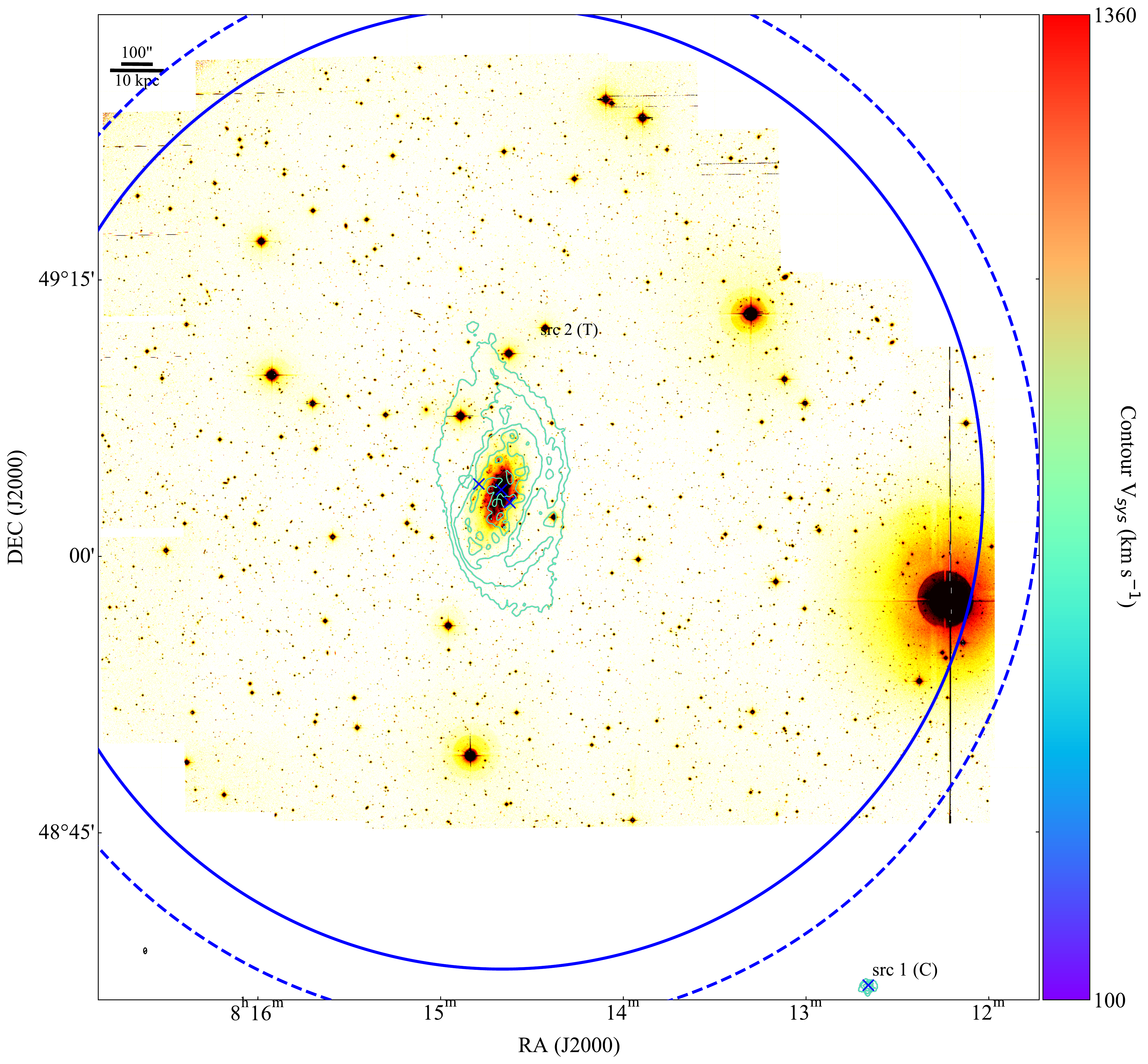} 
\caption{As Figure \ref{fig:Overview_n0925.pdf} but for the NGC\,2541 field.} 
\label{fig:Overview_n2541.pdf}
\end{figure*}
\begin{figure} 
\centering 
\includegraphics[width = 8 cm]{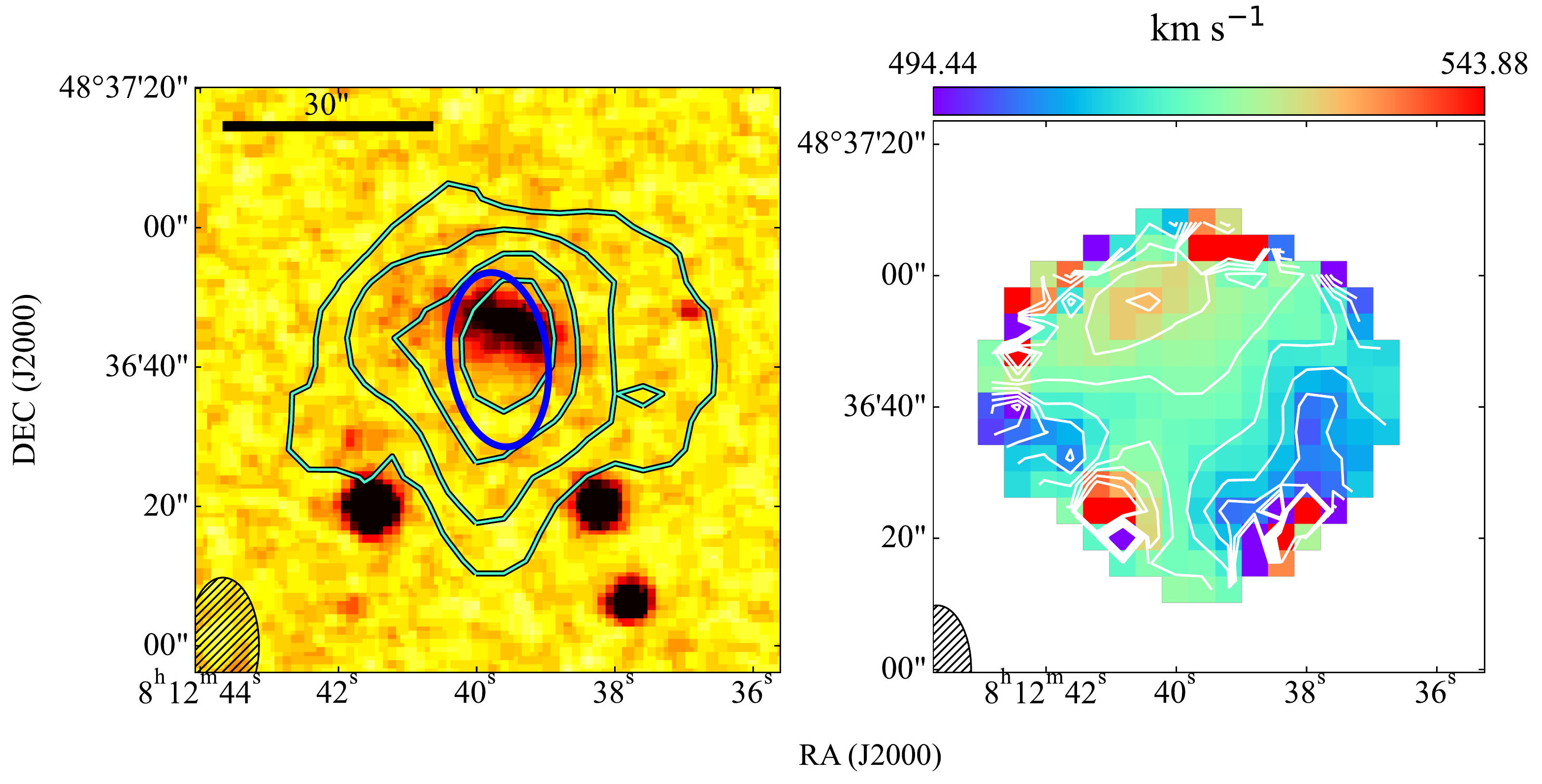} 
\caption{ As \ref{fig:n0925_image_src_4.pdf} but for source no. 1 in the cube of NGC\,2541. Contours for moment 0 are: 1.4, 5.5, 11.0, 16.5$\times 10^{20} {\rm cm}^{-2}$ but overlaid on a DDS 2 $R$-band image as it is outside the HALOSTARS field. Contours for the velocity field start at 494.44 km\,s$^{-1}$ and increase with 5.00 km s$^{-1}$. This source is identified as WISEA\,J081239.49+483645.3.} 
\label{fig:n2541_image_src_1.pdf}
\end{figure}

A companion is found at a projected distance of 120 kpc with a velocity difference of only $\sim 40$ \kms (See Figure \ref{fig:n2541_image_src_1.pdf}). Even though the galaxy is warped, it appears rather regular and undisturbed. The detected companion is on the edge of the cube and hence its detection is too faint in \HI\ to provide any meaningful analysis about its structure or kinematics. 
\subsubsection{NGC 3198} 
\begin{figure*}[htbp] 
\centering 
\includegraphics[width = 18 cm]{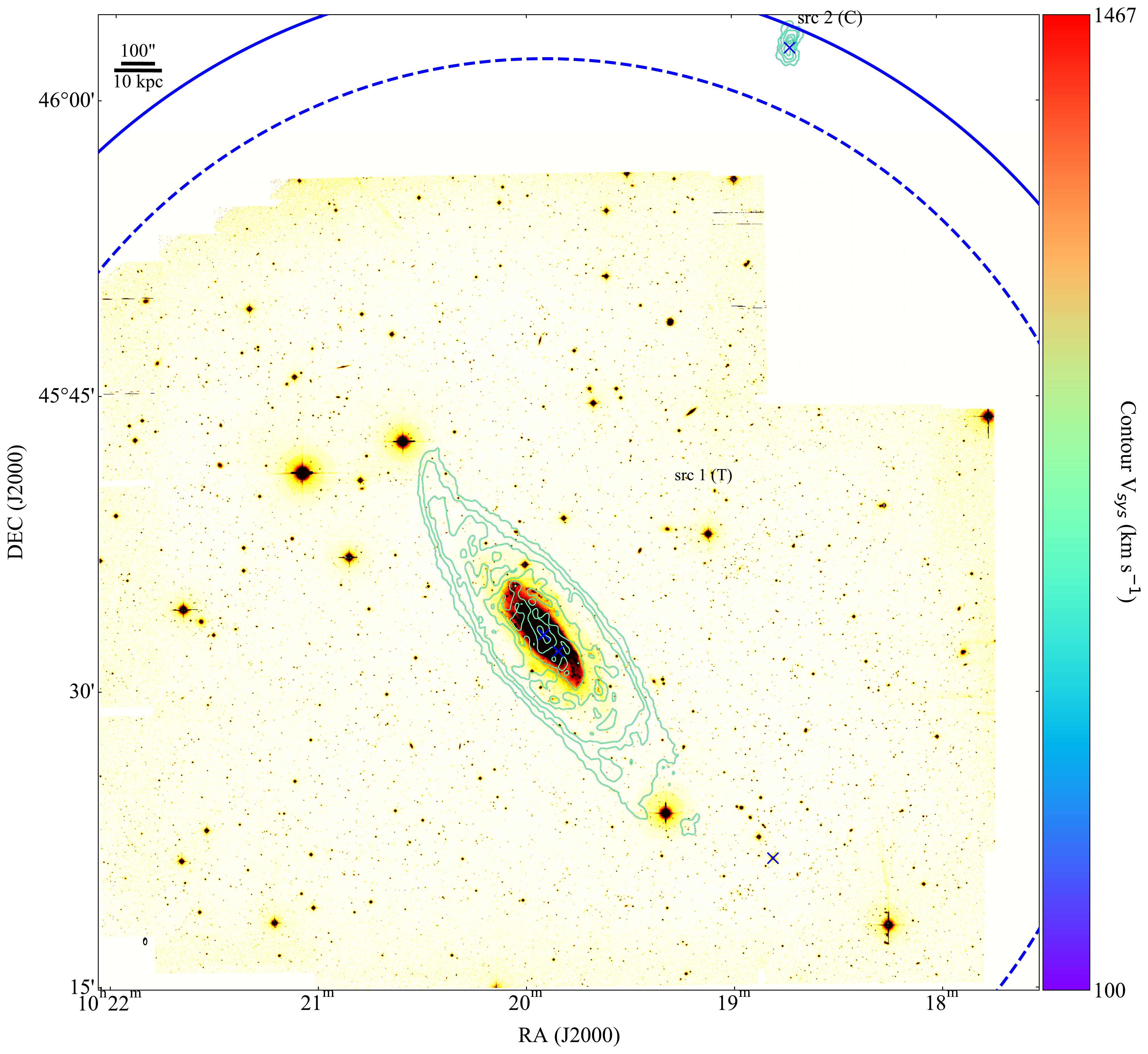} 
\caption{As Figure \ref{fig:Overview_n0925.pdf} but for the NGC\,3198 field.} 
\label{fig:Overview_n3198.pdf}
\end{figure*}
This galaxy is extensively investigated for its halo properties in \cite{Gentile2013}.  It is a rather regular galaxy  where we detect a single companion galaxy at a projected distance $\sim$ 140 kpc. The systemic velocity of this companion is aligned with the rotation of NGC\,3198's disk. 
\subsubsection{NGC 4258} 
\begin{figure*}[htbp] 
\centering 
\includegraphics[width = 18 cm]{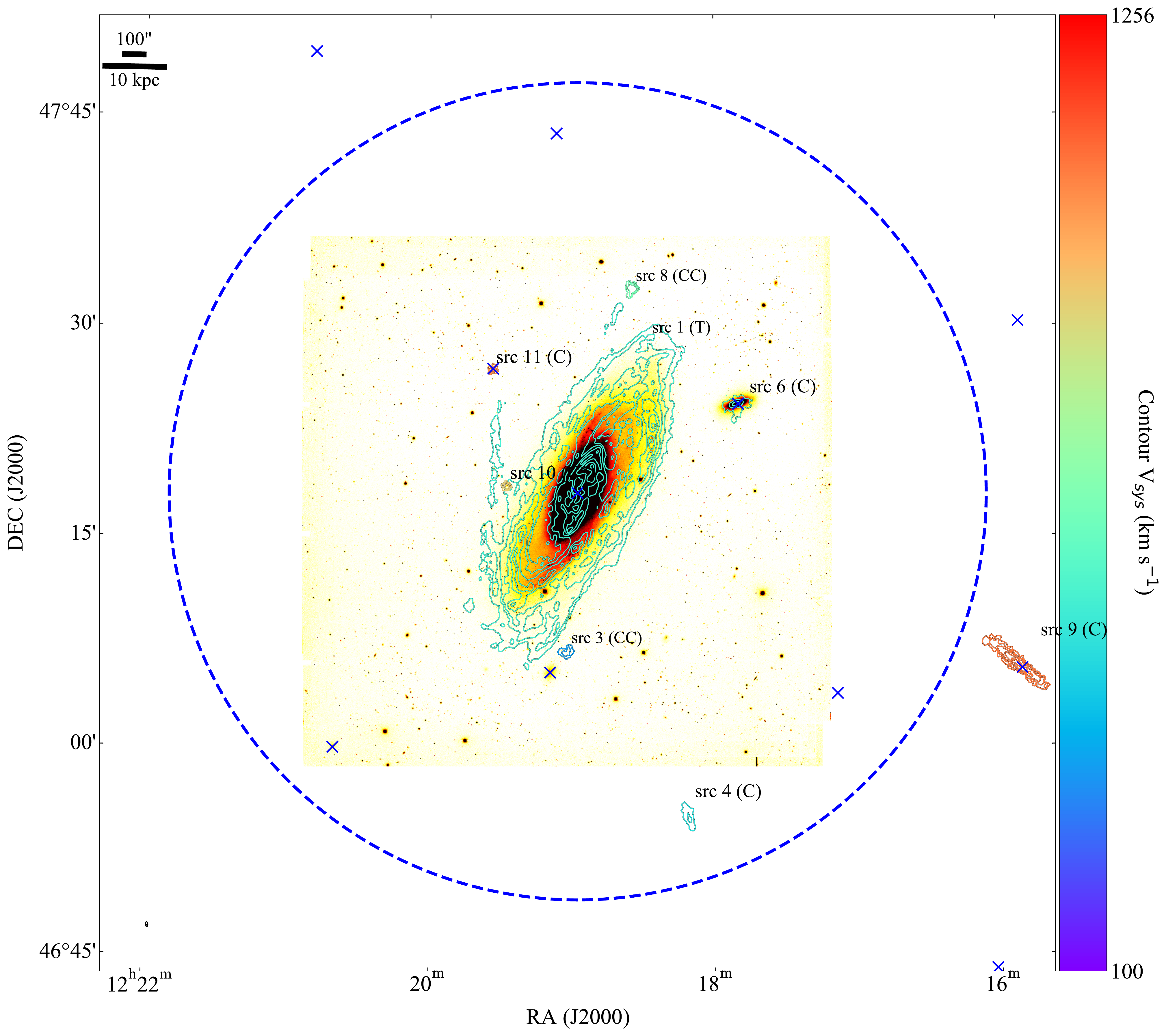} 
\caption{As Figure \ref{fig:Overview_n0925.pdf} but for the NGC\,4258 field overlaid on a KPNO $R$-band image.} 
\label{fig:Overview_n4258.pdf}
\end{figure*}
\begin{figure} 
\centering 
\includegraphics[width = 8 cm]{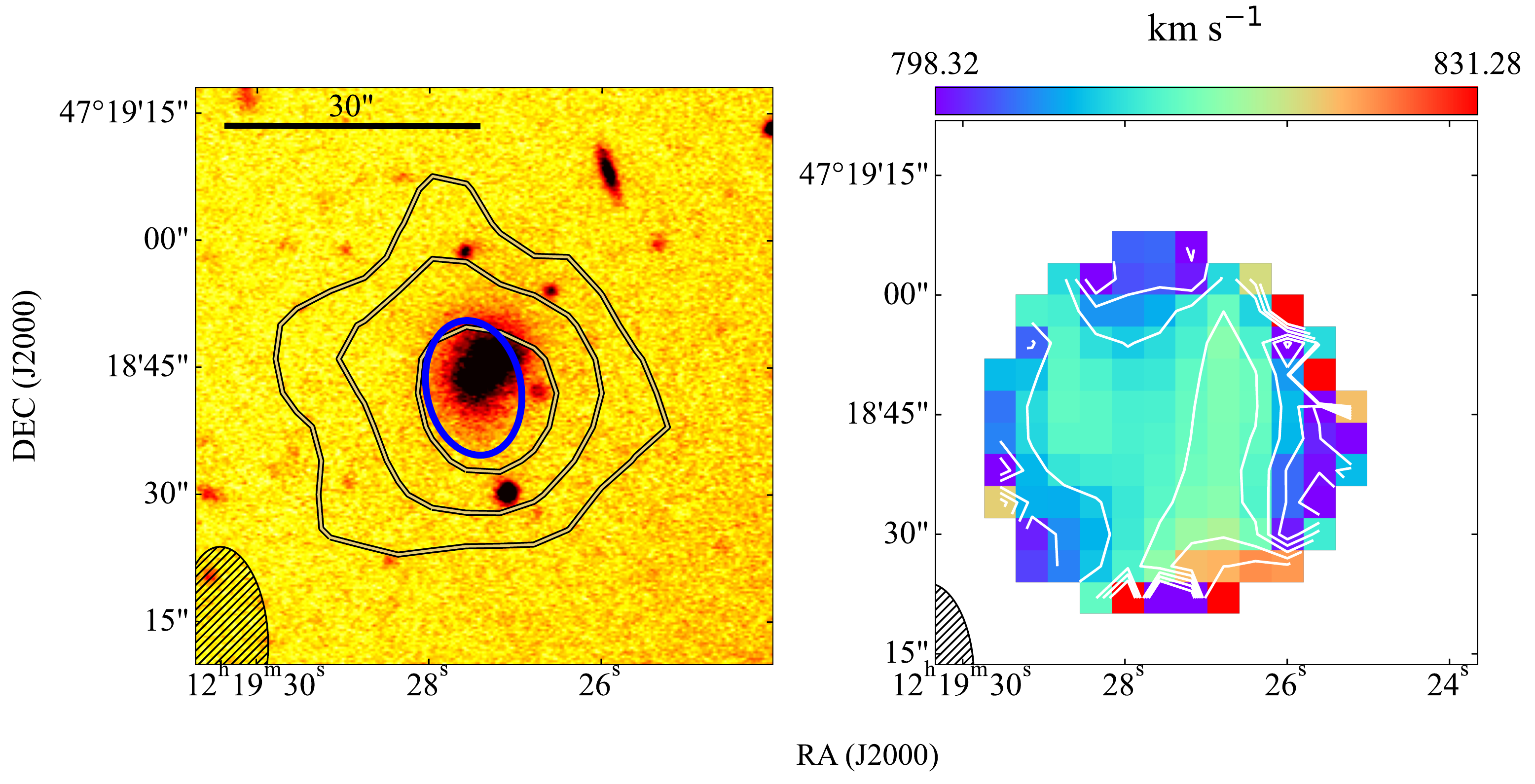} 
\caption{ As \ref{fig:n0925_image_src_4.pdf} but for source no. 10 in the cube of NGC\,4258. Contours for moment 0 are: 1.2, 4.8, 9.6$\times 10^{19} {\rm cm}^{-2}$ but overlaid on our KPNO $R$-band image. Contours for the velocity field start at 798.32 km\,s$^{-1}$ and increase with 5.00 km s$^{-1}$.} 
\label{fig:n4258_image_src_10.pdf}
\end{figure}

The field of NGC\,4258 is a rich field with several detections. The galaxy has four smaller companions which makes past or ongoing weak interaction very likely.  Besides the four companions the galaxy has several tails that run off from the main disk. These typically end in fragmented clouds. Sometimes these are connected to the disk, and merged with the galaxy in our source finding (see Figure \ref{fig:Overview_n4258.pdf} in the SE). In other instances, as in the case of our sources 3 and 8, these are distinct enough to be counted as individual detections. However, they all appear connected to the disk in some shape or form. As such it is impossible to tell whether they are infalling or being tidally stripped from the galaxy. All these detections follow the general rotation of the disk. Sources 9 and 11 are far beyond the velocity range of the disk however. \\
\indent For source 10 we could not spectroscopically confirm an optical counterpart however  our KPNO $R$ band image \ref{fig:n4258_image_src_10.pdf} shows a smudge of light. Hence it is classified as a companion. This source occurs close to the minor axis of the host and has a systemic velocity $\sim 75$ \kms, higher than the highest velocity in the disk of NGC\,4258, technically this means it is counter rotating. 
\subsubsection{NGC 4274}\label{ind:4274} 
\begin{figure*}[htbp] 
\centering 
\includegraphics[width = 18 cm]{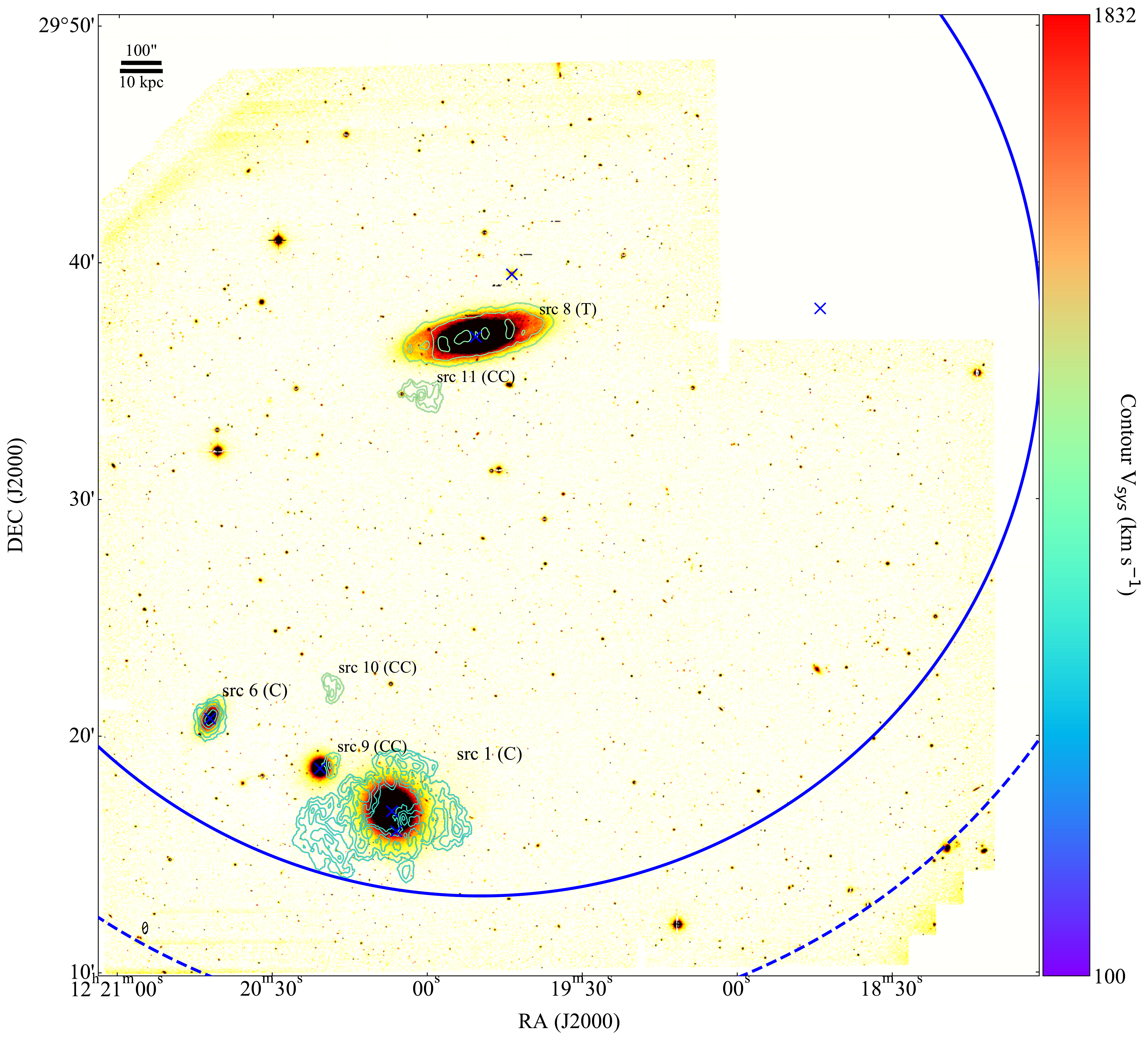} 
\caption{As Figure \ref{fig:Overview_n0925.pdf} but for the NGC\,4274 field.} 
\label{fig:Overview_n4274.pdf}
\end{figure*}
NGC\,4274 is another rich field with at least three companion galaxies, NGC\,4286, NGC\,4283 and NGC\,4278. All  are about 90\,kpc to the south of the target and much closer to each other than to NGC\,4274. Their systemic velocities and locations are aligned with the rotation in the disk of NGC\,4274.  The \HI\ in NGC\,4278 is sporadic and fragmented. However, this appears to be more an effect of sensitivity than the actual distribution. Some \HI\ is detected toward NGC\,4283 but it appears to be part of foreground emission coming from NGC\,4278.\\
\indent We find three cloud candidates in this field. Two (Source 9 and 10) are closer to the system of companion galaxies and are most likely tidal debris of interactions between these galaxies. Source 10 would be counter rotating when compared to the disk of NGC\,4274 but is co-rotating with the companion galaxies and source 9 co-rotating with NGC\,4274.\\
\indent Source 11 is  close to NGC\,4274, extended and has a hint of rotation. Unfortunately the spatial resolution of the HALOGAS observations is not sufficient to confirm this rotation. Even in deep-follow up observations there appears to be no optical counterpart and hence we refer to this object as {\it'Dark galaxy candidate'}. If truly a galaxy, possibly with an undetected optical counterpart, it would be a companion instead of a cloud. This cloud candidate is counter-rotating from the perspective of NGC\,4274's disk.
\subsubsection{NGC 4414} 
\begin{figure*}[htbp] 
\centering 
\includegraphics[width = 18 cm]{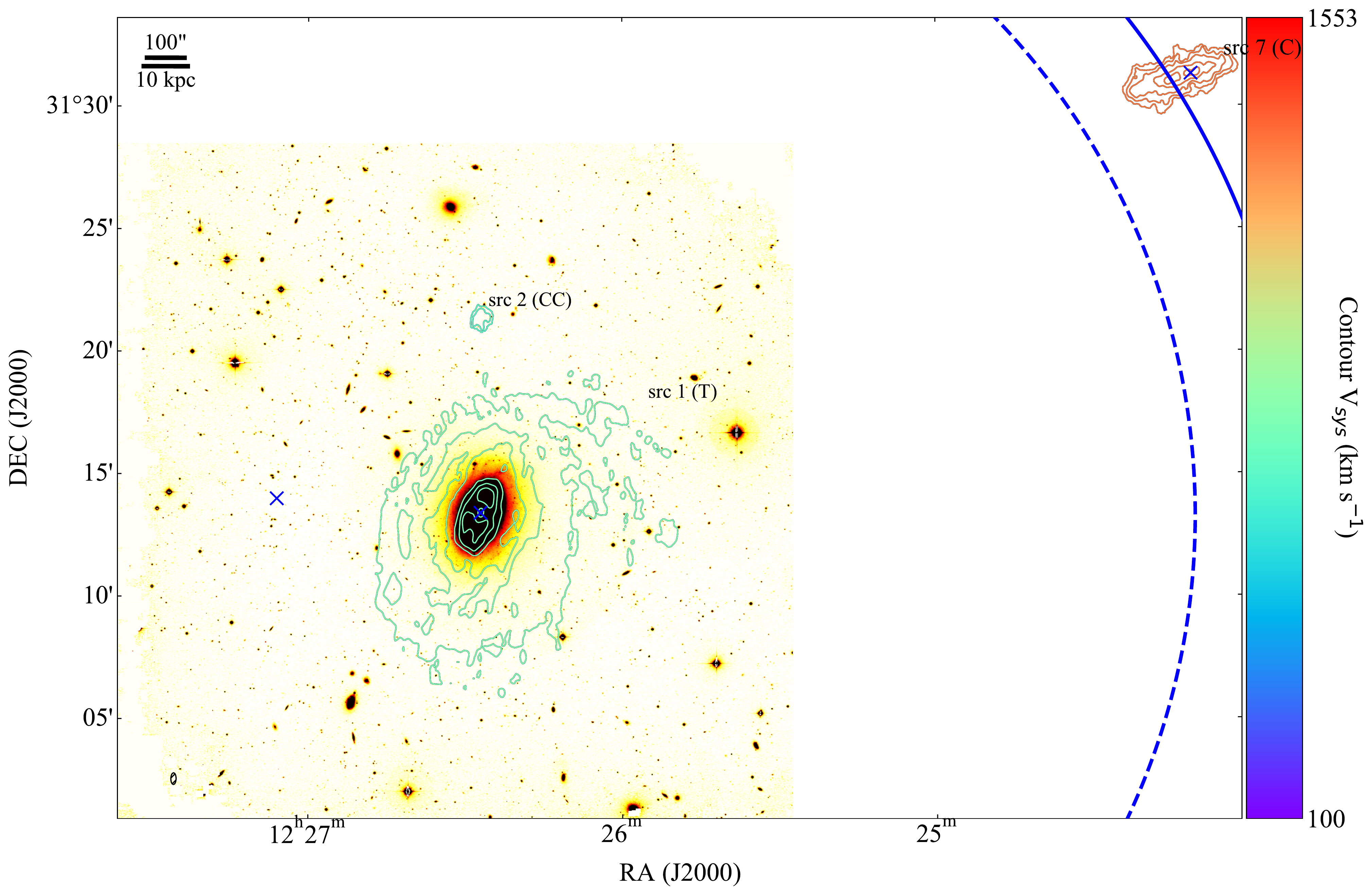} 
\caption{As Figure \ref{fig:Overview_n0925.pdf} but for the NGC\,4414 field.} 
\label{fig:Overview_n4414.pdf}
\end{figure*}
The single cloud candidate detected in this cube appears at the end of a tail of the target galaxy and is co-rotating. In a galaxy of such low inclination this is likely material falling into the disk and not ejected from it as most of the ejection mechanisms work in a vertical direction compared to the disk orientation. Besides the detected cloud candidate, the galaxy also displays a significant extension in \hi\ to the North-West. In our source finding this has merged with the main disk. NGC\,4414 is the galaxy with the highest SFR in the sample and the HALOGAS data are previously described in \cite{deBlok2014}.\\
\indent Besides the target and cloud candidate, we also detect NGC\,4359. This galaxy is significantly removed from the target and its systemic velocity is not in alignment with the rotation of NGC\,4414's disk. 
\subsubsection{NGC 4448} 
\begin{figure*}[htbp] 
\centering 
\includegraphics[width = 18 cm]{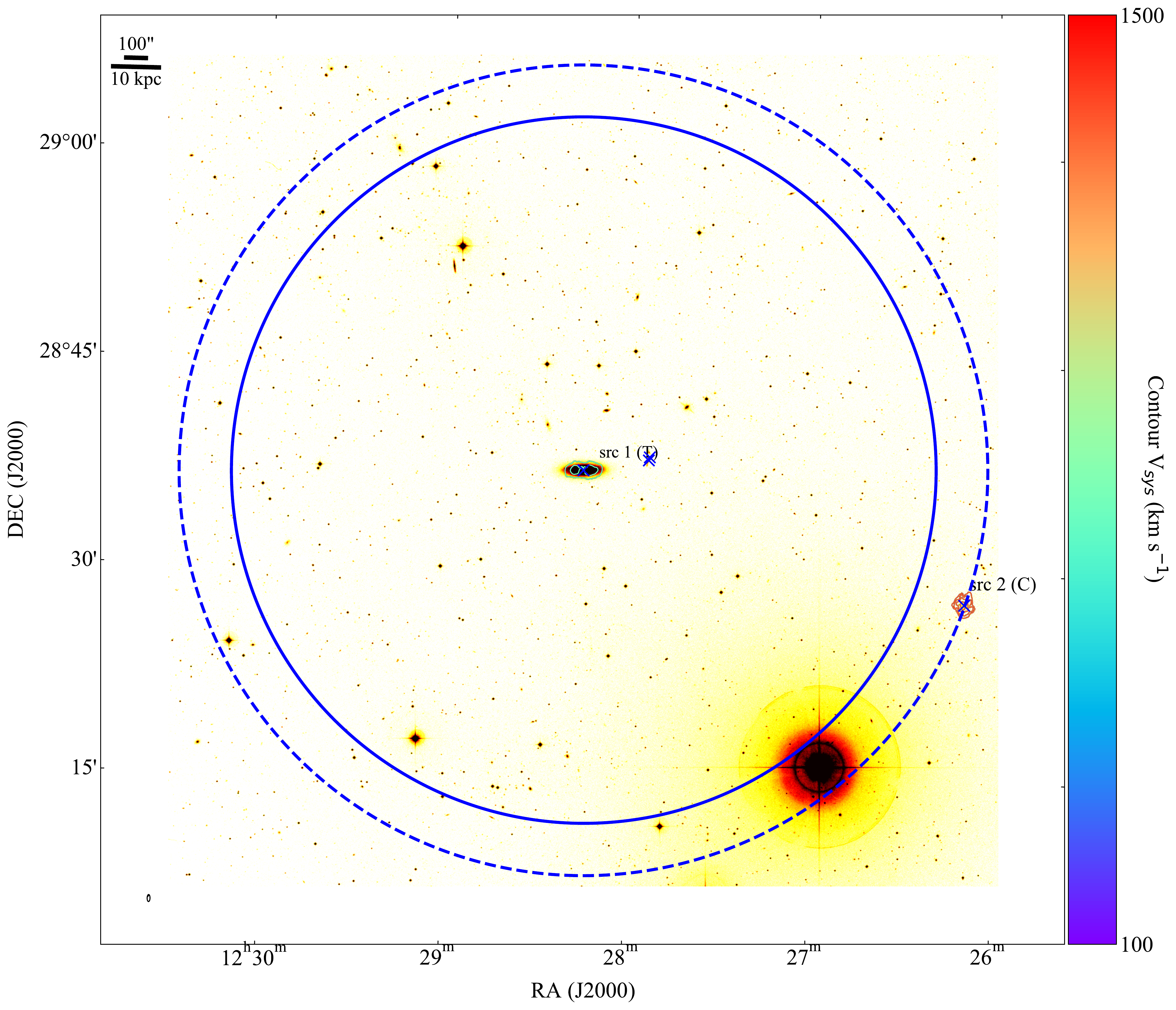} 
\caption{As Figure \ref{fig:Overview_n0925.pdf} but for the NGC\,4448 field overlaid on a DSS 2 $R$-band image.} 
\label{fig:Overview_n4448.pdf}
\end{figure*}
NGC\,4448 is an isolated, regularly rotating galaxy seen at a high inclination. The \hi\ distribution in the galaxy appears to be rather compact despite having an average line width (see Table \ref{table:hosts}). The optical disk extends beyond the \hi\ disk which gives the impression of the gas in the outer regions being stripped.\\
\indent The galaxy IC\,3334 is also detected in the data cube albeit with a significantly higher systemic velocity. 
\subsubsection{NGC 4559} 
\begin{figure*}[htbp] 
\centering 
\includegraphics[width = 18 cm]{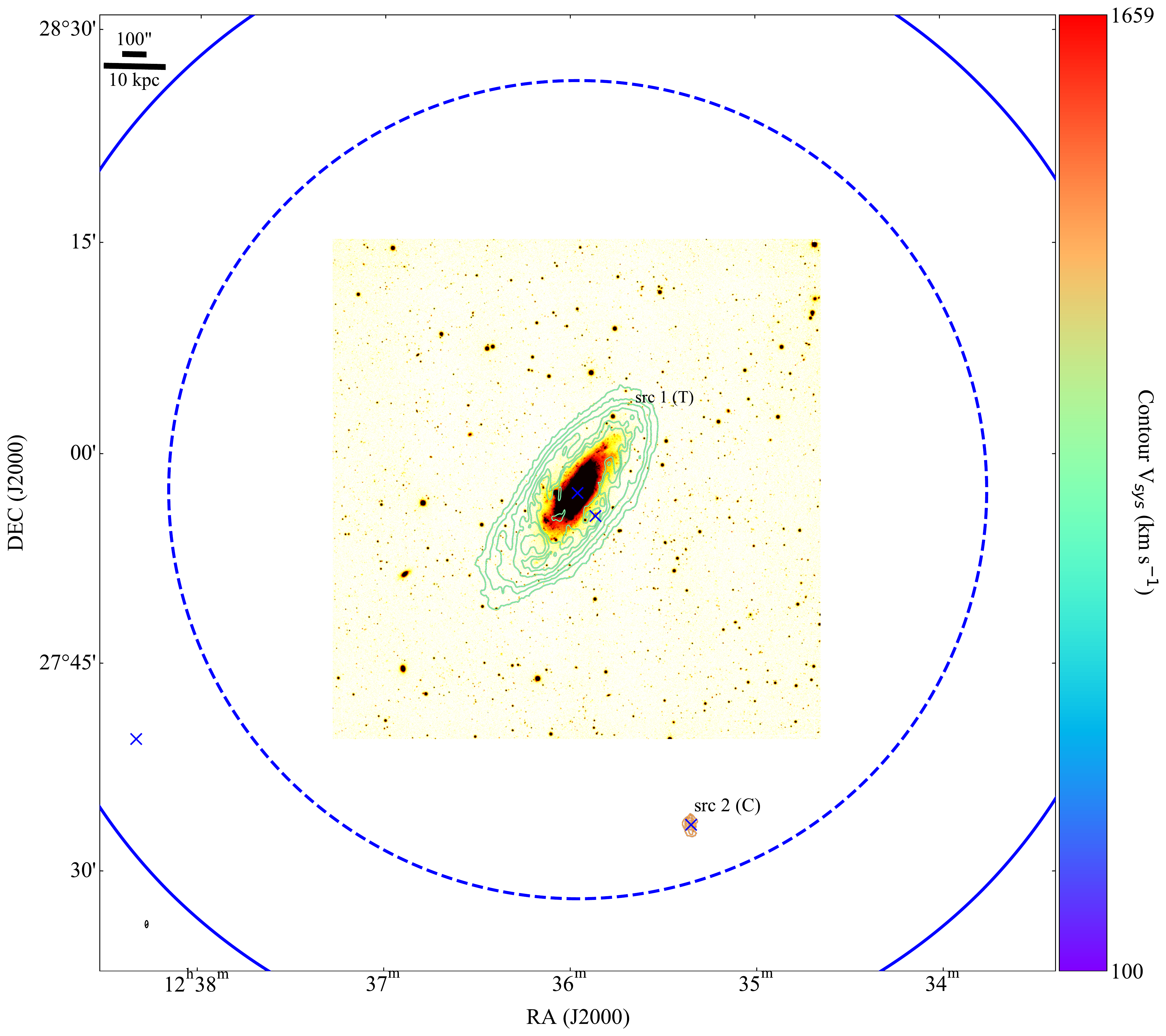} 
\caption{As Figure \ref{fig:Overview_n0925.pdf} but for the NGC\,4559 field overlaid on a KPNO $R$-band image.} 
\label{fig:Overview_n4559.pdf}
\end{figure*}
The HALOGAS data for NGC\,4559 are presented and discussed in detail in \cite{Vargas2017}. This is a moderately inclined galaxy without obvious peculiarities. We detect a single companion at a projected distance of 65 kpc. The companion shows no regular velocity structure and the \hi\ appears more extended than the optical from the WISE images. The difference in systemic velocity between this companion and NGC\,4559 is more than 400 \kms. 
\subsubsection{NGC 4565} 
\begin{figure*}[htbp] 
\centering 
\includegraphics[width = 18 cm]{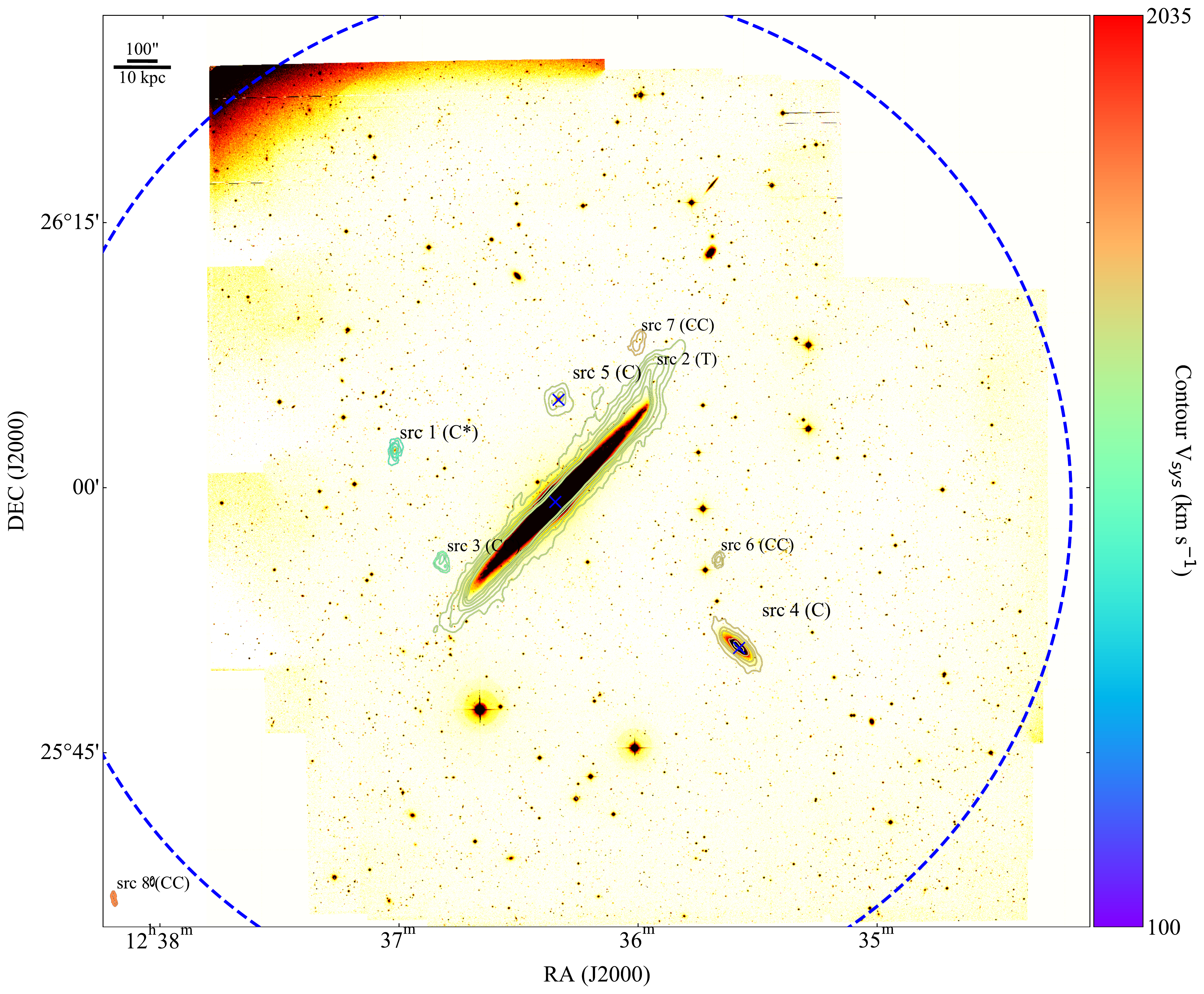} 
\caption{As Figure \ref{fig:Overview_n0925.pdf} but for the NGC\,4565 field.} 
\label{fig:Overview_n4565.pdf}
\end{figure*}
NGC\,4565 is a disturbed galaxy with three companions. We find two anomalous gas clouds in this system that are close to the galaxy and we classify them as cloud candidates. Additionally we find another two cloud candidates in this data cube as well. All companions and cloud candidates are co-rotating with the disk of the galaxy except for source 8 which is furthest removed in velocity as well as projected distance.\\
\indent Tidal debris is a possible origin for the co-rotating cloud candidates, given the environment of interaction. This observation is extensively described in \cite{Zschaechner2012}. 
\subsubsection{NGC 4631} 
\begin{figure*}[htbp] 
\centering 
\includegraphics[width = 18 cm]{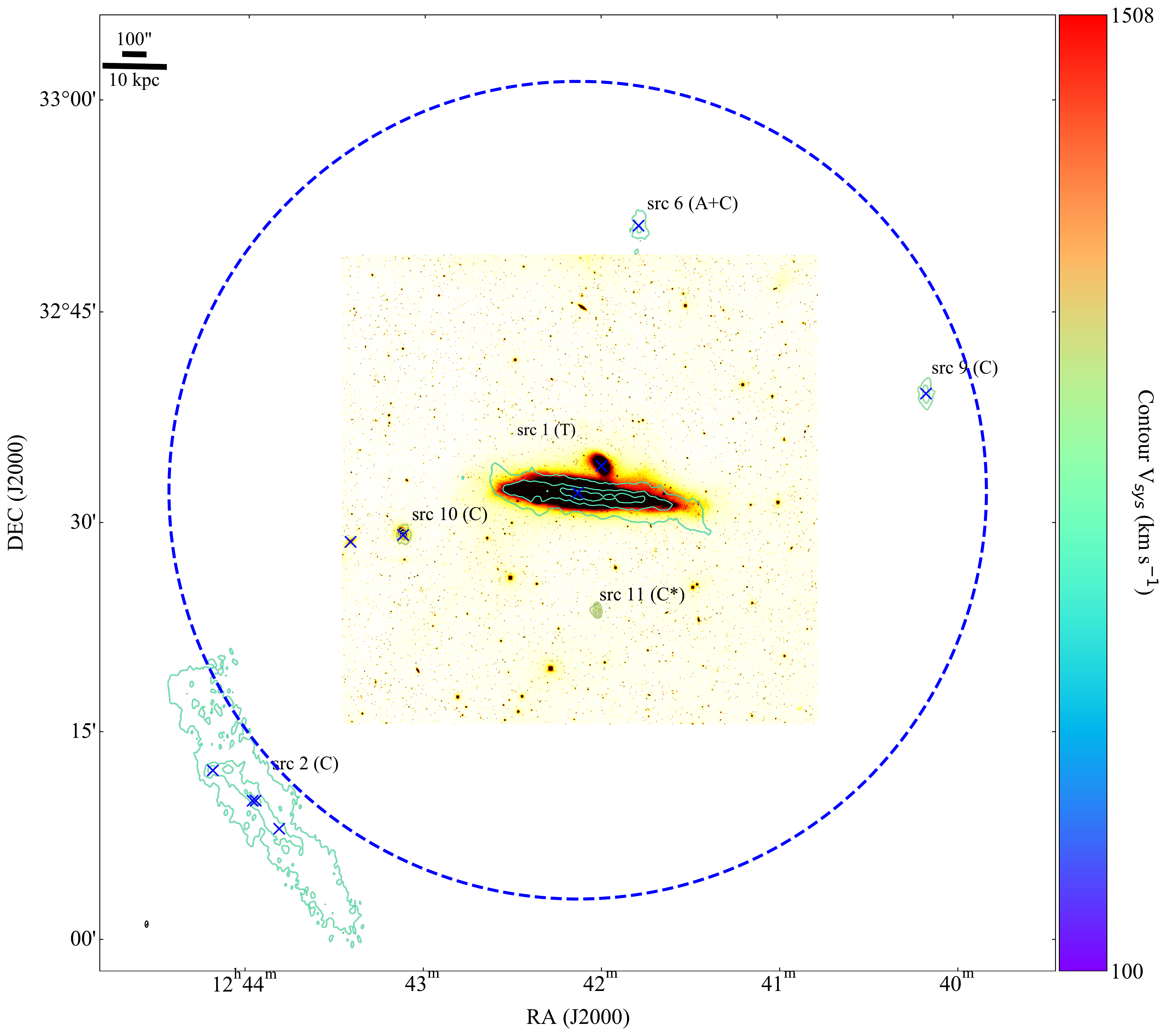} 
\caption{As Figure \ref{fig:Overview_n0925.pdf} but for the NGC\,4631 field overlaid on a KPNO $R$-band image.} 
\label{fig:Overview_n4631.pdf}
\end{figure*}
NGC\,4631 is a galaxy that is heavily interacting with its environment. This extended gas distribution was already seen in older observations \citep{Rand1994} and, as there is gas almost everywhere in the field of view, is complicated to "clean". Hence, significant side lobes of the dirty beam remain in the final data cube and we detect several artifacts.  Complicating matters even more, it appears that several real sources are situated on top of side lobe emission. They can be visually identified as they are so much brighter than the sidelobes but any velocity and flux measurements are highly unreliable. \\
\indent Besides the obvious companion NGC\,4656,  we detect  three companions which spatially coincide with the gas surrounding NGC\,4631 but are distinct in velocity space. Additionally, we detect a source that is further removed from the system. The latter (source 9) is the only companion which is not aligned with the rotational direction of the disk. \\
\indent As the galaxy is in an equal mass merger and due to the issues with the side lobes we exclude it from  the analysis presented in this paper. 
\subsubsection{NGC 5055} 
\begin{figure*}[htbp] 
\centering 
\includegraphics[width = 18 cm]{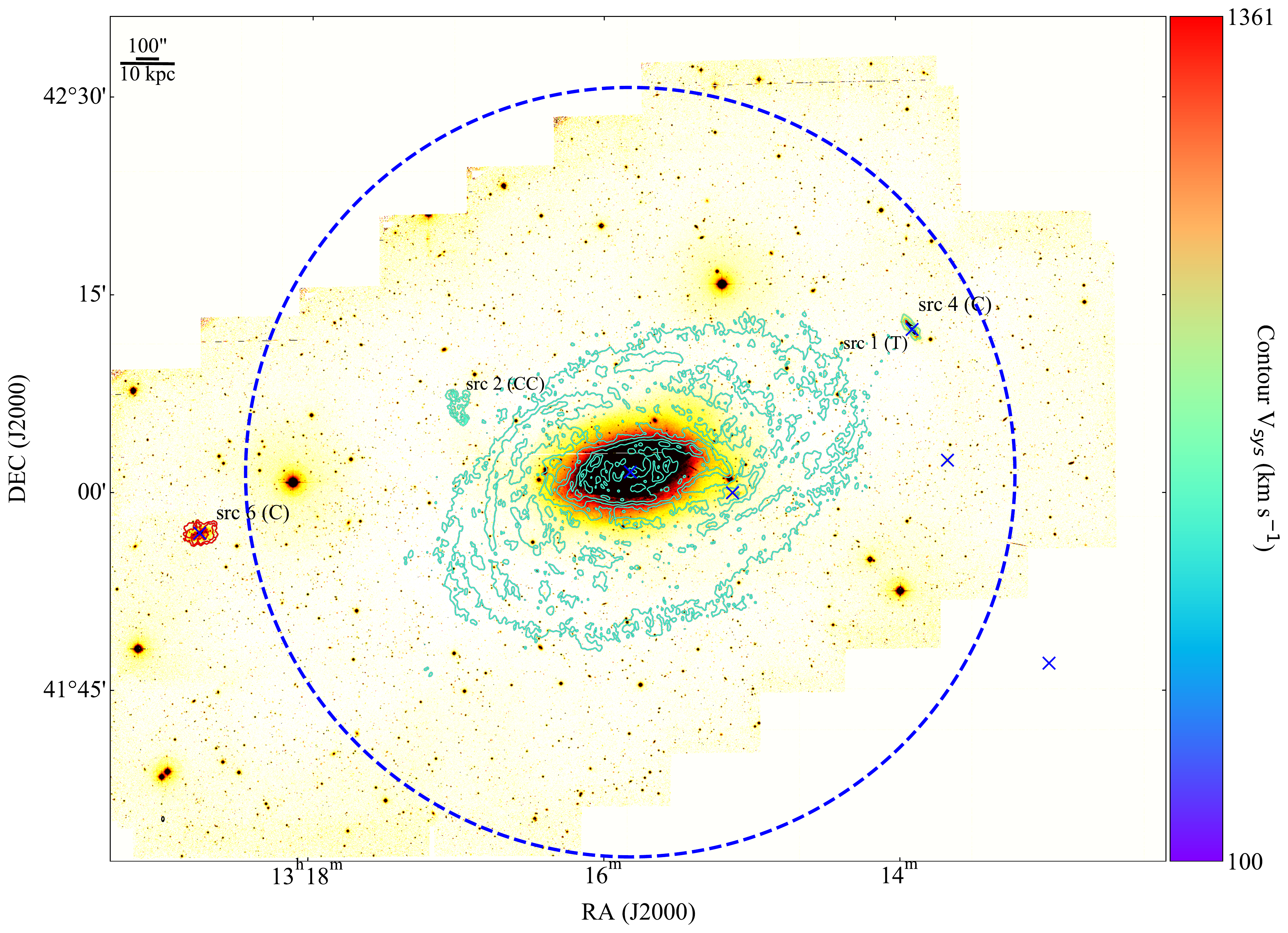} 
\caption{As Figure \ref{fig:Overview_n0925.pdf} but for the NGC\,5055 field.} 
\label{fig:Overview_n5055.pdf}
\end{figure*}
This is a large spiral galaxy. Besides the target we detect two companions, UGC\,8313 and UGC\,8365, and a gas cloud candidate which almost connects to the main disk. The cloud candidate and UGC\,8365 follow the velocity orientation of the disk, although the velocity difference between NGC\,5055 and UGC\,8365 is so large that it is doubtful they are in the same dynamical system. UGC\,8313's systemic velocity is such that it coincides with the velocities in the receding side of NGC\,5055 while its location is on the approaching side. 
\subsubsection{NGC 5585} 
\begin{figure*}[htbp] 
\centering 
\includegraphics[width = 18 cm]{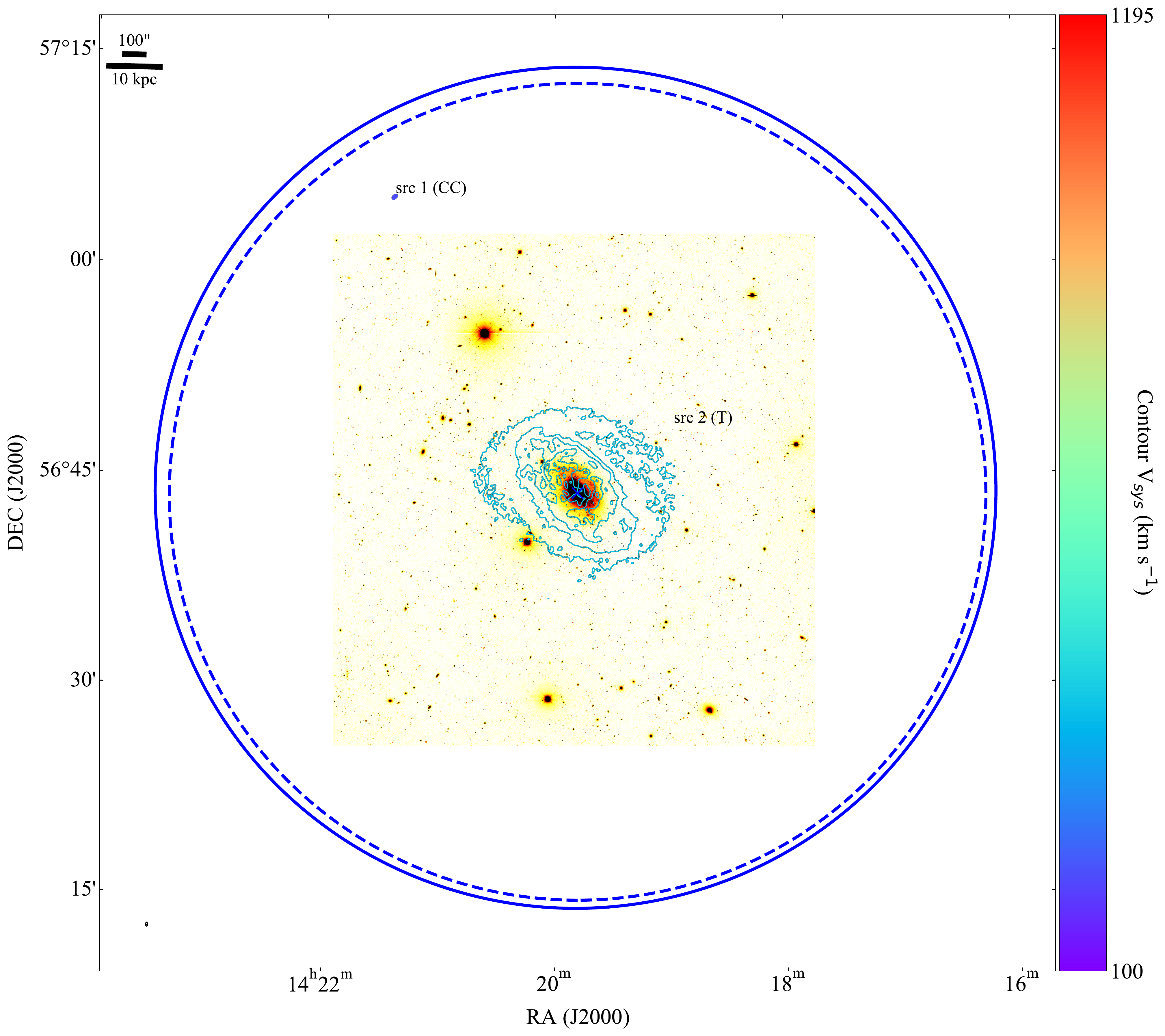} 
\caption{As Figure \ref{fig:Overview_n0925.pdf} but for the NGC\,5585 field overlaid on a KPNO $R$-band image.} 
\label{fig:Overview_n5585.pdf}
\end{figure*}
This galaxy is a spiral galaxy with an \hi\ disk that extends significantly beyond the optical detection in the DSS image and that is slightly warped. Besides the target, we detect a single cloud candidate at a projected distance of 58 kpc. The cloud candidate shows a hint of a velocity structure but as it is barely resolved this could be an artifact of the masking. If this cloud is part of the dynamical system describing NGC\,5585 it would be counter rotating. 
\subsubsection{UGC 4278} 
\begin{figure*}[htbp] 
\centering 
\includegraphics[width = 18 cm]{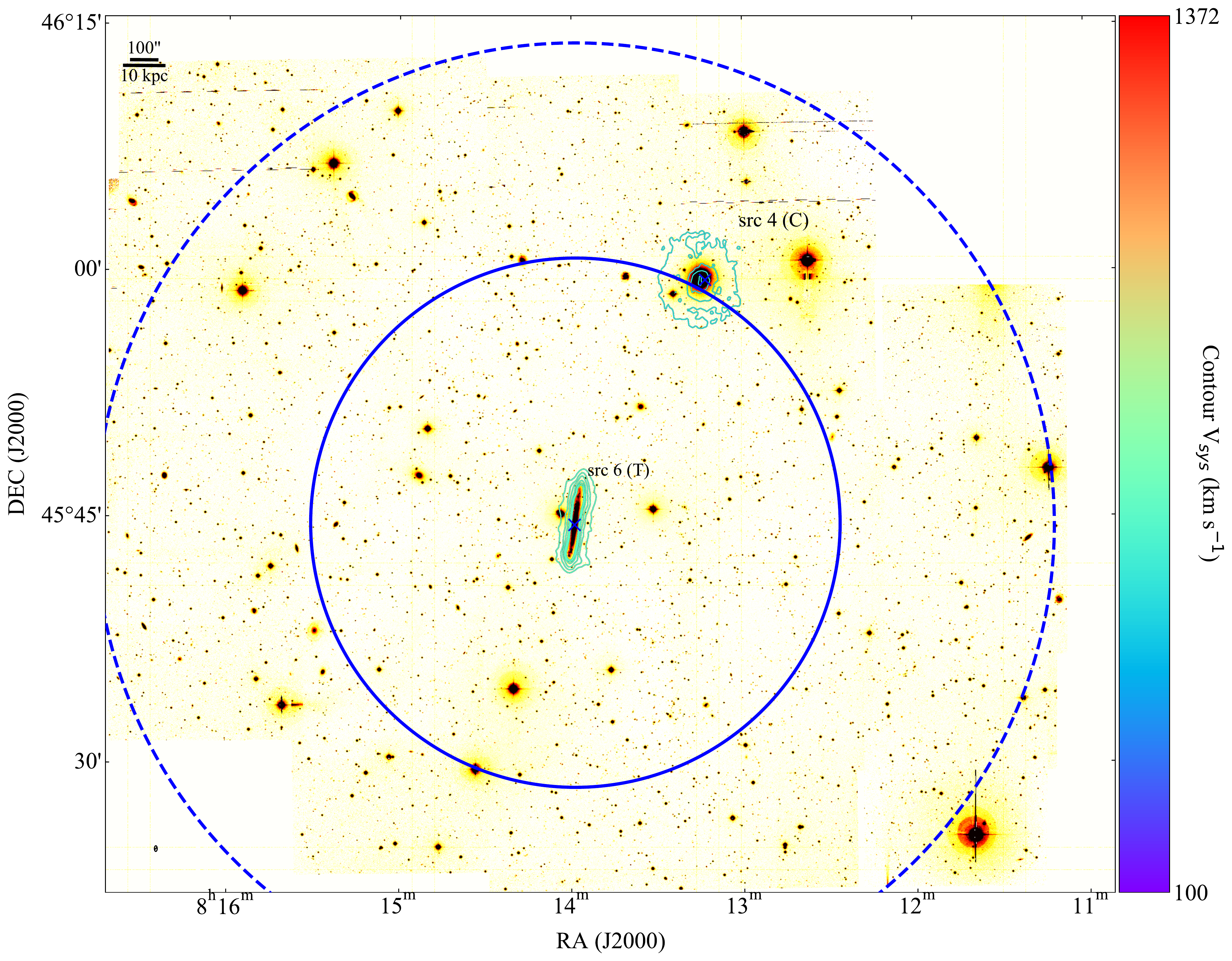} 
\caption{As Figure \ref{fig:Overview_n0925.pdf} but for the UGC\,4278 field.} 
\label{fig:Overview_u4278.pdf}
\end{figure*}
Besides the target galaxy we find one other source in the field that we deem to be real, the companion NGC\,2537. We do not detect anything in between the two sources and their disks appear regular. This leads us to believe that there is currently no interaction between the galaxies. If they were dynamically connected their velocity structure is such that NGC\,2537 would be counter rotating in the frame of UGC\,4278 but in the frame of NGC\,2357 the target would be co-rotating. 
\subsubsection{UGC 7774} 
\begin{figure*}[htbp] 
\centering 
\includegraphics[width = 18 cm]{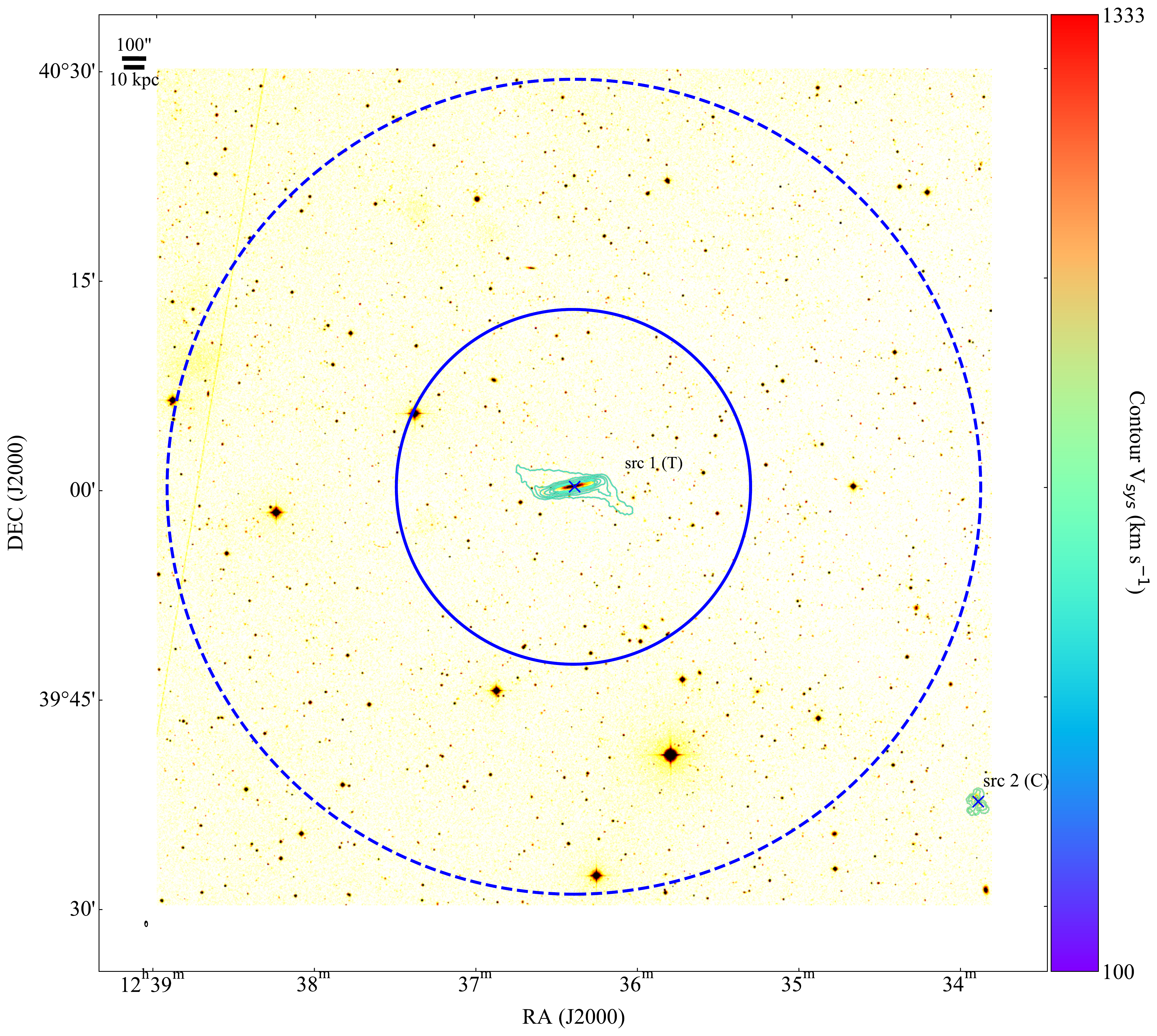} 
\caption{As Figure \ref{fig:Overview_n0925.pdf} but for the UGC\,7774 field overlaid on a DSS 2 $R$-band image.} 
\label{fig:Overview_u7774.pdf}
\end{figure*}
UGC\,7774 is an edge-on galaxy with one of the most striking examples of a warp known (see Figure \ref{fig:Overview_u7774.pdf}). We detect a single companion at a rather large distance from the galaxy (d$_{\rm proj} \sim$ 232 kpc). This companion does appear quite disturbed and hence it might be related to UGC\,7774's warp however a detailed analysis would be required to further this idea. Its systemic velocity is aligned with the  rotation in the western side of UGC\,7774.\\ 
\clearpage
\subsection{Cloud Candidates}
\begin{figure*} 
\centering 
\includegraphics[width = 18 cm]{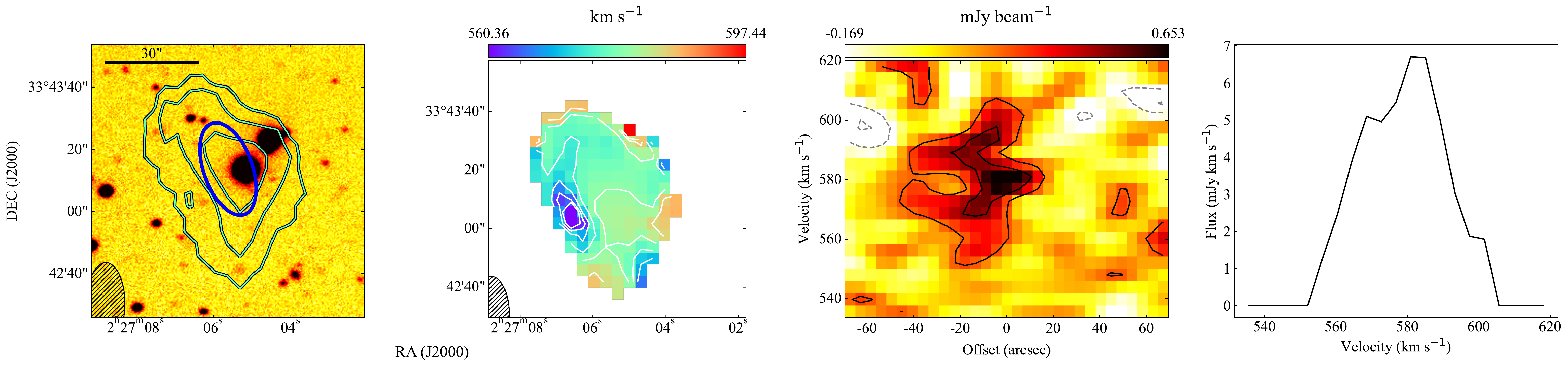} 
\caption{From left to right: Optical image overlaid with contours of the \hi moment 0 (intensity) map , moment 1 (velocity ) map,  PV-Diagram along major axis and line profile for source no. 2 in the cube of NGC\,0925. Contours for moment 0 are: 6.8, 27.2, 54.4$\times 10^{18} {\rm cm}^{-2}$  overlaid on our HALOSTARS $R$-band image. These are corrected with the same factor as the integrated flux in the various tables to account for the primary beam response. The blue ellipse indicates the \sofia\ fitted ellipse. Contours for the velocity field start at 560.36 km\,s$^{-1}$ and increase with 5.00 km\,s$^{-1}$. The PV-diagram is extracted at a PA of 206$^{\circ}$ with contours at -3, -1.5, 1.5, 3 $\sigma$, with $\sigma=0.160$ mJy beam$^{-1}$.} 
\label{fig:n0925_image_src_2.pdf}
\end{figure*}
\begin{figure*} 
\centering 
\includegraphics[width = 18 cm]{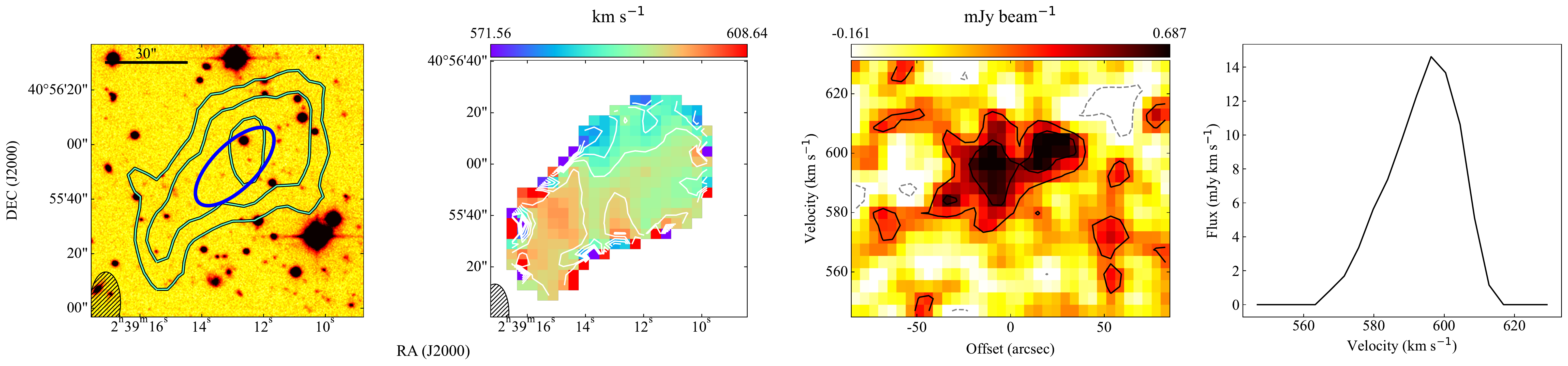} 
\caption{ As \ref{fig:n0925_image_src_2.pdf} but for source no. 2 in the cube of NGC\,1003. Contours for moment 0 are: 7.7, 31.0, 62.0$\times 10^{18} {\rm cm}^{-2}$. Contours for the velocity field start at 571.56 km\,s$^{-1}$ and increase with 5.00 km s$^{-1}$. The PV-diagram is extracted at a PA of 123$^{\circ}$ with contours at -1.5, 1.5, 3 $\sigma$, with $\sigma=0.182$ mJy beam$^{-1}$.} 
\label{fig:n1003_image_src_2.pdf}
\end{figure*}
\begin{figure*} 
\centering 
\includegraphics[width = 18 cm]{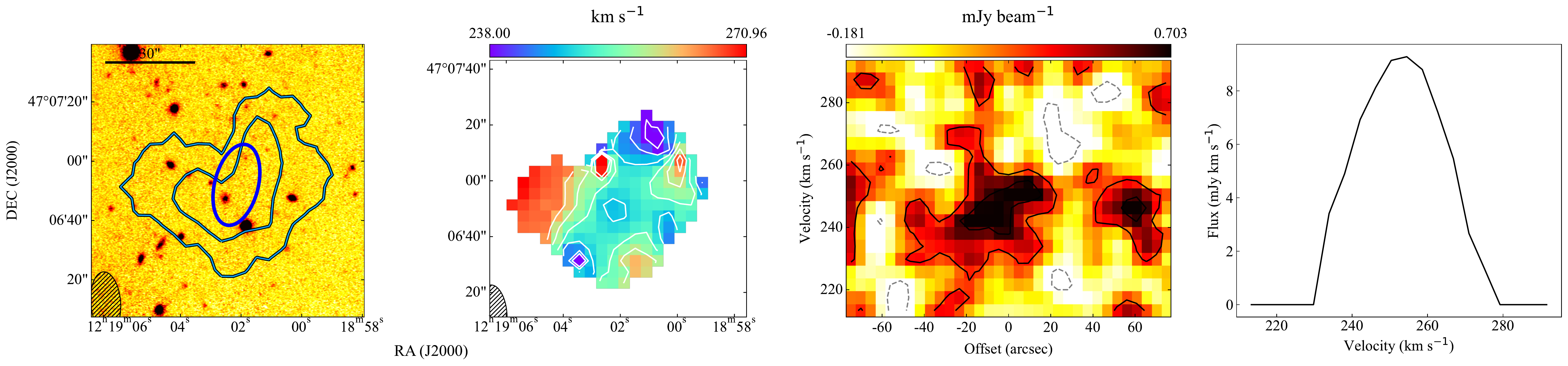} 
\caption{ As \ref{fig:n0925_image_src_2.pdf} but for source no. 3 in the cube of NGC\,4258. Contours for moment 0 are: 1.2, 5.0$\times 10^{19} {\rm cm}^{-2}$ but overlaid on our KPNO $R$-band image. Contours for the velocity field start at 238.00 km\,s$^{-1}$ and increase with 5.00 km s$^{-1}$. The PV-diagram is extracted at a PA of 148$^{\circ}$ with contours at -1.5, 1.5, 3 $\sigma$, with $\sigma=0.215$ mJy beam$^{-1}$.} 
\label{fig:n4258_image_src_3.pdf}
\end{figure*}
\begin{figure*} 
\centering 
\includegraphics[width = 18 cm]{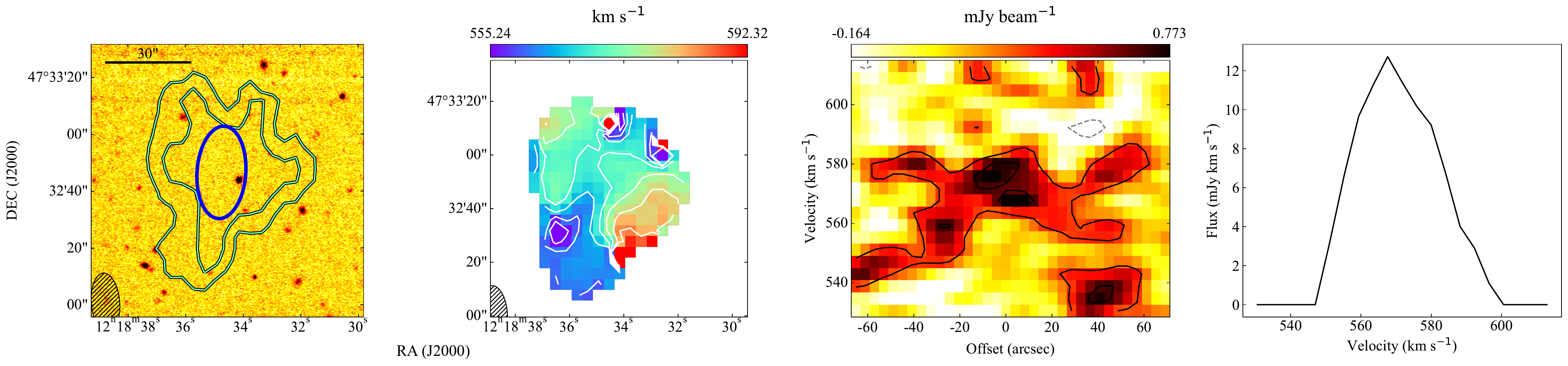} 
\caption{ As \ref{fig:n0925_image_src_2.pdf} but for source no. 8 in the cube of NGC\,4258. Contours for moment 0 are: 1.6, 6.5$\times 10^{19} {\rm cm}^{-2}$ but overlaid on our KPNO $R$-band image. Contours for the velocity field start at 555.24 km\,s$^{-1}$ and increase with 5.00 km s$^{-1}$. The PV-diagram is extracted at a PA of 349$^{\circ}$ with contours at -1.5, 1.5, 3 $\sigma$, with $\sigma=0.227$ mJy beam$^{-1}$.} 
\label{fig:n4258_image_src_8.pdf}
\end{figure*}
\begin{figure*} 
\centering 
\includegraphics[width = 18 cm]{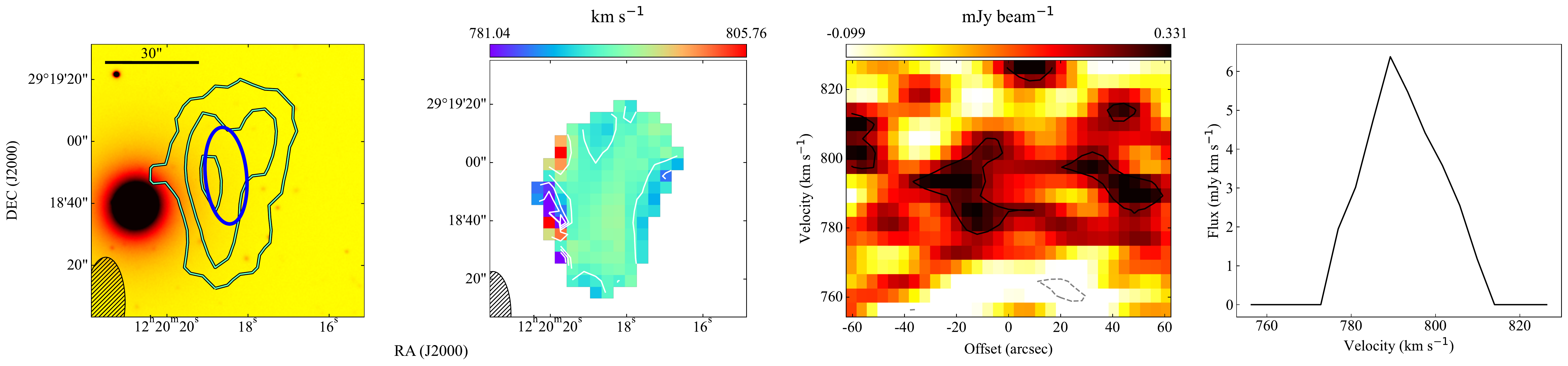} 
\caption{ As \ref{fig:n0925_image_src_2.pdf} but for source no. 9 in the cube of NGC\,4274. Contours for moment 0 are: 1.1, 4.5, 9.0$\times 10^{19} {\rm cm}^{-2}$. Contours for the velocity field start at 781.04 km\,s$^{-1}$ and increase with 5.00 km s$^{-1}$. The PV-diagram is extracted at a PA of 186$^{\circ}$ with contours at -1.5, 1.5 $\sigma$, with $\sigma=0.173$ mJy beam$^{-1}$.} 
\label{fig:n4274_image_src_9.pdf}
\end{figure*}
\begin{figure*} 
\centering 
\includegraphics[width = 18 cm]{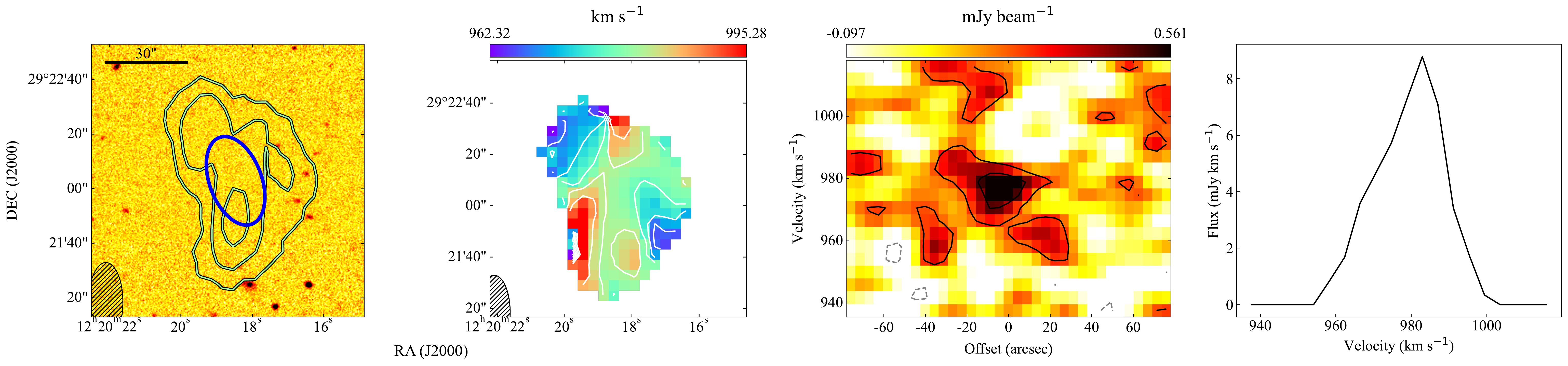} 
\caption{ As \ref{fig:n0925_image_src_2.pdf} but for source no. 10 in the cube of NGC\,4274. Contours for moment 0 are: 8.3, 33.1, 66.3$\times 10^{18} {\rm cm}^{-2}$. Contours for the velocity field start at 962.32 km\,s$^{-1}$ and increase with 5.00 km s$^{-1}$. The PV-diagram is extracted at a PA of 206$^{\circ}$ with contours at -1.5, 1.5, 3 $\sigma$, with $\sigma=0.161$ mJy beam$^{-1}$.} 
\label{fig:n4274_image_src_10.pdf}
\end{figure*}
\begin{figure*} 
\centering 
\includegraphics[width = 18 cm]{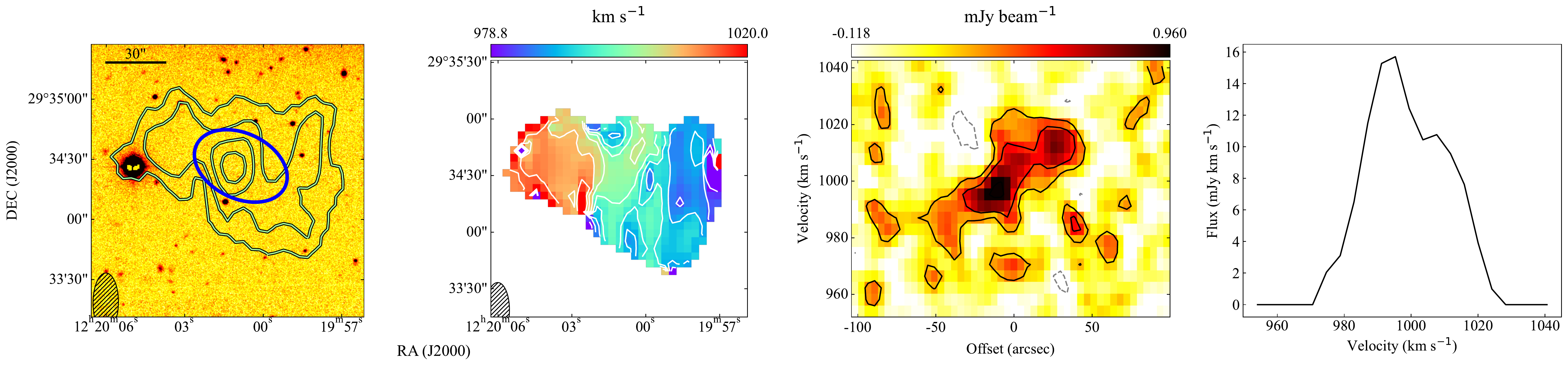} 
\caption{ As \ref{fig:n0925_image_src_2.pdf} but for source no. 11 in the cube of NGC\,4274. Contours for moment 0 are: 5.3, 21.2, 42.5, 63.7$\times 10^{18} {\rm cm}^{-2}$. Contours for the velocity field start at 978.80 km\,s$^{-1}$ and increase with 5.00 km s$^{-1}$. The PV-diagram is extracted at a PA of 71$^{\circ}$ with contours at -1.5, 1.5, 3, 6 $\sigma$, with $\sigma=0.172$ mJy beam$^{-1}$.} 
\label{fig:n4274_image_src_11.pdf}
\end{figure*}
\begin{figure*} 
\centering 
\includegraphics[width = 18 cm]{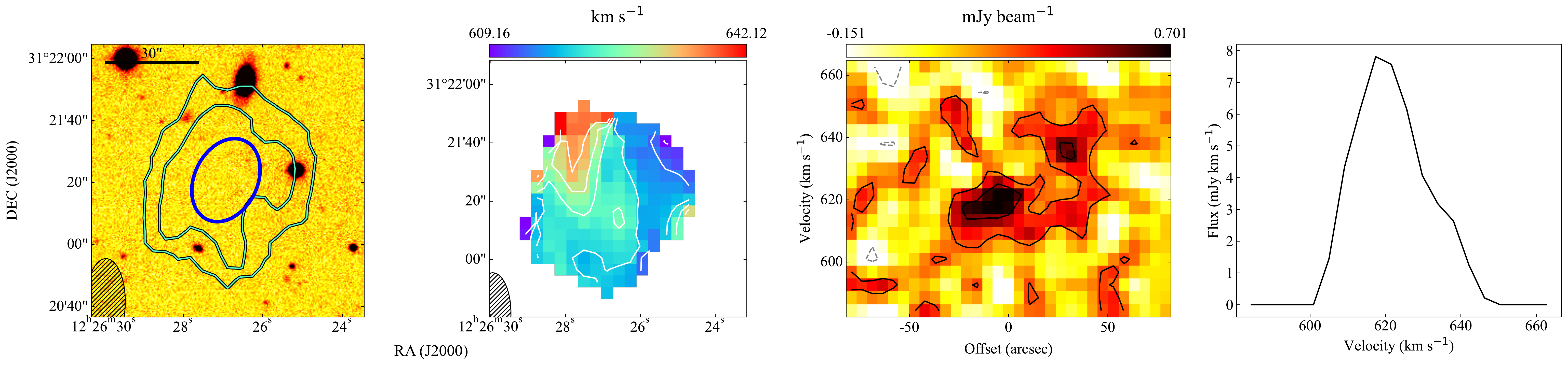} 
\caption{ As \ref{fig:n0925_image_src_2.pdf} but for source no. 2 in the cube of NGC\,4414. Contours for moment 0 are: 6.1, 24.2$\times 10^{18} {\rm cm}^{-2}$. Contours for the velocity field start at 609.16 km\,s$^{-1}$ and increase with 5.00 km s$^{-1}$. The PV-diagram is extracted at a PA of 139$^{\circ}$ with contours at -1.5, 1.5, 3 $\sigma$, with $\sigma=0.189$ mJy beam$^{-1}$.} 
\label{fig:n4414_image_src_2.pdf}
\end{figure*}
\begin{figure*} 
\centering 
\includegraphics[width = 18 cm]{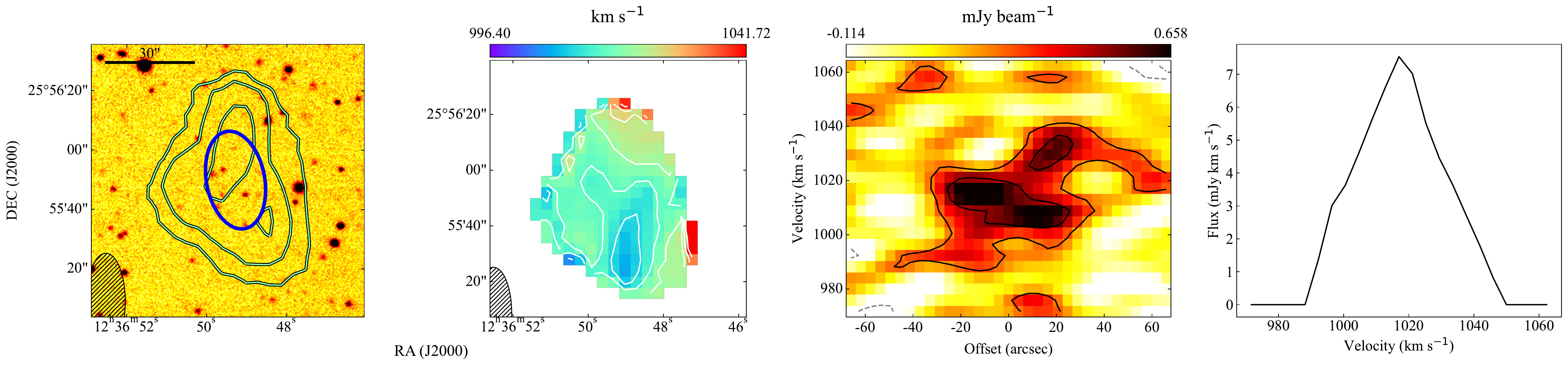} 
\caption{ As \ref{fig:n0925_image_src_2.pdf} but for source no. 3 in the cube of NGC\,4565. Contours for moment 0 are: 5.8, 23.1, 46.2$\times 10^{18} {\rm cm}^{-2}$. Contours for the velocity field start at 996.40 km\,s$^{-1}$ and increase with 5.00 km s$^{-1}$. The PV-diagram is extracted at a PA of 13$^{\circ}$ with contours at -1.5, 1.5, 3 $\sigma$, with $\sigma=0.170$ mJy beam$^{-1}$.} 
\label{fig:n4565_image_src_3.pdf}
\end{figure*}
\begin{figure*} 
\centering 
\includegraphics[width = 18 cm]{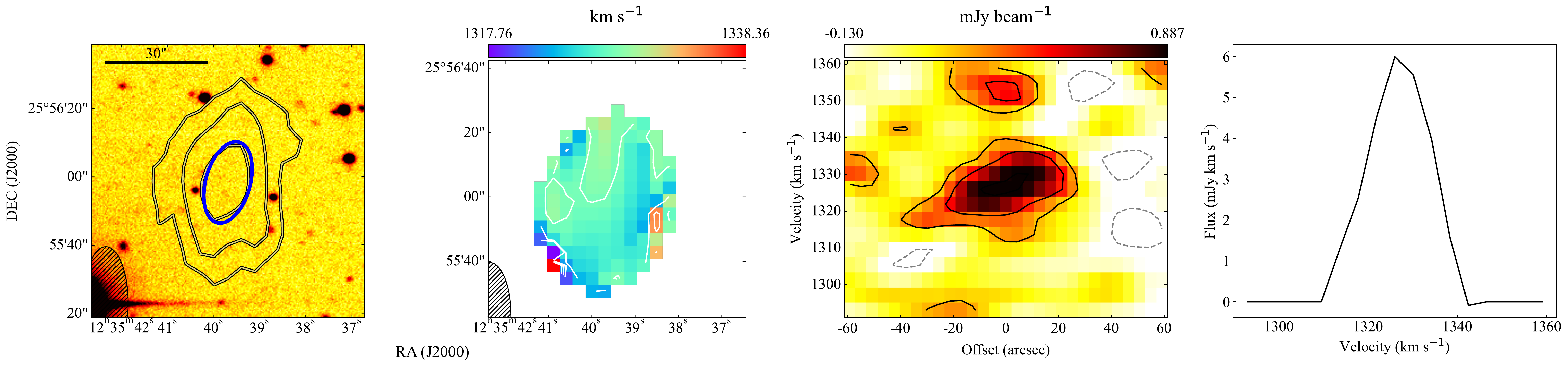} 
\caption{ As \ref{fig:n0925_image_src_2.pdf} but for source no. 6 in the cube of NGC\,4565. Contours for moment 0 are: 5.0, 19.9, 39.8$\times 10^{18} {\rm cm}^{-2}$. Contours for the velocity field start at 1317.76 km\,s$^{-1}$ and increase with 5.00 km s$^{-1}$. The PV-diagram is extracted at a PA of 341$^{\circ}$ with contours at -1.5, 1.5, 3, 6 $\sigma$, with $\sigma=0.156$ mJy beam$^{-1}$.} 
\label{fig:n4565_image_src_6.pdf}
\end{figure*}
\begin{figure*} 
\centering 
\includegraphics[width = 18 cm]{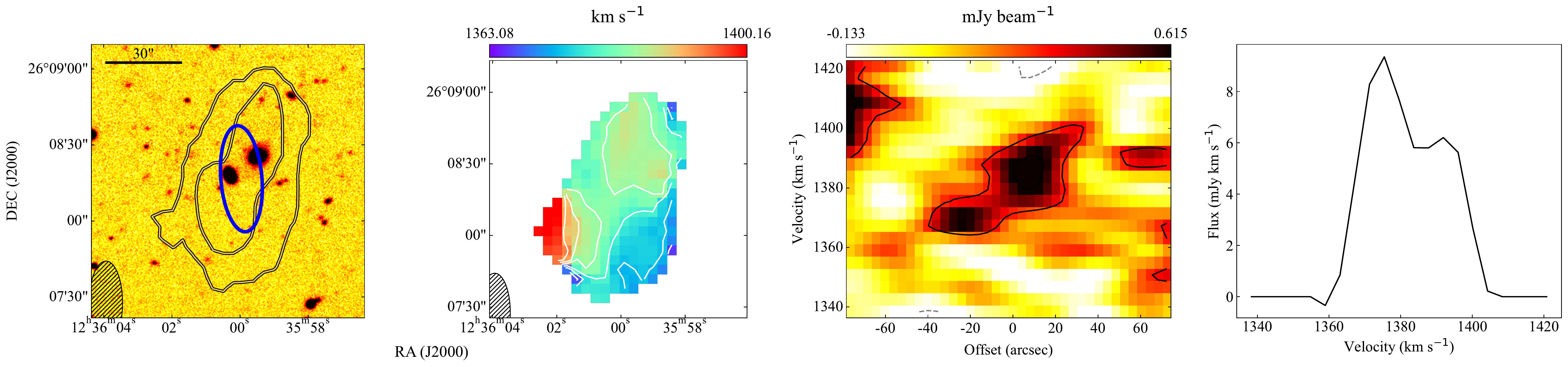} 
\caption{ As \ref{fig:n0925_image_src_2.pdf} but for source no. 7 in the cube of NGC\,4565. Contours for moment 0 are: 7.3, 29.0$\times 10^{18} {\rm cm}^{-2}$. Contours for the velocity field start at 1363.08 km\,s$^{-1}$ and increase with 5.00 km s$^{-1}$. The PV-diagram is extracted at a PA of 5$^{\circ}$ with contours at -1.5, 1.5, 3 $\sigma$, with $\sigma=0.211$ mJy beam$^{-1}$.} 
\label{fig:n4565_image_src_7.pdf}
\end{figure*}
\begin{figure*} 
\centering 
\includegraphics[width = 18 cm]{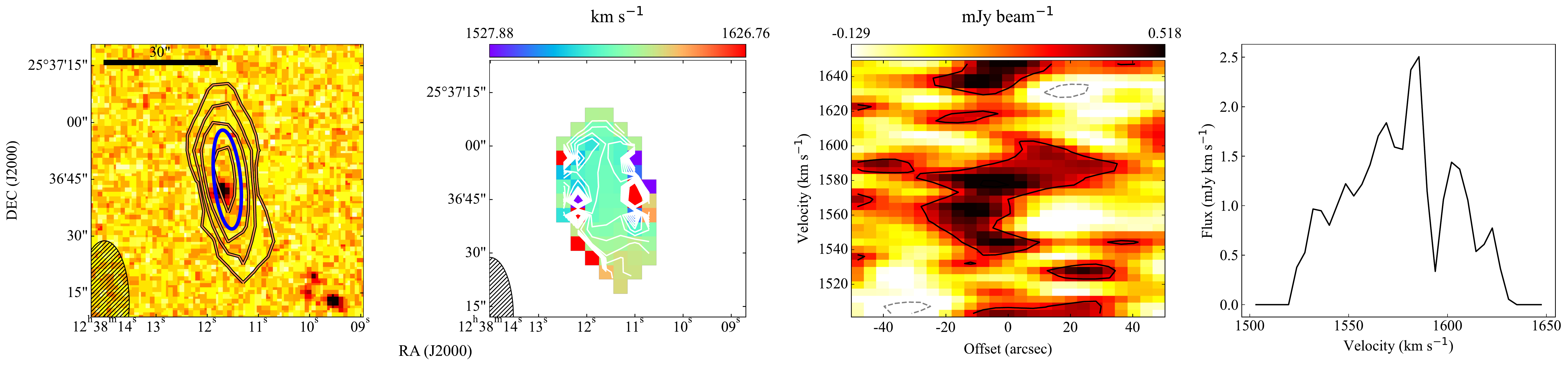} 
\caption{ As \ref{fig:n0925_image_src_2.pdf} but for source no. 8 in the cube of NGC\,4565. Contours for moment 0 are: 9.6, 38.5, 77.0, 153.9$\times 10^{19} {\rm cm}^{-2}$ but overlaid on a DDS 2 $R$-band image as it is outside the HALOSTARS field. Contours for the velocity field start at 1527.88 km\,s$^{-1}$ and increase with 5.00 km s$^{-1}$. The PV-diagram is extracted at a PA of 186$^{\circ}$ with contours at -1.5, 1.5, 3 $\sigma$, with $\sigma=0.193$ mJy beam$^{-1}$.} 
\label{fig:n4565_image_src_8.pdf}
\end{figure*}
\begin{figure*} 
\centering 
\includegraphics[width = 18 cm]{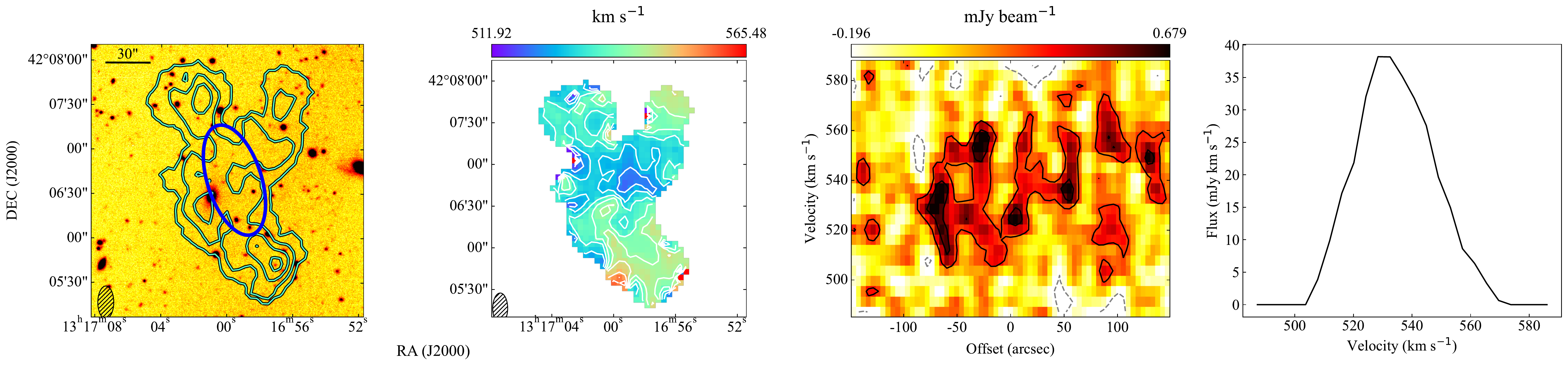} 
\caption{ As \ref{fig:n0925_image_src_2.pdf} but for source no. 2 in the cube of NGC\,5055. Contours for moment 0 are: 1.2, 4.8, 9.6, 14.4$\times 10^{19} {\rm cm}^{-2}$. Contours for the velocity field start at 511.92 km\,s$^{-1}$ and increase with 5.00 km s$^{-1}$. The PV-diagram is extracted at a PA of 209$^{\circ}$ with contours at -3, -1.5, 1.5, 3 $\sigma$, with $\sigma=0.187$ mJy beam$^{-1}$.} 
\label{fig:n5055_image_src_2.pdf}
\end{figure*}
\begin{figure*} 
\centering 
\includegraphics[width = 18 cm]{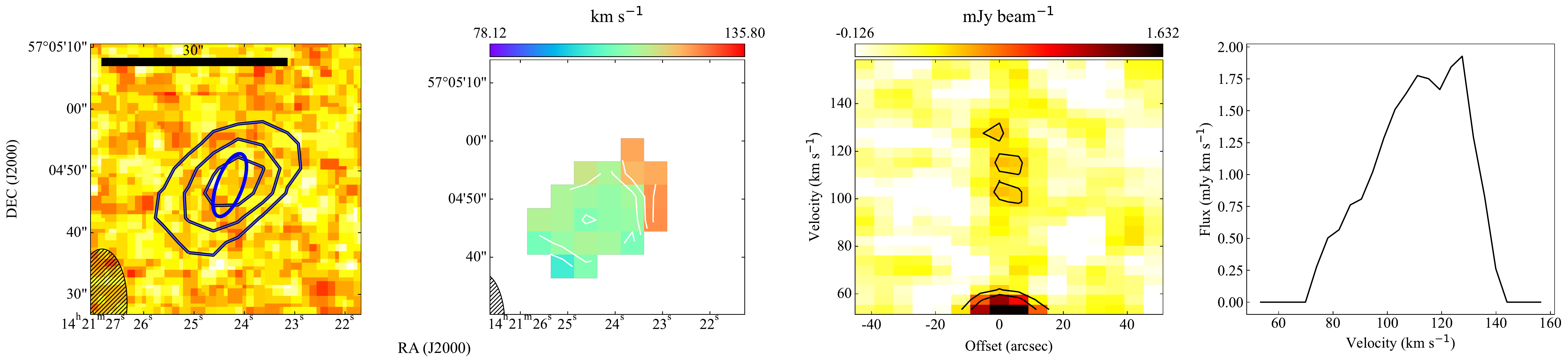} 
\caption{ As \ref{fig:n0925_image_src_2.pdf} but for source no. 1 in the cube of NGC\,5585. Contours for moment 0 are: 6.1, 24.2, 48.4$\times 10^{19} {\rm cm}^{-2}$ but overlaid on a DDS 2 $R$-band image as it is outside the KPNO field. Contours for the velocity field start at 78.12 km\,s$^{-1}$ and increase with 5.00 km s$^{-1}$. The PV-diagram is extracted at a PA of 135$^{\circ}$ with contours at , 1.5, 3, 6 $\sigma$, with $\sigma=0.274$ mJy beam$^{-1}$.} 
\label{fig:n5585_image_src_1.pdf}
\end{figure*}
\clearpage
\subsection{Companions}
\begin{figure} 
\centering 
\includegraphics[width = 8 cm]{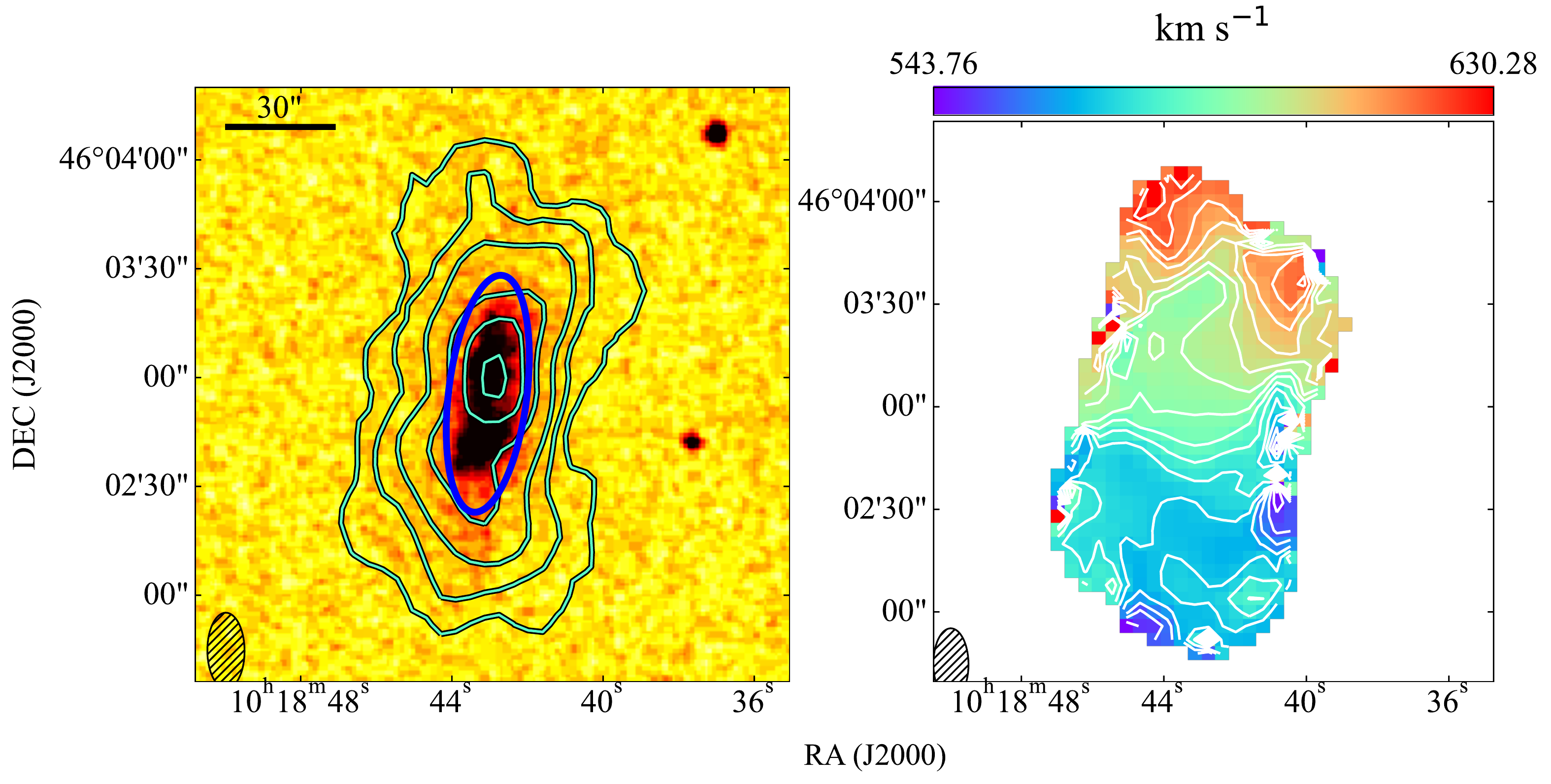} 
\caption{ As \ref{fig:n0925_image_src_4.pdf} but for source no. 2 in the cube of NGC\,3198. Contours for moment 0 are: 1.1, 4.2, 8.5, 16.9, 25.4, 33.8$\times 10^{20} {\rm cm}^{-2}$ but overlaid on a DDS 2 $R$-band image as it is outside the HALOSTARS field. Contours for the velocity field start at 543.76 km\,s$^{-1}$ and increase with 5.00 km s$^{-1}$. This source is identified as VV\,834\,NED02.} 
\label{fig:n3198_image_src_2.pdf}
\end{figure}
\begin{figure} 
\centering 
\includegraphics[width = 8 cm]{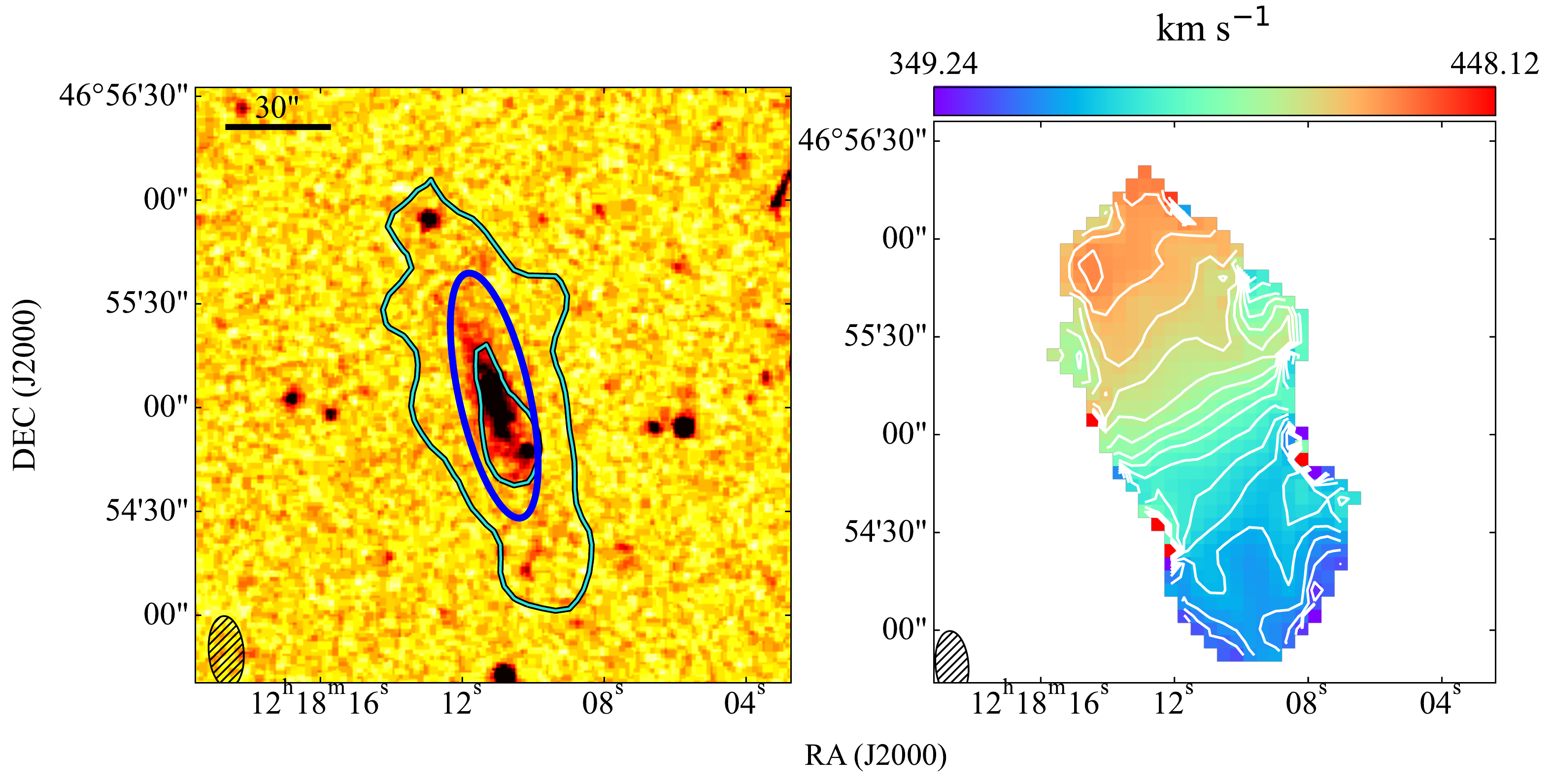} 
\caption{ As \ref{fig:n0925_image_src_4.pdf} but for source no. 4 in the cube of NGC\,4258. Contours for moment 0 are: 3.4, 13.7$\times 10^{20} {\rm cm}^{-2}$ but overlaid on a DDS 2 $R$-band image as it is outside the KPNO field. Contours for the velocity field start at 349.24 km\,s$^{-1}$ and increase with 5.00 km s$^{-1}$.} 
\label{fig:n4258_image_src_4.pdf}
\end{figure}
\begin{figure} 
\centering 
\includegraphics[width = 8 cm]{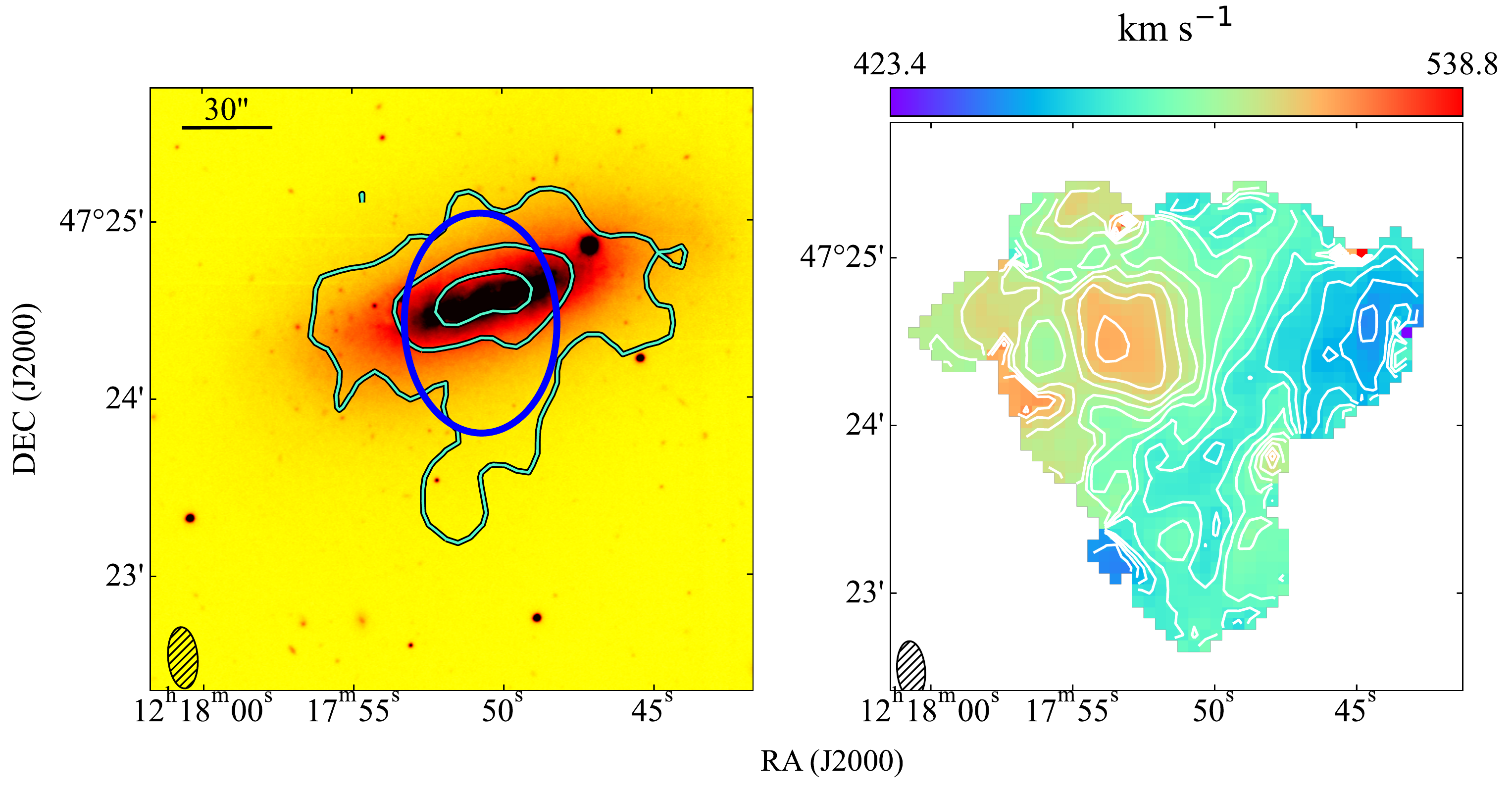} 
\caption{ As \ref{fig:n0925_image_src_4.pdf} but for source no. 6 in the cube of NGC\,4258. Contours for moment 0 are: 1.3, 5.3, 10.5$\times 10^{20} {\rm cm}^{-2}$ but overlaid on our KPNO $R$-band image. Contours for the velocity field start at 423.40 km\,s$^{-1}$ and increase with 5.77 km s$^{-1}$. This source is identified as NGC\,4248.} 
\label{fig:n4258_image_src_6.pdf}
\end{figure}
\begin{figure} 
\centering 
\includegraphics[width = 8 cm]{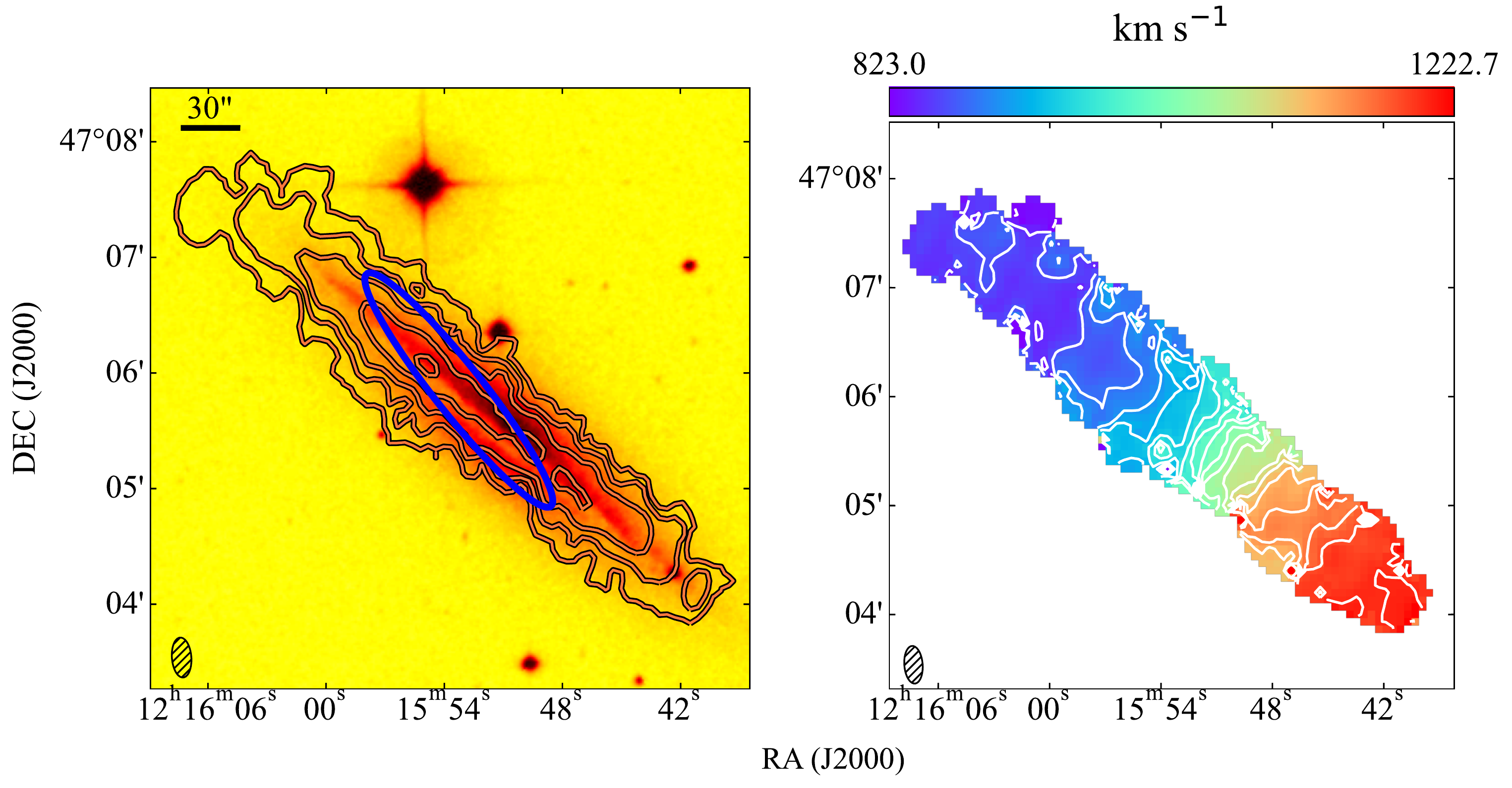} 
\caption{ As \ref{fig:n0925_image_src_4.pdf} but for source no. 9 in the cube of NGC\,4258. Contours for moment 0 are: 2.2, 8.8, 17.6, 35.2, 52.8$\times 10^{20} {\rm cm}^{-2}$ but overlaid on a DDS 2 $R$-band image as it is outside the KPNO field. Contours for the velocity field start at 823.04 km\,s$^{-1}$ and increase with 19.98 km s$^{-1}$.} 
\label{fig:n4258_image_src_9.pdf}
\end{figure}
\begin{figure} 
\centering 
\includegraphics[width = 8 cm]{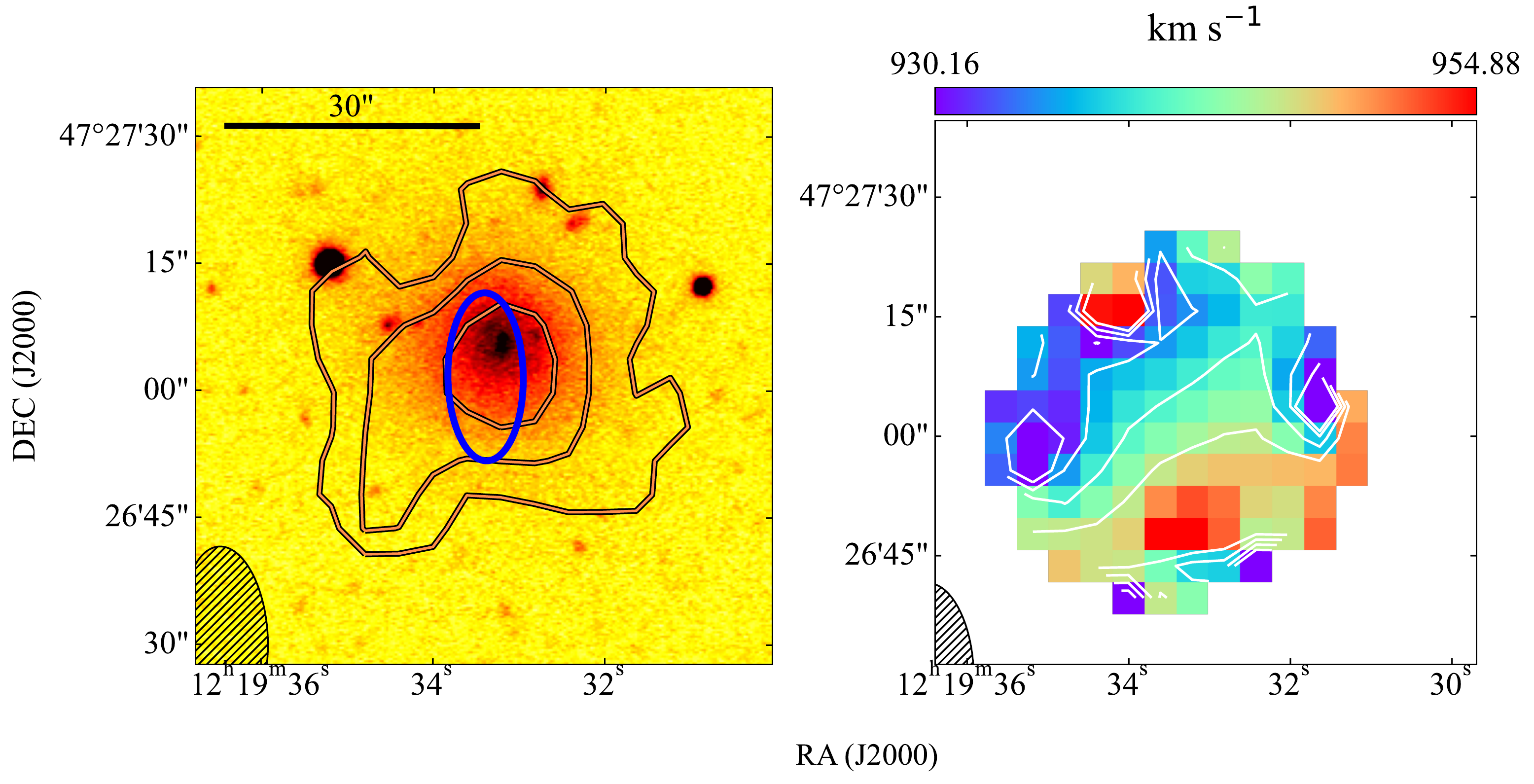} 
\caption{ As \ref{fig:n0925_image_src_4.pdf} but for source no. 11 in the cube of NGC\,4258. Contours for moment 0 are: 1.1, 4.3, 8.6$\times 10^{19} {\rm cm}^{-2}$ but overlaid on our KPNO $R$-band image. Contours for the velocity field start at 930.16 km\,s$^{-1}$ and increase with 5.00 km s$^{-1}$.} 
\label{fig:n4258_image_src_11.pdf}
\end{figure}
\begin{figure} 
\centering 
\includegraphics[width = 8 cm]{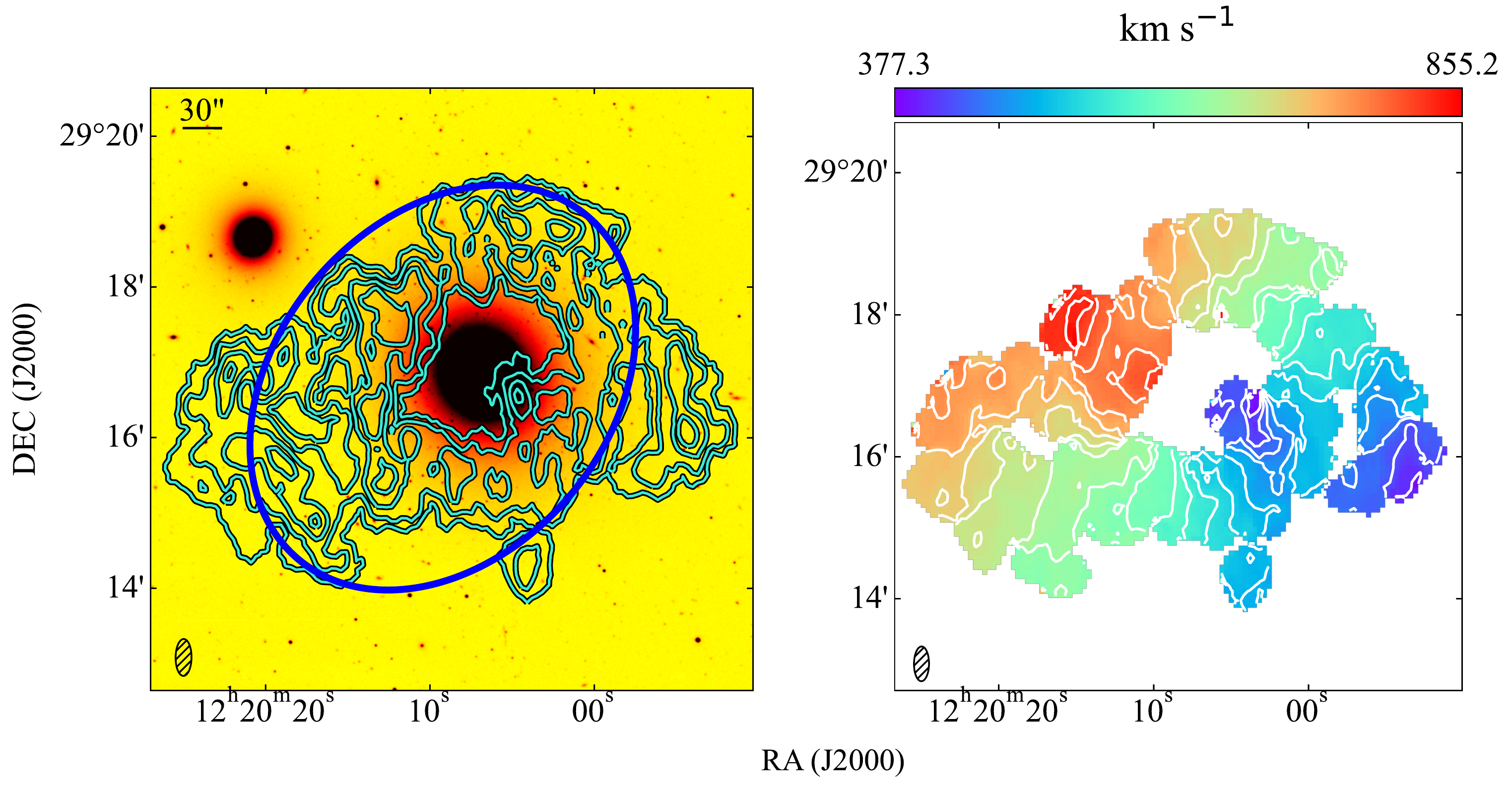} 
\caption{ As \ref{fig:n0925_image_src_4.pdf} but for source no. 1 in the cube of NGC\,4274. Contours for moment 0 are: 1.6, 6.6, 13.1, 19.7$\times 10^{19} {\rm cm}^{-2}$. Contours for the velocity field start at 377.28 km\,s$^{-1}$ and increase with 23.90 km s$^{-1}$.} 
\label{fig:n4274_image_src_1.pdf}
\end{figure}
\begin{figure} 
\centering 
\includegraphics[width = 8 cm]{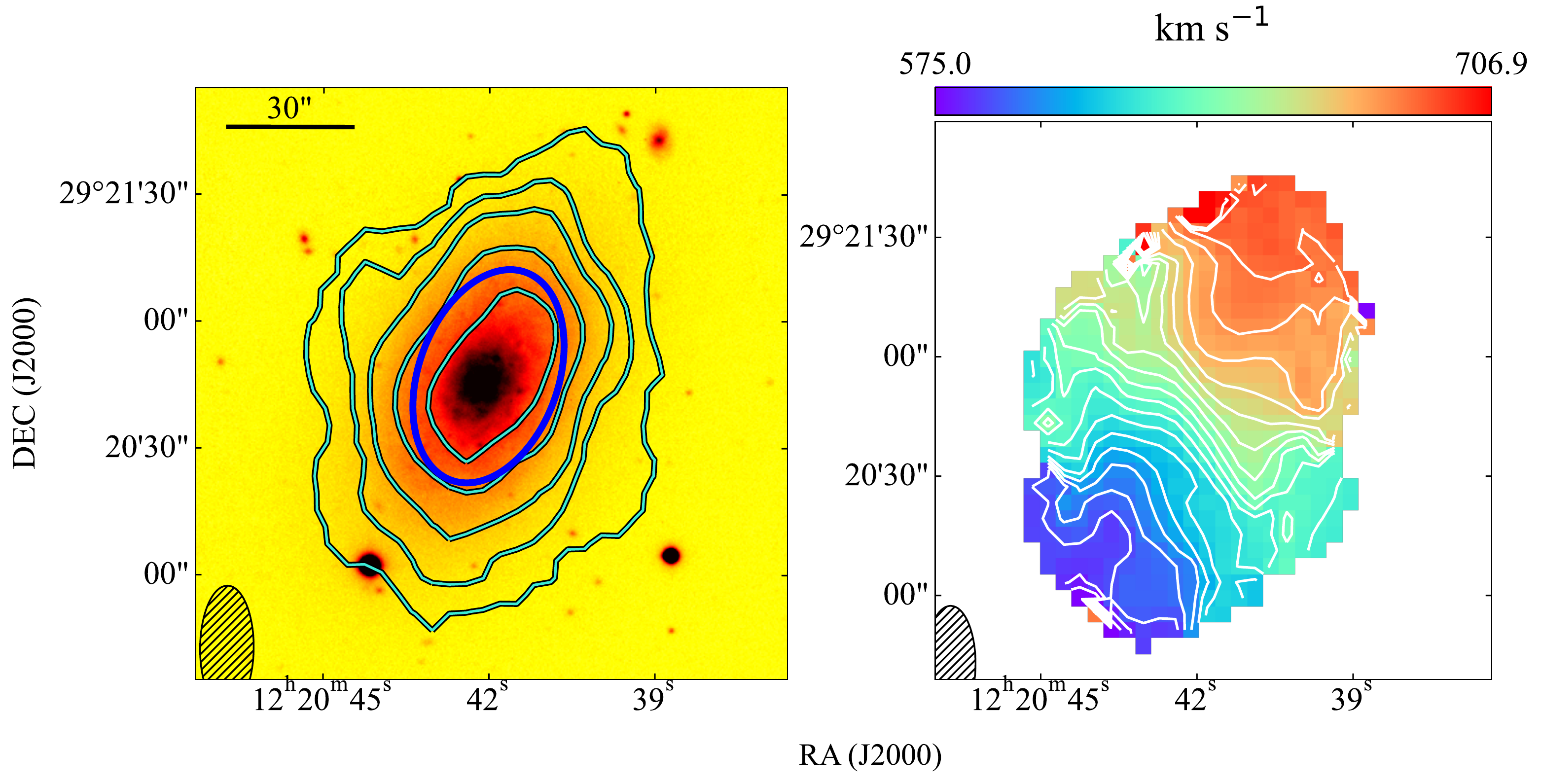} 
\caption{ As \ref{fig:n0925_image_src_4.pdf} but for source no. 6 in the cube of NGC\,4274. Contours for moment 0 are: 1.5, 6.2, 12.4, 24.7, 37.1$\times 10^{19} {\rm cm}^{-2}$. Contours for the velocity field start at 575.04 km\,s$^{-1}$ and increase with 6.59 km s$^{-1}$. This source is identified as NGC\,4286.} 
\label{fig:n4274_image_src_6.pdf}
\end{figure}
\begin{figure} 
\centering 
\includegraphics[width = 8 cm]{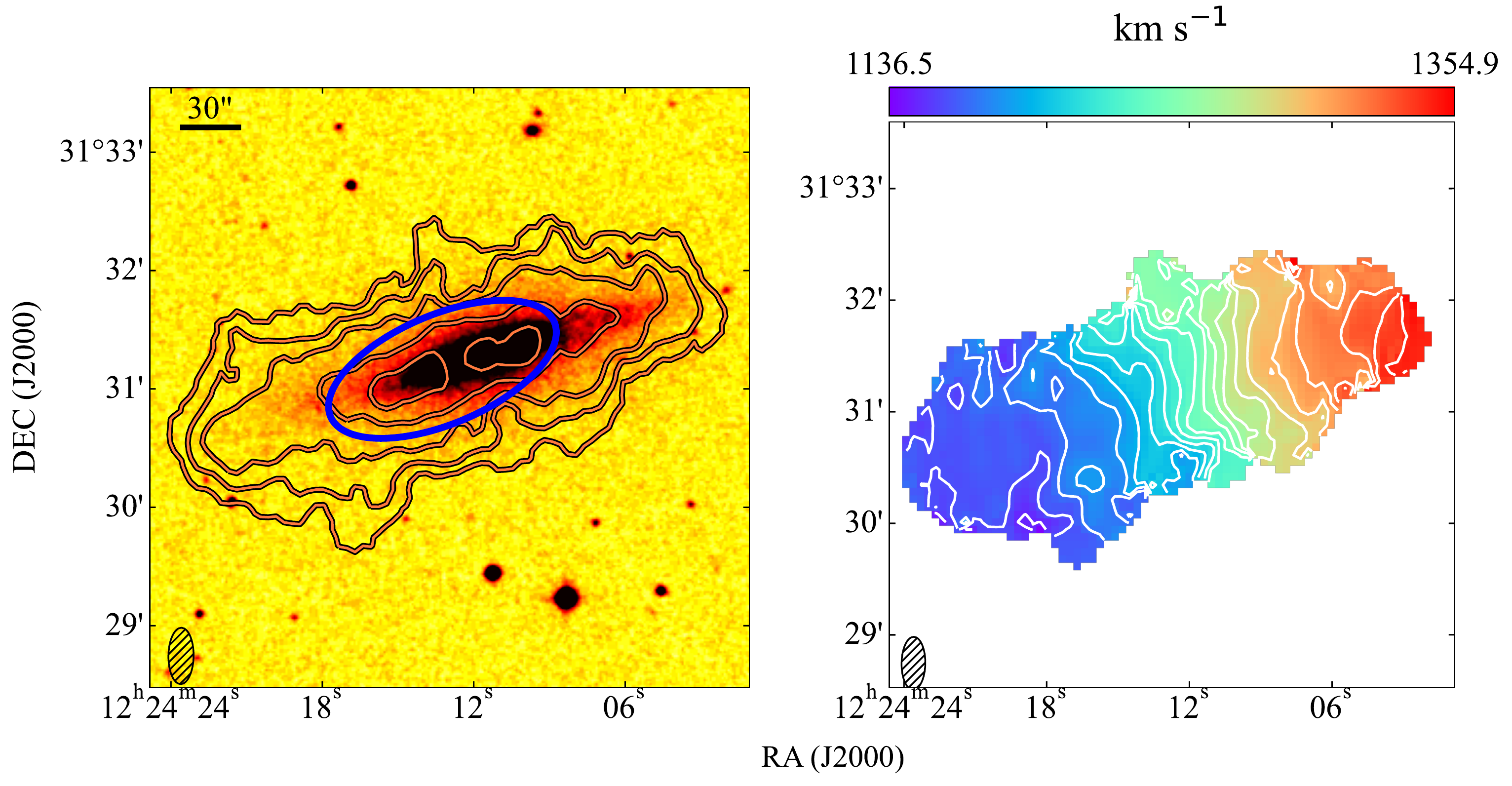} 
\caption{ As \ref{fig:n0925_image_src_4.pdf} but for source no. 7 in the cube of NGC\,4414. Contours for moment 0 are: 1.2, 4.8, 9.6, 19.2, 28.8$\times 10^{20} {\rm cm}^{-2}$ but overlaid on a DDS 2 $R$-band image as it is outside the HALOSTARS field. Contours for the velocity field start at 1136.52 km\,s$^{-1}$ and increase with 10.92 km s$^{-1}$. This source is identified as NGC\,4359.} 
\label{fig:n4414_image_src_7.pdf}
\end{figure}
\begin{figure} 
\centering 
\includegraphics[width = 8 cm]{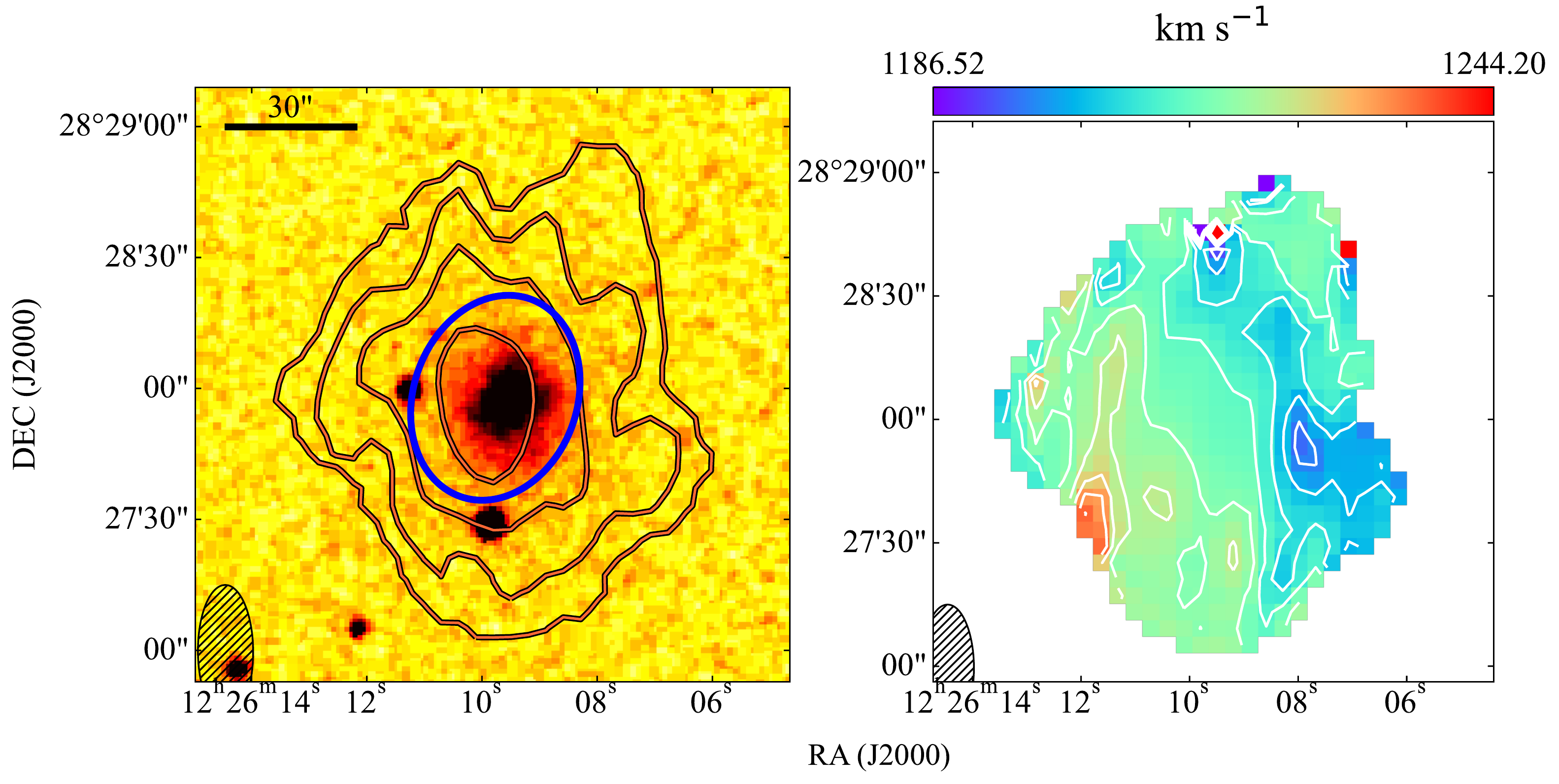} 
\caption{ As \ref{fig:n0925_image_src_4.pdf} but for source no. 2 in the cube of NGC\,4448. Contours for moment 0 are: 4.0, 15.9, 31.8, 63.6$\times 10^{19} {\rm cm}^{-2}$ but overlaid on a DDS 2 $R$-band image. Contours for the velocity field start at 1186.52 km\,s$^{-1}$ and increase with 5.00 km s$^{-1}$. This source is identified as IC\,3334.} 
\label{fig:n4448_image_src_2.pdf}
\end{figure}
\begin{figure} 
\centering 
\includegraphics[width = 8 cm]{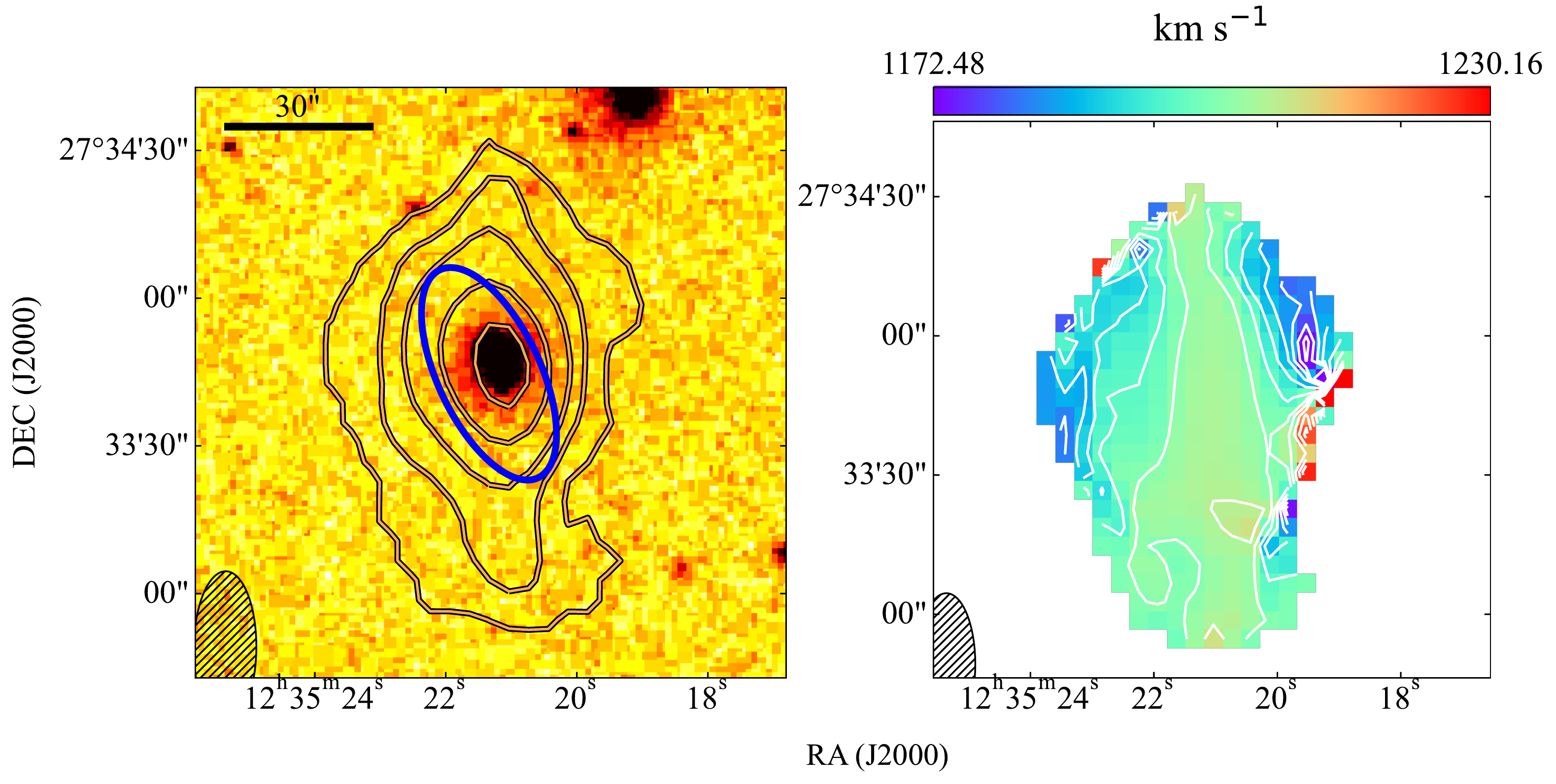} 
\caption{ As \ref{fig:n0925_image_src_4.pdf} but for source no. 2 in the cube of NGC\,4559. Contours for moment 0 are: 2.4, 9.4, 18.9, 37.7, 56.6$\times 10^{19} {\rm cm}^{-2}$ but overlaid on a DDS 2 $R$-band image as it is outside the KPNO field. Contours for the velocity field start at 1172.48 km\,s$^{-1}$ and increase with 5.00 km s$^{-1}$. This source is identified as WISEA\,J123521.10+273342.9.} 
\label{fig:n4559_image_src_2.pdf}
\end{figure}
\begin{figure} 
\centering 
\includegraphics[width = 8 cm]{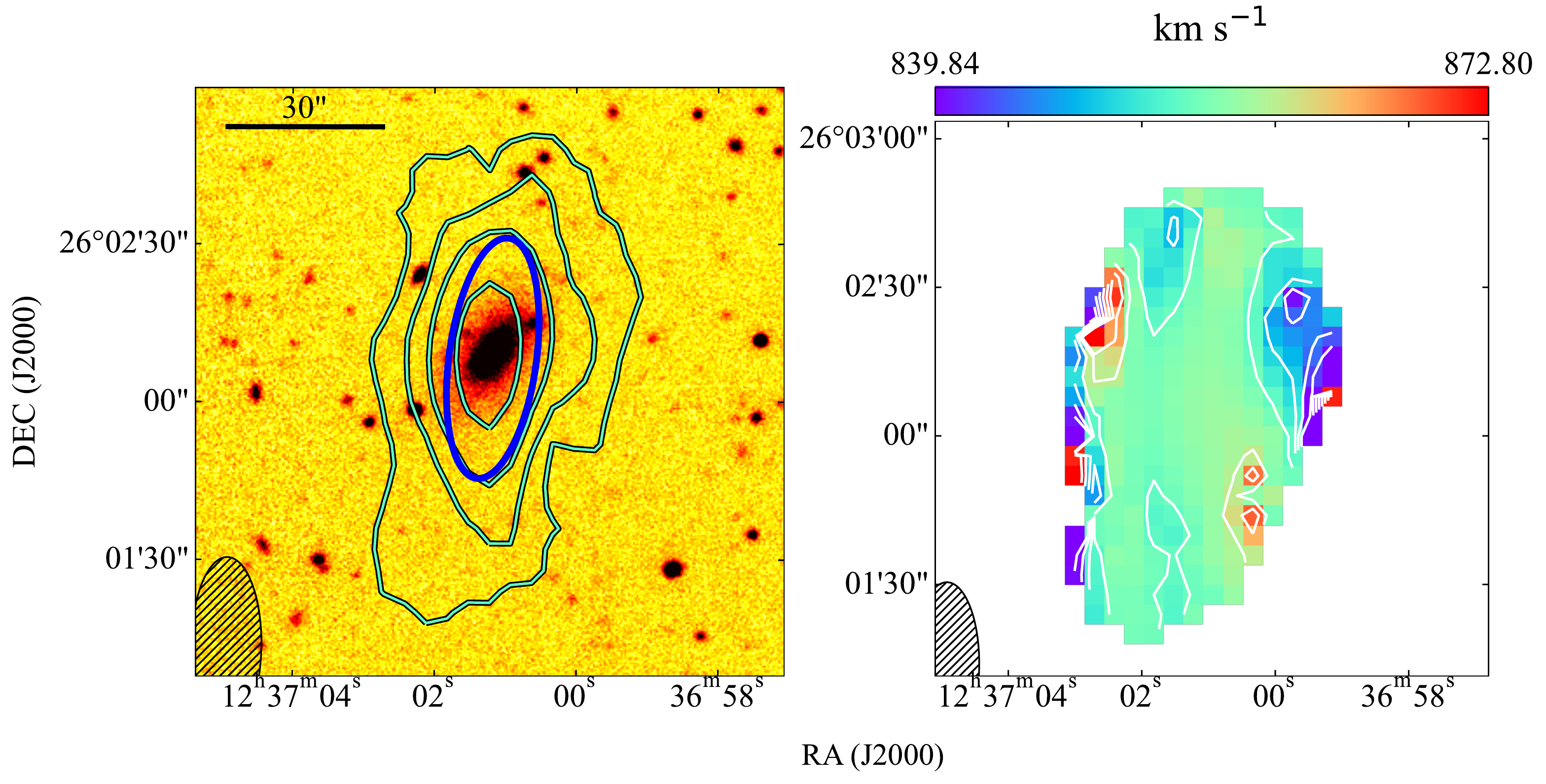} 
\caption{ As \ref{fig:n0925_image_src_4.pdf} but for source no. 1 in the cube of NGC\,4565. Contours for moment 0 are: 5.9, 23.6, 47.2, 94.3$\times 10^{18} {\rm cm}^{-2}$. Contours for the velocity field start at 839.84 km\,s$^{-1}$ and increase with 5.00 km s$^{-1}$.} 
\label{fig:n4565_image_src_1.pdf}
\end{figure}
\begin{figure} 
\centering 
\includegraphics[width = 8 cm]{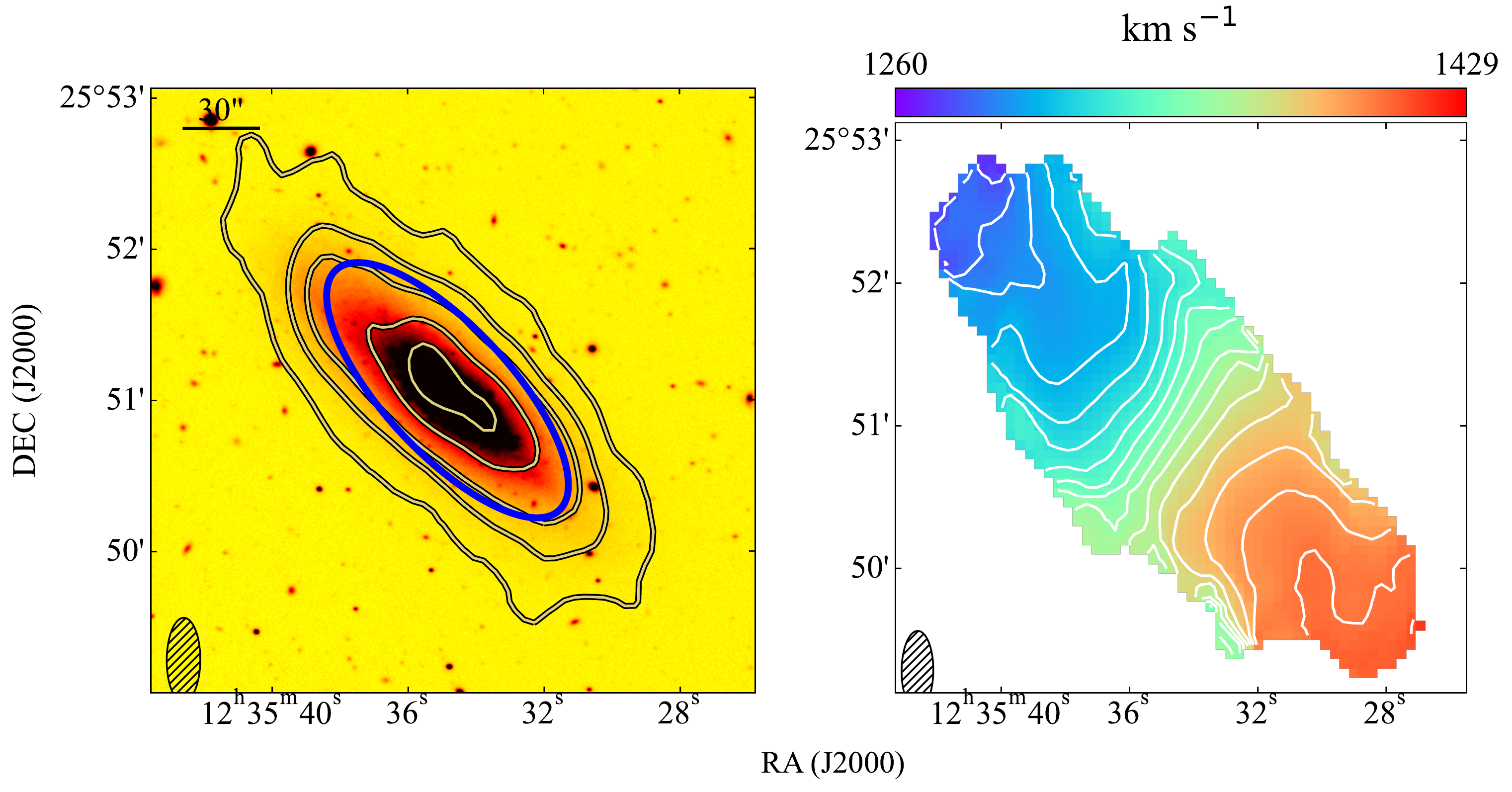} 
\caption{ As \ref{fig:n0925_image_src_4.pdf} but for source no. 4 in the cube of NGC\,4565. Contours for moment 0 are: 6.7, 26.8, 53.6, 107.1, 160.7$\times 10^{19} {\rm cm}^{-2}$. Contours for the velocity field start at 1260.08 km\,s$^{-1}$ and increase with 8.45 km s$^{-1}$. This source is identified as NGC\,4562.} 
\label{fig:n4565_image_src_4.pdf}
\end{figure}
\begin{figure} 
\centering 
\includegraphics[width = 8 cm]{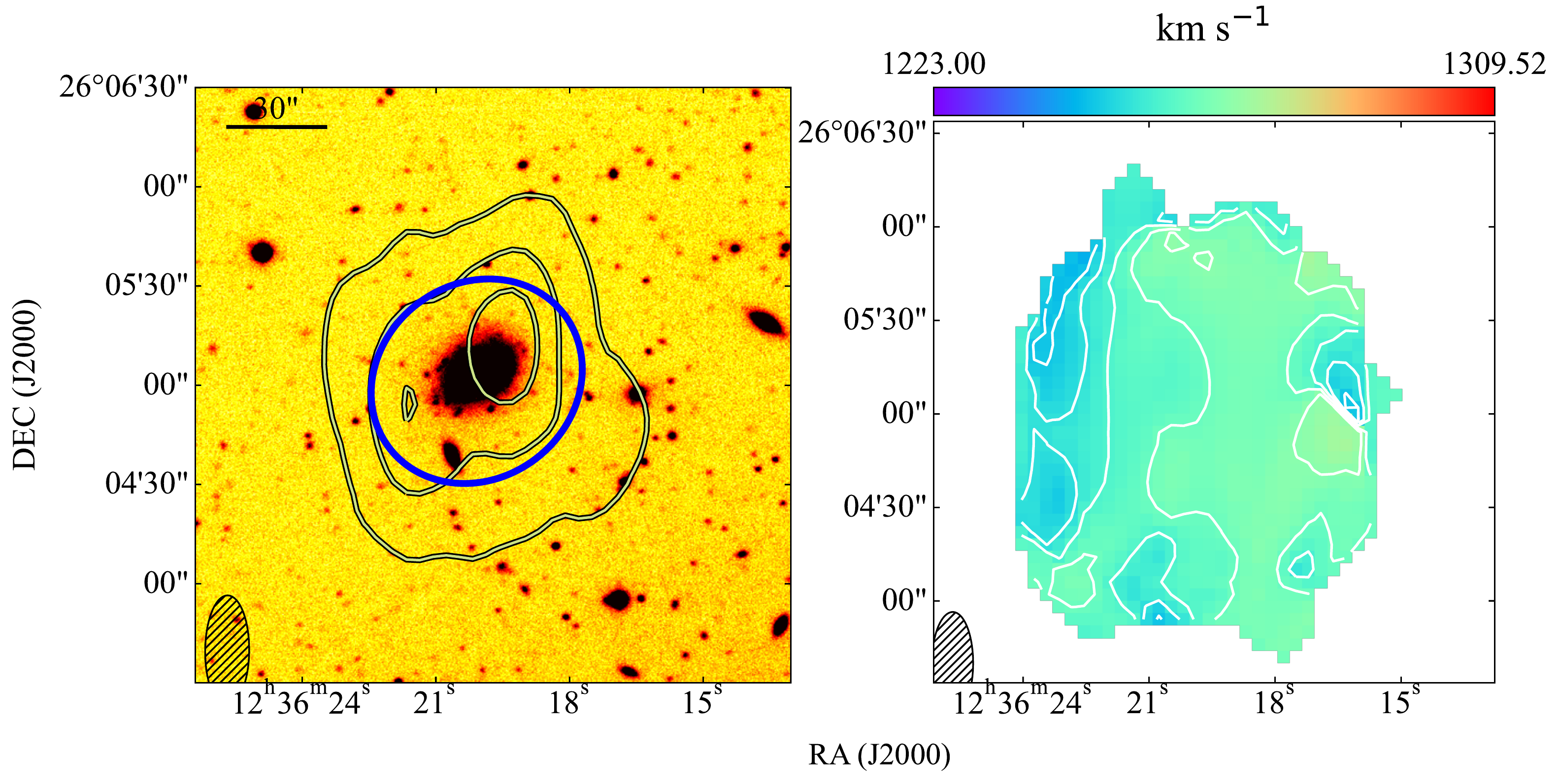} 
\caption{ As \ref{fig:n0925_image_src_4.pdf} but for source no. 5 in the cube of NGC\,4565. Contours for moment 0 are: 5.4, 21.6, 43.3$\times 10^{19} {\rm cm}^{-2}$. Contours for the velocity field start at 1223.00 km\,s$^{-1}$ and increase with 5.00 km s$^{-1}$. This source is identified as IC\,3571.} 
\label{fig:n4565_image_src_5.pdf}
\end{figure}
\begin{figure} 
\centering 
\includegraphics[width = 8 cm]{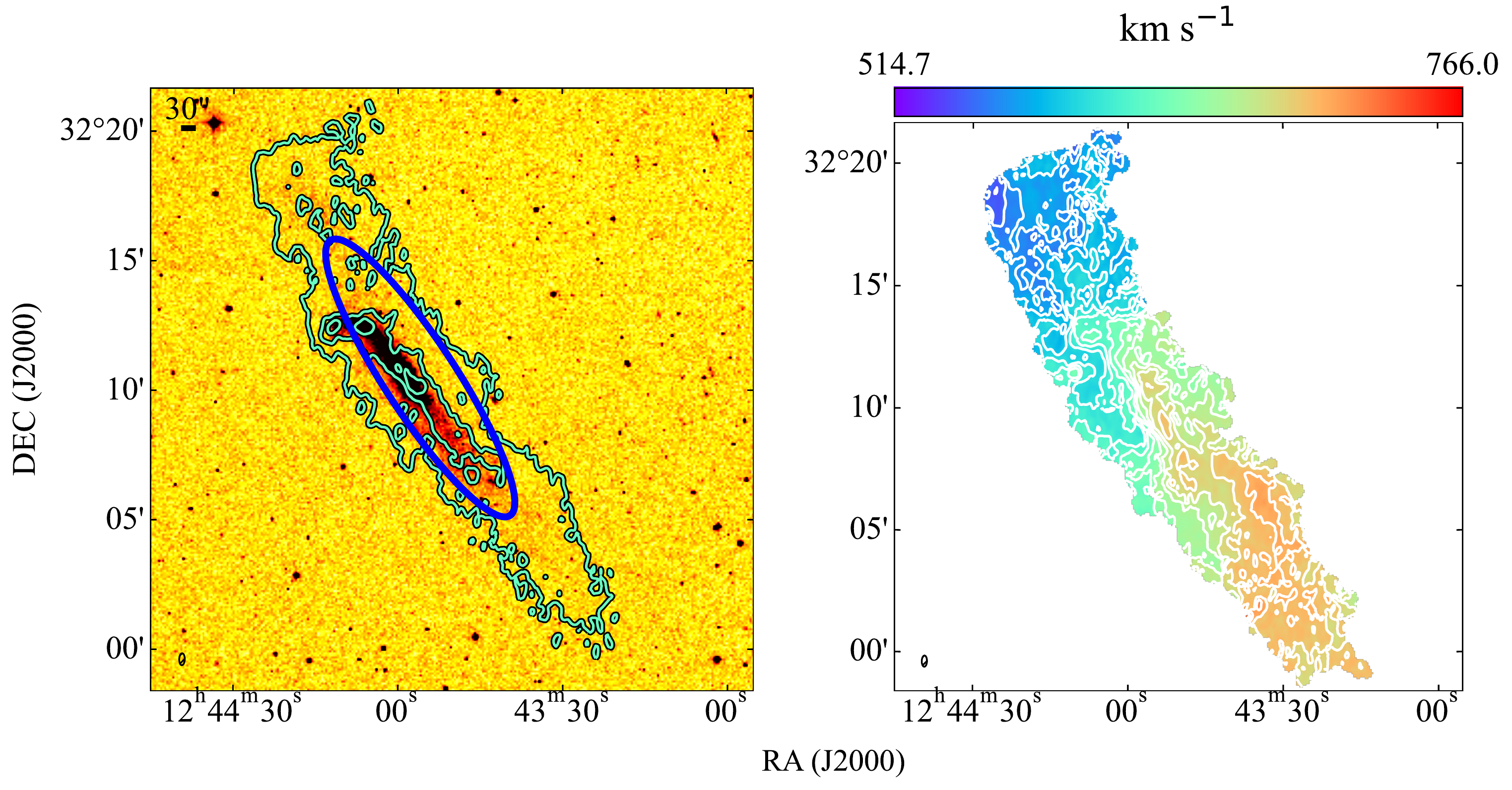} 
\caption{ As \ref{fig:n0925_image_src_4.pdf} but for source no. 2 in the cube of NGC\,4631. Contours for moment 0 are: 7.1, 28.6, 57.2$\times 10^{20} {\rm cm}^{-2}$ but overlaid on a DDS 2 $R$-band image as it is outside the KPNO field. Contours for the velocity field start at 514.72 km\,s$^{-1}$ and increase with 12.57 km s$^{-1}$. This source is identified as NGC\,4656.} 
\label{fig:n4631_image_src_2.pdf}
\end{figure}
\begin{figure} 
\centering 
\includegraphics[width = 8 cm]{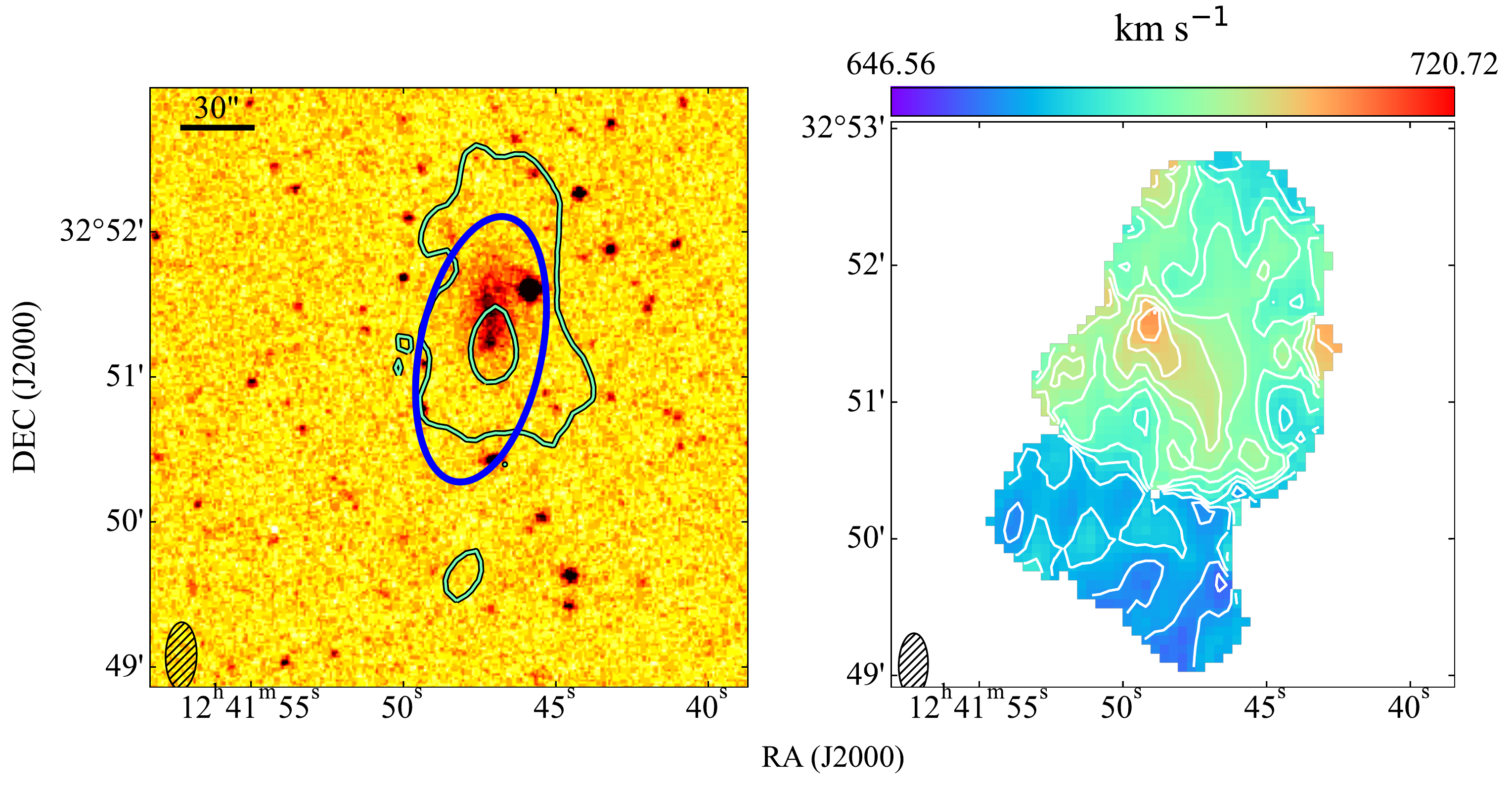} 
\caption{ As \ref{fig:n0925_image_src_4.pdf} but for source no. 6 in the cube of NGC\,4631. Contours for moment 0 are: 1.4, 5.4$\times 10^{20} {\rm cm}^{-2}$ but overlaid on a DDS 2 $R$-band image as it is outside the KPNO field. Contours for the velocity field start at 646.56 km\,s$^{-1}$ and increase with 5.00 km s$^{-1}$. This source is identified as SDSS\,J124146.99+325124.8.} 
\label{fig:n4631_image_src_6.pdf}
\end{figure}
\begin{figure} 
\centering 
\includegraphics[width = 8 cm]{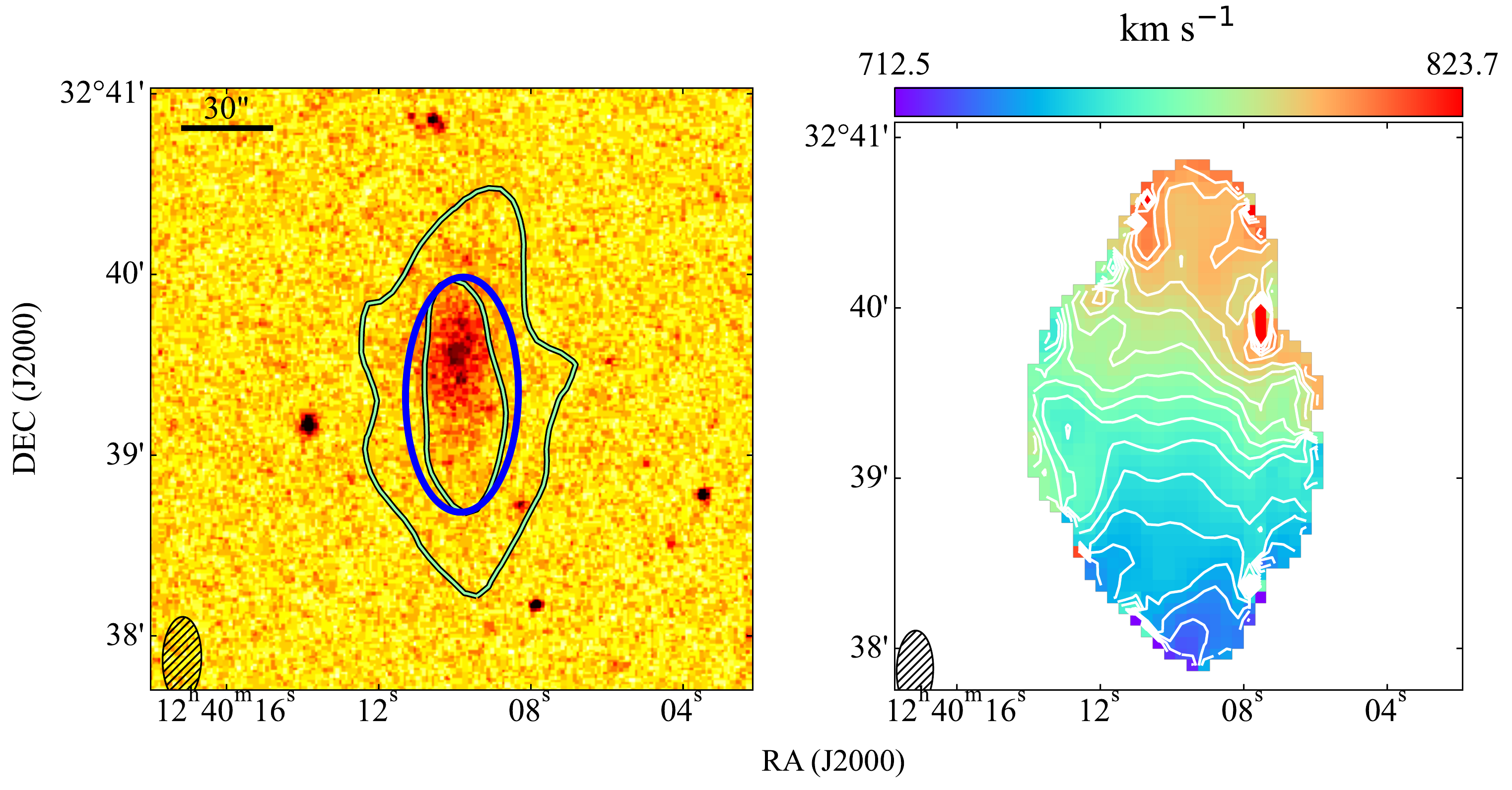} 
\caption{ As \ref{fig:n0925_image_src_4.pdf} but for source no. 9 in the cube of NGC\,4631. Contours for moment 0 are: 2.9, 11.5$\times 10^{20} {\rm cm}^{-2}$ but overlaid on a DDS 2 $R$-band image as it is outside the KPNO field. Contours for the velocity field start at 712.48 km\,s$^{-1}$ and increase with 5.56 km s$^{-1}$. This source is identified as SDSS\,J124010.08+323930.4.} 
\label{fig:n4631_image_src_9.pdf}
\end{figure}
\begin{figure} 
\centering 
\includegraphics[width = 8 cm]{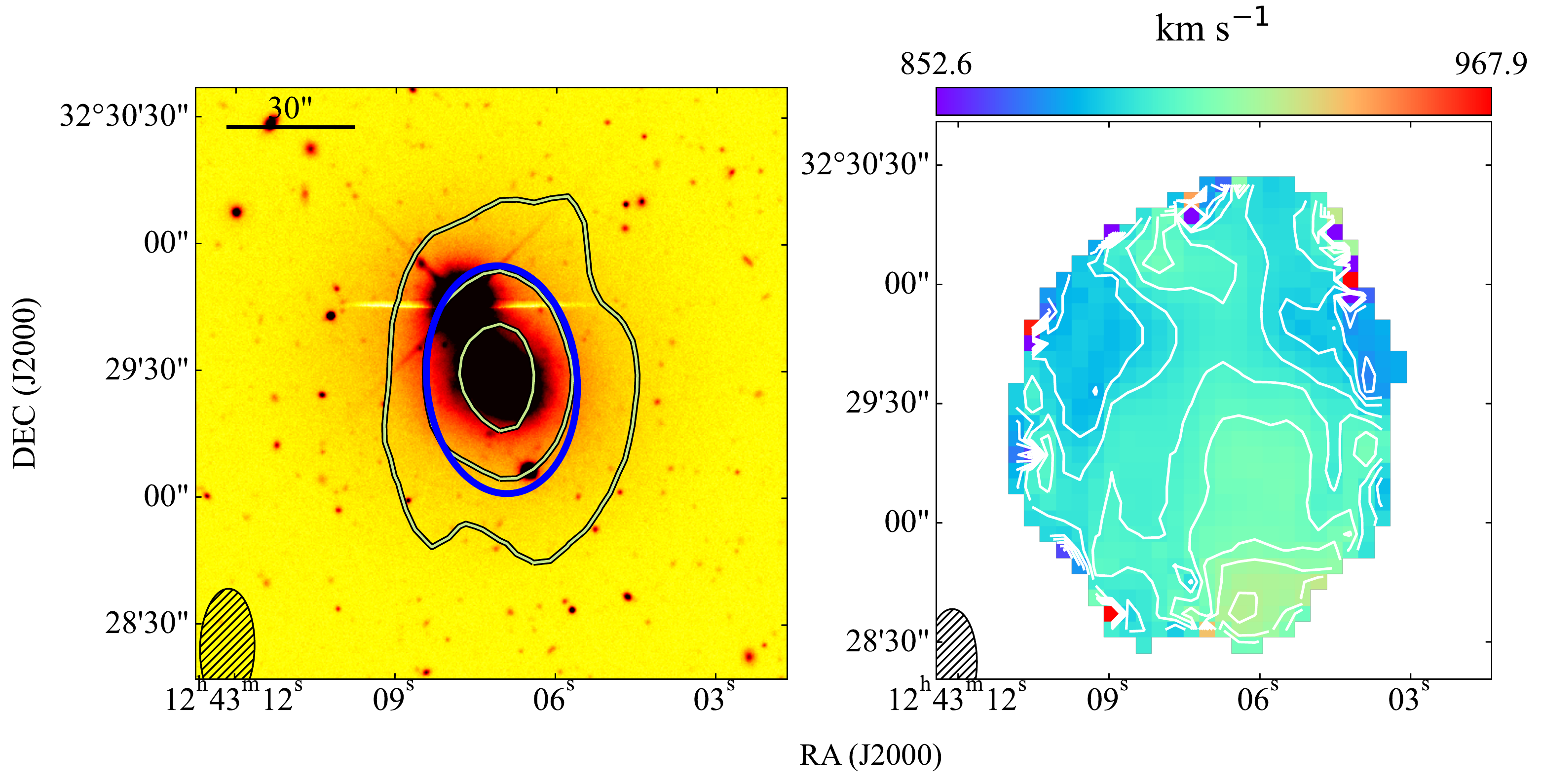} 
\caption{ As \ref{fig:n0925_image_src_4.pdf} but for source no. 10 in the cube of NGC\,4631. Contours for moment 0 are: 10.0, 40.0, 79.9$\times 10^{19} {\rm cm}^{-2}$ but overlaid on our KPNO $R$-band image. Contours for the velocity field start at 852.56 km\,s$^{-1}$ and increase with 5.77 km s$^{-1}$. This source is identified as MCG\,+06-28-022.} 
\label{fig:n4631_image_src_10.pdf}
\end{figure}
\begin{figure} 
\centering 
\includegraphics[width = 8 cm]{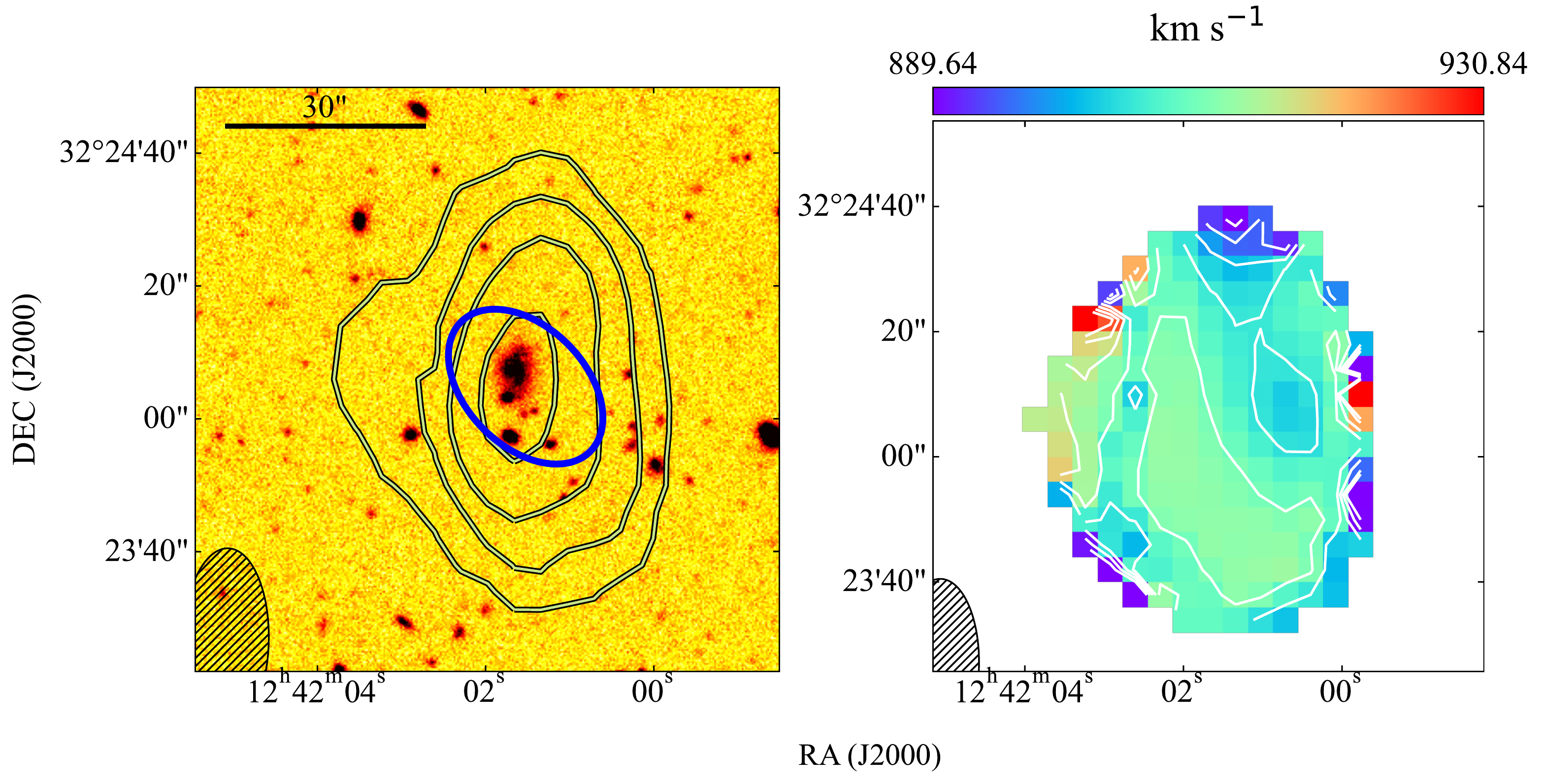} 
\caption{ As \ref{fig:n0925_image_src_4.pdf} but for source no. 11 in the cube of NGC\,4631. Contours for moment 0 are: 7.3, 29.1, 58.2, 116.3$\times 10^{18} {\rm cm}^{-2}$ but overlaid on our KPNO $R$-band image. Contours for the velocity field start at 889.64 km\,s$^{-1}$ and increase with 5.00 km s$^{-1}$.} 
\label{fig:n4631_image_src_11.pdf}
\end{figure}
\begin{figure} 
\centering 
\includegraphics[width = 8 cm]{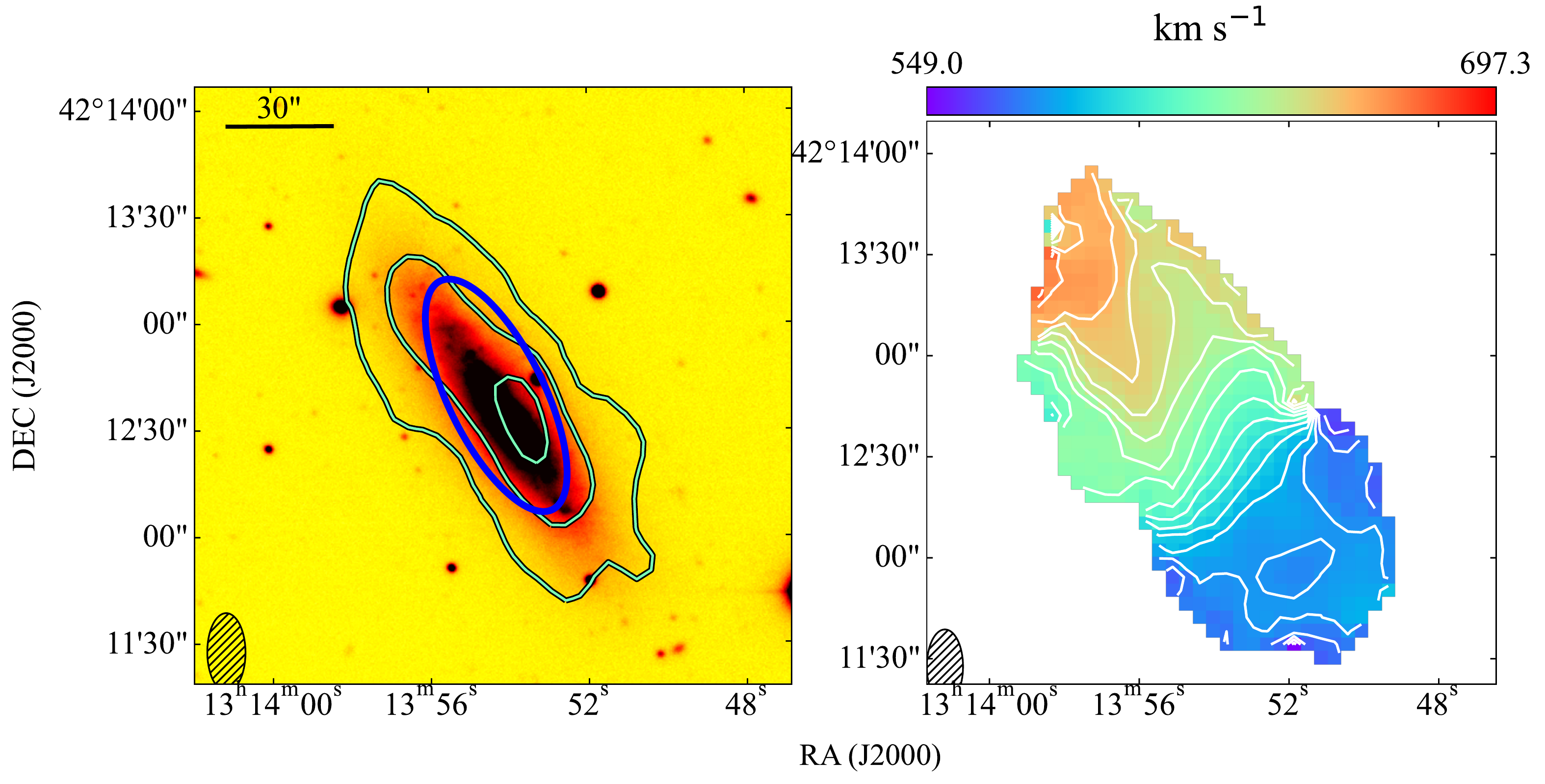} 
\caption{ As \ref{fig:n0925_image_src_4.pdf} but for source no. 4 in the cube of NGC\,5055. Contours for moment 0 are: 3.2, 12.9, 25.9$\times 10^{20} {\rm cm}^{-2}$. Contours for the velocity field start at 549.00 km\,s$^{-1}$ and increase with 7.42 km s$^{-1}$. This source is identified as UGC\,8313.} 
\label{fig:n5055_image_src_4.pdf}
\end{figure}
\begin{figure} 
\centering 
\includegraphics[width = 8 cm]{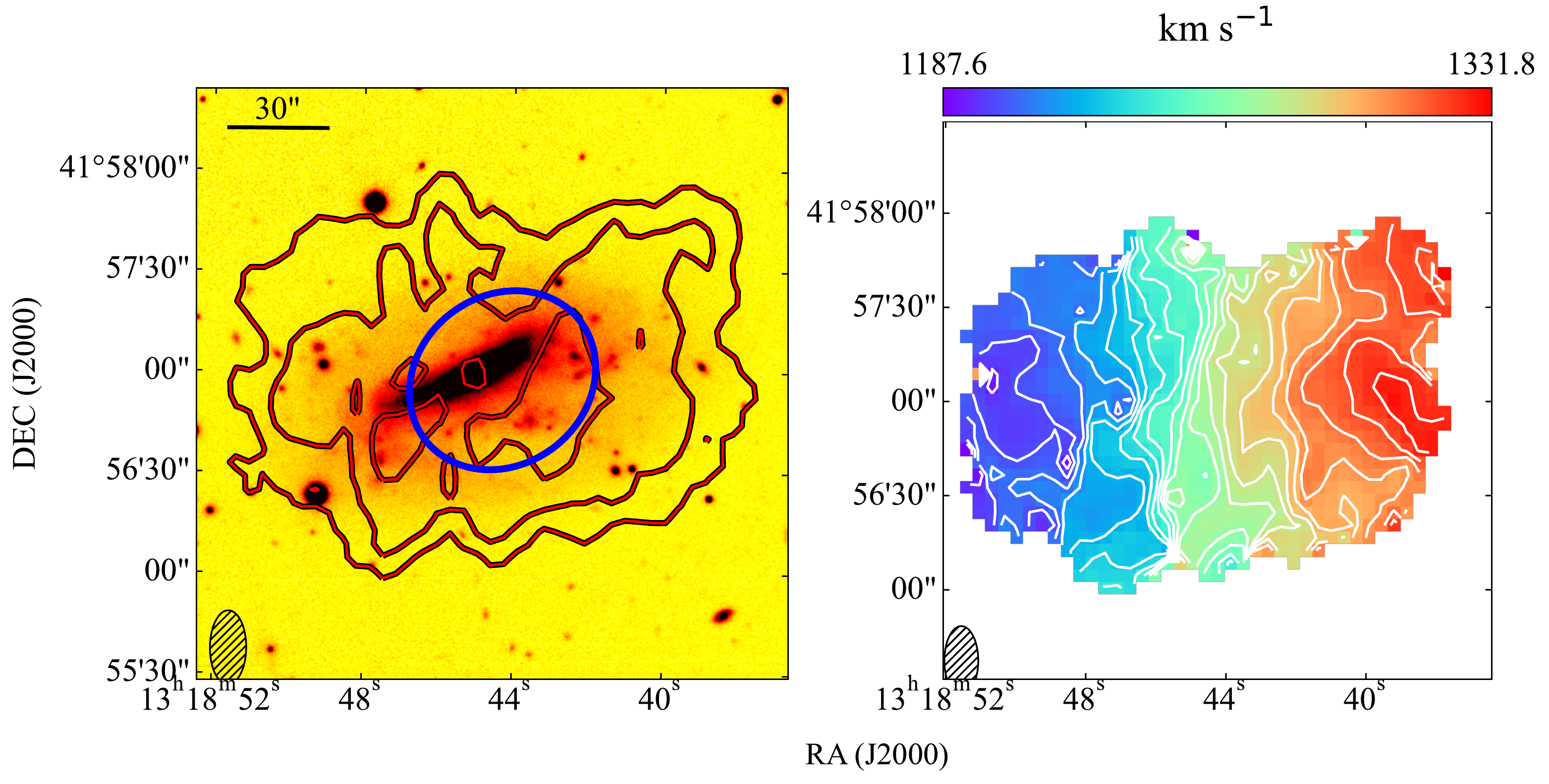} 
\caption{ As \ref{fig:n0925_image_src_4.pdf} but for source no. 6 in the cube of NGC\,5055. Contours for moment 0 are: 1.3, 5.3, 10.7$\times 10^{20} {\rm cm}^{-2}$. Contours for the velocity field start at 1187.60 km\,s$^{-1}$ and increase with 7.21 km s$^{-1}$. This source is identified as UGC\,8365.} 
\label{fig:n5055_image_src_6.pdf}
\end{figure}
\begin{figure} 
\centering 
\includegraphics[width = 8 cm]{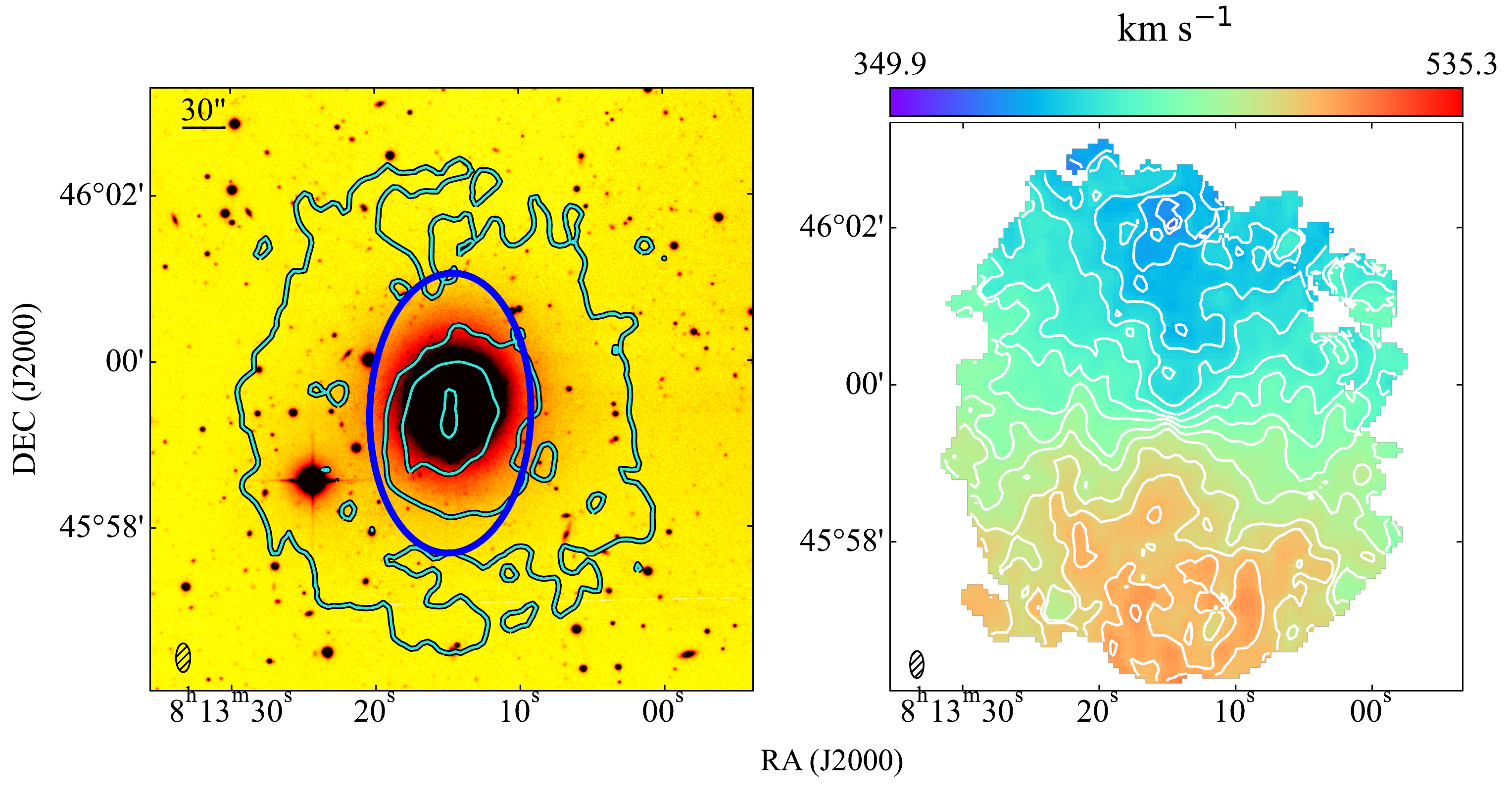} 
\caption{ As \ref{fig:n0925_image_src_4.pdf} but for source no. 4 in the cube of UGC\,4278. Contours for moment 0 are: 1.3, 5.1, 10.2$\times 10^{20} {\rm cm}^{-2}$. Contours for the velocity field start at 349.88 km\,s$^{-1}$ and increase with 9.27 km s$^{-1}$. This source is identified as NGC\,2537.} 
\label{fig:u4278_image_src_4.pdf}
\end{figure}
\begin{figure} 
\centering 
\includegraphics[width = 8 cm]{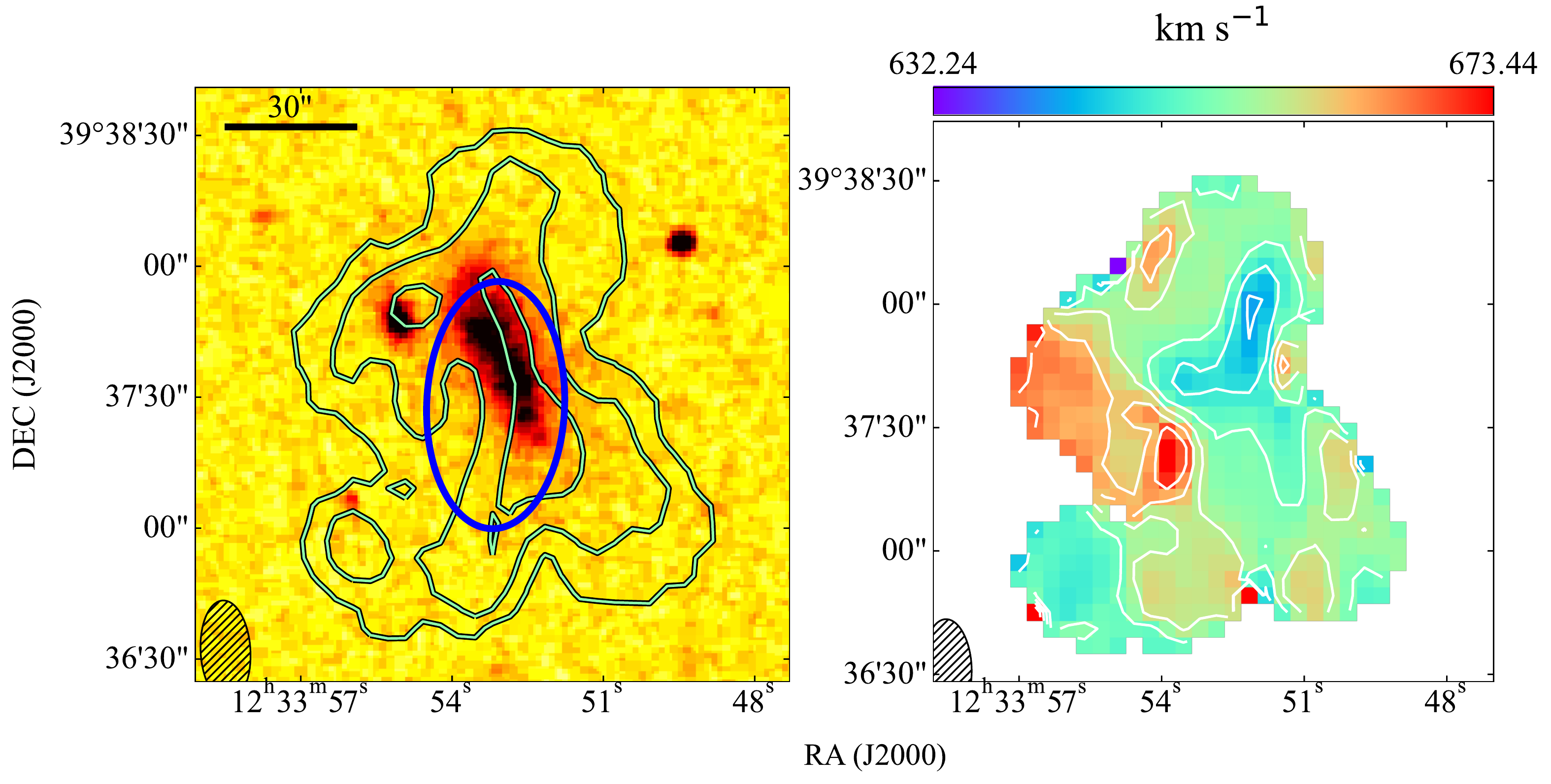} 
\caption{ As \ref{fig:n0925_image_src_4.pdf} but for source no. 2 in the cube of UGC\,7774. Contours for moment 0 are: 2.2, 8.9, 17.8$\times 10^{20} {\rm cm}^{-2}$ but overlaid on a DDS 2 $R$-band image. Contours for the velocity field start at 632.24 km\,s$^{-1}$ and increase with 5.00 km s$^{-1}$. This source is identified as MCG\,+07-26-024.} 
\label{fig:u7774_image_src_2.pdf}
\end{figure}
\end{appendix}
\end{document}